\pdfoutput=1
\PassOptionsToPackage{unicode}{hyperref}
\PassOptionsToPackage{hyphens}{url}
\documentclass[
]{article}
\usepackage{amsmath,amssymb}
\usepackage{iftex}
\ifPDFTeX
  \usepackage[T1]{fontenc}
  \usepackage[utf8]{inputenc}
  \usepackage{textcomp} 
\else 
  \usepackage{unicode-math} 
  \defaultfontfeatures{Scale=MatchLowercase}
  \defaultfontfeatures[\rmfamily]{Ligatures=TeX,Scale=1}
\fi
\usepackage{lmodern}
\ifPDFTeX\else
\fi
\IfFileExists{upquote.sty}{\usepackage{upquote}}{}
\IfFileExists{microtype.sty}{
  \usepackage[]{microtype}
  \UseMicrotypeSet[protrusion]{basicmath} 
}{}
\makeatletter
\@ifundefined{KOMAClassName}{
  \IfFileExists{parskip.sty}{%
    \usepackage{parskip}
  }{
    \setlength{\parindent}{0pt}
    \setlength{\parskip}{6pt plus 2pt minus 1pt}}
}{
  \KOMAoptions{parskip=half}}
\makeatother
\usepackage{xcolor}
\usepackage{graphicx}
\makeatletter
\def\maxwidth{\ifdim\Gin@nat@width>\linewidth\linewidth\else\Gin@nat@width\fi}
\def\maxheight{\ifdim\Gin@nat@height>\textheight\textheight\else\Gin@nat@height\fi}
\makeatother
\setkeys{Gin}{width=\maxwidth,height=\maxheight,keepaspectratio}
\makeatletter
\def\fps@figure{htbp}
\makeatother
\usepackage{amsmath,amssymb}

\usepackage{booktabs}
\usepackage{longtable}
\usepackage{tabularx}
\usepackage{adjustbox}
\usepackage{pdflscape}
\usepackage{makecell}

\usepackage{etoolbox}

\usepackage{graphicx}
\usepackage{float}

\usepackage{caption}
\usepackage{hyperref}
\usepackage{xurl}  
\hypersetup{colorlinks=true, linkcolor=blue!60!black, citecolor=blue!60!black, urlcolor=blue!60!black}

\usepackage{geometry}
\usepackage{fancyhdr}
\usepackage{needspace}
\BeforeBeginEnvironment{longtable}{\needspace{8\baselineskip}}
\BeforeBeginEnvironment{figure}{\needspace{5\baselineskip}}
\makeatletter
\def\fps@figure{H}
\makeatother

\usepackage{fancyvrb}

\usepackage{indentfirst}
\BeforeBeginEnvironment{longtable}{\noindent}
\BeforeBeginEnvironment{tabular}{\noindent}
\BeforeBeginEnvironment{tabularx}{\noindent}
\BeforeBeginEnvironment{table}{\noindent}
\BeforeBeginEnvironment{figure}{\noindent}
\BeforeBeginEnvironment{verbatim}{\noindent}
\BeforeBeginEnvironment{Verbatim}{\noindent}

\DeclareUnicodeCharacter{2192}{\ensuremath{\rightarrow}}
\DeclareUnicodeCharacter{2194}{\ensuremath{\leftrightarrow}}
\DeclareUnicodeCharacter{2212}{\ensuremath{-}}
\DeclareUnicodeCharacter{03C1}{\ensuremath{\rho}}
\ifLuaTeX
  \usepackage{selnolig}  
\fi
\usepackage{bookmark}
\IfFileExists{xurl.sty}{\usepackage{xurl}}{} 
\hypersetup{
  pdftitle={Mechanism-Resolved Interface Momentum Transfer in Immersed-Boundary Lattice Boltzmann Simulations},
  pdfauthor={Hongju Jo},
  pdfsubject={Physics of Fluids --- Mechanism-Resolved IB-LBM Simulations},
  pdfkeywords={immersed boundary method; lattice Boltzmann method; interface momentum transfer; distribution-function
correction; particle sedimentation; differential-density wake-interaction; collision model; central-moment
multiple-relaxation-time},
  hidelinks,
  pdfcreator={LaTeX via pandoc}}

\title{Mechanism-Resolved Interface Momentum Transfer in Immersed-Boundary Lattice Boltzmann Simulations}
\author{Hongju Jo}
\date{}

\begin{document}
\maketitle

\noindent Independent Researcher, Seoul, Republic of Korea\\
\noindent \emph{a)} Author to whom correspondence should be addressed: jhjoo3217@yonsei.ac.kr

\smallskip

\noindent
{\raggedright The following article has been accepted by \emph{Physics of Fluids}.
After it is published, it will be found at
\mbox{\url{https://doi.org/10.1063/5.0336036}}.\par}

\subsection{Abstract}\label{abstract}

Immersed-boundary lattice Boltzmann (IB-LBM) simulations of cylinder and particle flows are usually compared by ranking
boundary-enforcement schemes, kernels, and collision models. Such comparisons can obscure the mechanism that controls
each response. Here, canonical fixed-cylinder, oscillating-cylinder, and sedimenting-particle cases are used as
controlled probes of Eulerian--Lagrangian interface momentum transfer. For prescribed bodies, differences among direct
forcing (DF), multi-direct forcing (MDF), and distribution-function correction (DFC) are most clearly discriminated by
local no-slip fidelity rather than by a universal drag ranking. For DFC, the kernel-dependent drag-coefficient ranking
reversal between the hat and Peskin 4-point kernels is associated with the spatial redistribution of the marker-resolved
correction and the resulting near-boundary slip and pressure deviation, rather than the total correction magnitude
alone. For sedimenting particles, single-particle settling provides a moving-body baseline, whereas two-particle
differential-density wake-interaction sedimentation shows the wake-exposed light particle to be comparatively more
sensitive to the explicit internal-mass correction in the force evaluation, a configuration-dependent, finite-window
moving-boundary closure response. Targeted two-relaxation-time (TRT) and central-moment multiple-relaxation-time
(CM-MRT) collision controls remain secondary to the boundary and closure mechanisms within the tested regimes
(prescribed-body comparisons at \(\mathrm{Re} \leq 200\), two-dimensional, and the reported moving-particle cases). The
resulting picture replaces a universal scheme ranking with a mechanism-resolved interpretation of IB-LBM interface
momentum transfer.

\noindent \textbf{Keywords}: immersed boundary method; lattice Boltzmann method; interface momentum transfer;
distribution-function correction; particle sedimentation; differential-density wake-interaction; collision model;
central-moment multiple-relaxation-time

\subsection{1. Introduction}\label{introduction}

The immersed boundary method (IBM), first proposed by Peskin {[}1{]} for cardiac flow simulations, has become a standard
technique for handling complex and moving geometries on fixed Cartesian grids. By decoupling the fluid domain from the
solid boundary, IBM avoids the computational expense of body-conforming mesh generation while accommodating arbitrary
geometries and large structural deformations. Reviews of the method's development and variants are provided by Mittal
and Iaccarino {[}2{]}, Peskin {[}3{]}, Sotiropoulos and Yang {[}4{]}, Huang and Tian {[}5{]}, and Verzicco {[}6{]}.

IBM has been widely combined with the lattice Boltzmann method (LBM), whose data-local collision--streaming algorithm
maps naturally to GPU architectures {[}7,8{]}. As a mesoscopic solver derived from kinetic theory {[}9{]}, the LBM
evolves discrete distribution functions on a regular lattice rather than solving the Navier--Stokes equations directly,
which simplifies both parallelization and boundary treatment. Within the IB-LBM framework, regularized delta functions
couple the two representations: fluid quantities are sampled at Lagrangian boundary points, and the resulting boundary
forces are returned to the Eulerian lattice. Early IB-LBM formulations were developed by Feng and Michaelides
{[}10,11{]} and Niu et al.~{[}7{]}. Lai and Peskin {[}12{]} established formal second-order accuracy for IBM with
reduced numerical viscosity, providing a theoretical reference point for the convergence analysis used in this study.

Among the various IBM formulations, a primary distinction exists between penalty-based (feedback forcing) approaches
{[}13{]} and direct computation of the boundary force. In the \emph{direct forcing} (DF) approach {[}14{]}, the boundary
force at each Lagrangian marker is obtained in a single step from the difference between the desired and interpolated
velocities. This one-pass cycle --- interpolate, compute force, spread --- is the lowest-cost variant among the three
families considered here. \emph{Multi-direct forcing} (MDF) {[}15,16{]} re-applies the interpolation--force--spreading
cycle \(N\) times within each time step, so that each successive pass corrects the residual mismatch left by the
previous one. The cumulative force converges toward the value that would enforce exact no-slip, with diminishing
increments as \(N\) increases. Breugem {[}17{]} combined multi-direct forcing with a retraction of the Lagrangian grid
toward the particle interior, which raises the spatial order of grid convergence of the direct-forcing IBM from first to
second order. \emph{Distribution-function correction} (DFC) {[}18{]} takes a different route: instead of adding a
macroscopic body force, it adjusts the post-streaming distribution functions at boundary-adjacent nodes so that their
moments recover the desired wall velocity. The target state is derived from a bounce-back relation evaluated at the
Lagrangian markers, and a single correction pass redistributes the resulting difference back to the Eulerian lattice.
This distribution-level correction is conceptually akin to the partially saturated cells approach of Noble and
Torczynski {[}19{]}, where the boundary condition is embedded in the distribution functions rather than applied as an
external force.

Despite the growing adoption of these methods, systematic comparisons under identical numerical conditions are rare.
Most studies introduce a new method and compare it against one predecessor, often with different grid resolutions,
domain sizes, or flow conditions. Kang and Hassan {[}20{]} provided a systematic comparison of direct-forcing IB-LBM
variants for stationary boundaries, but their scope was confined to direct-forcing interface schemes --- including an
implicit diffuse scheme that is exactly the iterative multi-direct forcing of Wang et al.~{[}16{]} (with the number of
forcings NF up to 20) --- and did not extend to distribution-function correction (DFC) or to freely moving particles.
Since then, the individual methods have continued to develop: MDF has been extended with second-order spatial accuracy
and grid retraction {[}17{]}; DFC has been applied to thermal flows {[}21{]} and further simplified for computational
efficiency {[}22{]}; recent assessments of direct-forcing IB-LBM have systematically quantified spatial accuracy and
spurious force characteristics {[}23{]}; and broader surveys of LBM and IB-LBM applications in fluid--structure
interaction are provided by Wang, Liu, and Rajamuni {[}24{]}. Parallel developments span largely orthogonal directions:
refined direct-forcing variants such as the accelerated multi-direct forcing of Suzuki, Falagkaris, Krüger, and Inamuro
{[}25{]} that aim to suppress local slip and internal residual at the immersed interface, adaptive octree
grid-refinement combined with MRT collision {[}26{]} and GPU-accelerated implementations {[}27{]} that extend the
accessible parameter range, central-moment and cumulant collision operators {[}28{]} that target the relaxation-side
response, and applications to turbulent fluid--structure interaction {[}29{]}. Within the slip-suppression axis, Gsell
and Favier {[}30{]} traced the boundary slip error of direct-forcing schemes to the non-reciprocity of the interpolation
and spreading operators, showed that it scales with the Courant number in explicit implementations, and proposed an a
priori rescaling of the IB force that suppresses slip and penetration errors at no additional computational cost. The
marker-level no-slip residual quantified throughout the present work therefore remains an actively studied accuracy
metric. Rather than extending any of these axes individually, the present work organizes the canonical benchmark cases,
under matched numerical conditions, as controlled probes of the momentum-transfer mechanism that each direction is
intended to address.

A complicating factor is the LBM collision model. Even for the same IBM formulation, the choice among
Bhatnagar--Gross--Krook / single-relaxation-time (BGK / SRT) {[}9{]}, two-relaxation-time (TRT) {[}31{]}, and
multiple-relaxation-time (MRT) {[}32{]} collision operators can noticeably affect accuracy and stability, particularly
at moderate-to-high Reynolds numbers. Lallemand and Luo {[}33{]} demonstrated that the single relaxation time in BGK
inherently couples hydrodynamic and non-hydrodynamic mode damping, producing viscosity-dependent errors that grow as
\(\tau\) approaches the stability limit. Luo et al.~{[}34{]} showed that both TRT and MRT outperform BGK across the
canonical benchmarks they tested. A recent discrete-setting asymptotic analysis by Wissocq and Sagaut {[}35{]} refines
this picture: separating consistency errors inherited from the kinetic description from purely numerical discretization
errors, they relate the low dissipation of BGK to the structure of its Taylor expansion, attribute the over-dissipation
and instabilities of regularized and fixed-relaxation MRT models in the low-viscosity regime to a hyperviscous
degeneracy, and find no a priori advantage of one moment basis over another. The collision operator therefore remains an
accuracy-relevant choice even for well-resolved flows. Following these observations, the present manuscript treats
SRT-BGK as a primary baseline against which TRT and central-moment MRT (CM-MRT) controls are evaluated for the same IBM
formulation, holding the boundary-enforcement scheme, kernel, and discretization fixed.

A second cross-cutting axis is how each IB-LBM scheme treats the \emph{internal-mass / fictitious-fluid} contribution
that arises because the boundary-only IBM force omits the fictitious-fluid mass inside a moving solid. Suzuki and
Inamuro {[}36{]} systematize the available treatments through a four-scheme taxonomy. The implementation of this
taxonomy in the present three-way IB-LBM benchmark --- which schemes are evaluated, which serves as the reference
baseline, and which lie outside scope --- is detailed in Sec.~2.7; it underpins the framing of the Sec.~5
single-particle / wake-interaction analysis and the per-scheme spread reported in Appendix B Table B.5.

The present work does not seek a universally best boundary-enforcement scheme. Its premise is that IB-LBM accuracy in
any given benchmark depends on which interface momentum-transfer mechanism the benchmark activates: prescribed bodies
test local no-slip fidelity at the boundary; the DFC kernel reversal exposes how the correction is redistributed at the
immersed interface; freely sedimenting particles reveal the wake-exposed internal-mass closure sensitivity; and the
collision-model controls probe and bound the relaxation-side response of the lattice operator. The same canonical cases
that are usually presented as a method-ranking benchmark are used here as controlled probes of these four mechanisms.
Each mechanism is supported by quantitative diagnostics, including local slip and leakage, the internal residual, the
marker-resolved correction magnitude, near-boundary pressure and velocity residuals, particle-bound acceleration
proxies, and case-specific collision spreads.

These mechanisms organize the contributions reported below: for prescribed bodies, the differences among DF, MDF, and
DFC are most clearly discriminated by local no-slip fidelity rather than by a universal drag ranking, and the
kernel-dependent sign reversal of DFC at \(\mathrm{Re}\geq 100\) is consistent with the spatial distribution of the
correction field as one signature of the redistribution mechanism (per-case \(\lambda_k\) non-uniformity and the
underlying anisotropic boundary geometry are documented in Sec. 4). For freely moving particles, the two-particle
differential-density wake-interaction configuration shows the wake-exposed light particle to be comparatively more
sensitive to the explicit internal-mass correction than the isolated single-particle baseline: removing the explicit
correction shifts the light-particle peak Reynolds number by \(-16.4\%\) relative to the explicit-history baseline (a
\(13.8\%\) underprediction of the Majumder reference value), whereas the within-run heavy-particle comparator remains at
or below \(3.0\%\) on every tested boundary-enforcement scheme, and two isolated-particle controls without two-particle
wake interaction record only small finite-window contrasts. Together with trajectory phase diagnostics, the
particle-bound acceleration proxy, and a Majumder-aligned cross-check, these controls strengthen the association with
the post-exchange two-particle configuration (Sec. 5.2). The signed contrast has the same direction in the DF / MDF /
DFC comparisons on both the \(50D\) and the \(60D\)-extended domains, and its magnitude attenuates on the longer domain
for every scheme (Appendix C Sec. C.2). Targeted TRT and CM-MRT controls under the same IBM framework, together with a
\(60D\)-extended sedimentation-channel domain check (Appendix C Sec. C.2), keep the collision-model spread secondary to
the boundary, correction, and closure-sensitivity contrasts within the tested regimes (prescribed-body cases at
\(\mathrm{Re} \leq 200\), two-dimensional, and the reported moving-particle cases) and reference-defined peak metrics
(Sec. 6).

Sec. 2 documents the numerical formulation and the local diagnostic framework. Sec. 3 reports the prescribed-body
regime. Sec. 4 reports the correction-dominated regime (DFC). Sec. 5 reports the moving-particle regime, separating the
single-particle baseline (Sec.~5.1) from the two-particle wake-interaction (wake-exposed closure-sensitivity) analysis
(Sec.~5.2). Sec. 6 reports the targeted collision-model controls. Sec. 7 synthesizes the four mechanisms, names the
limitations, and identifies future directions including the integration of implicit velocity correction (IVC) schemes
with the DF / MDF / DFC family compared here. Sec. 8 states the conclusions. Detailed numerical formulation, method ×
kernel × collision case matrices, grid and domain analyses, the Taylor--Green convergence study, and raw sedimentation
traces are compiled in Appendices A--E.

\subsection{2. Numerical Formulation and Diagnostics}\label{numerical-formulation-and-diagnostics}

\subsubsection{2.1 Lattice Boltzmann Method}\label{lattice-boltzmann-method}

The D2Q9 lattice Boltzmann equation with the BGK collision operator reads {[}9{]}

\[f_i(\mathbf{x} + \mathbf{e}_i \Delta t,\, t + \Delta t) = f_i(\mathbf{x}, t) - \frac{1}{\tau} \left[ f_i(\mathbf{x}, t) - f_i^{\mathrm{eq}}(\mathbf{x}, t) \right] + \Delta t \, F_i, \tag{1}\]

\noindent where \(f_i\) is the distribution function for velocity direction \(i\), \(\mathbf{e}_i\) the discrete velocity,
\(\tau\) the relaxation time related to the kinematic viscosity by \(\nu = c_s^2 (\tau - 1/2) \Delta t\), and \(F_i\)
the discrete forcing term. Eq. (1) is the general DnQq form, here specialized to the D2Q9 stencil
\(\{\mathbf{e}_i\}_{i=0}^{8}\) with \(\mathbf{e}_0 = \mathbf{0}\), \(\mathbf{e}_{1\text{–}4}\) axial, and
\(\mathbf{e}_{5\text{–}8}\) diagonal. The equilibrium distribution function is

\[f_i^{\mathrm{eq}} = w_i \rho \left[ 1 + \frac{\mathbf{e}_i \cdot \mathbf{u}}{c_s^2} + \frac{(\mathbf{e}_i \cdot \mathbf{u})^2}{2 c_s^4} - \frac{\mathbf{u} \cdot \mathbf{u}}{2 c_s^2} \right], \tag{2}\]

\par\needspace{6\baselineskip}

with weights \(w_0 = 4/9\), \(w_{i} = 1/9\) for \(i \in \{1,2,3,4\}\) (axial), \(w_{i} = 1/36\) for
\(i \in \{5,6,7,8\}\) (diagonal), and lattice speed of sound \(c_s = 1/\sqrt{3}\). The forcing term follows the
second-order Hermite scheme of Guo et al.~{[}37{]}:

\nopagebreak[4]\par

\[F_i = \left(1 - \frac{1}{2\tau}\right) w_i \left[ \frac{\mathbf{e}_i - \mathbf{u}}{c_s^2} + \frac{(\mathbf{e}_i \cdot \mathbf{u})}{c_s^4} \mathbf{e}_i \right] \cdot \mathbf{F}, \tag{3}\]

\noindent where \(\mathbf{F}\) is the body force returned by the IBM; the per-direction discrete forcing \(F_i\) that appears in
Eq. (1) is given by Eq. (3). Equation (3) is the BGK (single-\(\tau\)) forcing prefactor; under the two-relaxation-time
operator (Sec. 2.2) the symmetric and antisymmetric parts of the source relax with \((1 - s_e/2)\) and \((1 - s_o/2)\),
respectively, and under the central-moment operator it is recast in central-moment space as \(|\mathbf{R}\rangle\) in
Eq. (5). The macroscopic density and velocity are recovered as

\[\rho = \sum_i f_i, \qquad \rho \mathbf{u} = \sum_i f_i \mathbf{e}_i + \frac{\Delta t}{2} \mathbf{F}. \tag{4}\]

The \(\Delta t / 2\) correction in the velocity moment is consistent with the Guo forcing convention and is required for
the correct macroscopic limit of the IBM-driven flow.

\subsubsection{2.2 Collision Operators}\label{collision-operators}

The present work uses three collision operators, stated explicitly so that the targeted controls of Sec.~6 are
unambiguous.

\emph{BGK / single-relaxation-time (SRT)} {[}9{]}: a single relaxation parameter \(\tau\) governs all non-conserved
moments. This is the baseline operator used throughout Sec.~3--Sec.~5 unless stated otherwise.

\emph{Two-relaxation-time (TRT)} {[}31{]}: even and odd non-conserved moments relax with independent rates \(s_e\) and
\(s_o\), and the \emph{magic parameter} \(\Lambda_{eo} = \Lambda_e \, \Lambda_o\) --- formed from the kinetic
coefficients \(\Lambda_e = 1/s_e - 1/2\) and \(\Lambda_o = 1/s_o - 1/2\) --- controls boundary-layer regularity
(Ginzburg, Verhaeghe, and d'Humières {[}31{]}). The BGK operator is recovered as the TRT subclass
\(\Lambda_e = \Lambda_o\). Among the values cataloged by {[}31{]}, \(\Lambda_{eo} = 3/16\) fixes the exact location of a
bounce-back no-slip wall while \(\Lambda_{eo} = 1/4\) has favorable stability properties for the mass-conservation
equation. Throughout Sec. 6 we use \(\Lambda_{eo} = 1/4\) following Ginzburg et al.~{[}31{]} as a relaxation-side
stability choice; the exact wall-location property at \(\Lambda_{eo} = 3/16\) is specific to bounce-back walls and does
not transfer to the regularized immersed-boundary interface.

\emph{Central-moment multiple-relaxation-time (CM-MRT)} in the D2Q9 form: adapted from the central-moment
lattice-Boltzmann framework of De Rosis, Huang, and Coreixas {[}38, Appendix B{]}, the collision is performed on the
nine central moments \(k_{pq} = \sum_i f_i\,\bar{e}_{ix}^{\,p}\,\bar{e}_{iy}^{\,q}\) built from the peculiar velocities
\(\bar{\mathbf{e}}_i = \mathbf{e}_i - \mathbf{u}\) and ordered as
\((k_{00}, k_{10}, k_{01}, k_{20}, k_{02}, k_{11}, k_{21}, k_{12}, k_{22})\), with the Galilean-invariant equilibrium
central moments and the Guo et al.~{[}37{]} forcing recast in central-moment space ({[}38{]}, Appendix B). The collision
step is

\[|\mathbf{k}^*\rangle = (I - \Lambda) |\mathbf{k}\rangle + \Lambda |\mathbf{k}^{\mathrm{eq}}\rangle + (I - \Lambda/2) |\mathbf{R}\rangle, \tag{5}\]

\noindent with relaxation matrix \(\Lambda = \mathrm{diag}(0, 0, 0, \omega, \omega, \omega, s_q, s_q, s_\varepsilon)\): the lowest
three moments (\(k_{00}, k_{10}, k_{01}\)) are not relaxed (\(k_{00}\) is the conserved density, while the first-order
central moments \(k_{10}\) and \(k_{01}\) are not relaxed but carry the Guo forcing source --- under the half-force
velocity convention of Eq. (4) their pre-collision values are \(k_{10} = -F_x\Delta t/2\) and
\(k_{01} = -F_y\Delta t/2\), which the source term \(|\mathbf{R}\rangle\) maps to the post-collision momentum
\(k_{10}^* = +F_x\Delta t/2\) and \(k_{01}^* = +F_y\Delta t/2\) (De Rosis et al.~{[}38{]}, Appendix B)), the three
second-order moments (\(k_{20}, k_{02}, k_{11}\)) share the relaxation rate \(\omega = (\nu / c_s^2 + 1/2)^{-1}\) that
controls the kinematic viscosity, and the third- and fourth-order moments relax with \(s_q = 1.2\) and
\(s_\varepsilon = 1.4\), respectively. These higher-order rates are stability-tuned free parameters of the
multiple-relaxation framework; the central-moment construction itself --- the peculiar-velocity shift, the collision in
central-moment space, and the Guo et al.~{[}37{]} forcing recast in central-moment form --- is adapted from {[}38,
Appendix B{]}. The forcing in central-moment space, \(|\mathbf{R}\rangle\) ({[}38{]}), reduces to the Guo et
al.~{[}37{]} forcing in the second-order Hermite limit, as stated explicitly in {[}38{]}; in the lattice units adopted
here (\(\Delta t = 1\)), \(|\mathbf{R}\rangle\) is the per-step central-moment momentum source. The CM-MRT operator is a
central-moment formulation; the relation between two-dimensional cumulant and central-moment formulations is reviewed in
{[}38{]}, and cumulant collision models are treated here as a separate extension.

\subsubsection{2.3 Immersed-Boundary Coupling}\label{immersed-boundary-coupling}

The immersed boundary is discretized as a set of \(N_b\) Lagrangian marker points \(\{\mathbf{X}_k\}\) on the solid
surface. The coupling between the Eulerian and Lagrangian frames is achieved through the regularized delta function
\(D_h\).

\emph{Symbol convention.} Lower-case \(\mathbf{x}\) denotes the Eulerian grid coordinate; upper-case \(\mathbf{X}_k\)
denotes the \(k\)-th Lagrangian marker position. Lower-case \(\mathbf{f}\) denotes the Eulerian forcing density (per
unit volume) --- the force exerted \emph{on the fluid} by the boundary --- and upper-case \(\mathbf{F}_L\) denotes the
Lagrangian force evaluated at marker \(k\). Lower-case \(\mathbf{u}\) denotes the Eulerian velocity and upper-case
\(\mathbf{U}\) denotes the interpolated Lagrangian velocity at the marker (Eq. (7)), while the desired (prescribed or
moving-body) marker velocity is denoted \(\mathbf{U}_d\). The symbol \(h \equiv \Delta x\) is the Eulerian lattice
spacing.

In two dimensions,

\[D_h(\mathbf{x} - \mathbf{X}_k) = h^{-2} \, \phi\!\left(\frac{x - X_k}{h}\right) \phi\!\left(\frac{y - Y_k}{h}\right), \tag{6}\]

\noindent where \(\phi\) is a one-dimensional kernel function defined in Sec.~2.4. \emph{Interpolation} (Eulerian → Lagrangian)
and \emph{spreading} (Lagrangian → Eulerian) are then

\[\mathbf{U}(\mathbf{X}_k) = \sum_{\mathbf{x}} \mathbf{u}(\mathbf{x}) \, D_h(\mathbf{x} - \mathbf{X}_k) \, h^2, \tag{7}\]

\[\mathbf{f}(\mathbf{x}) = \sum_k \mathbf{F}_L(\mathbf{X}_k) \, D_h(\mathbf{x} - \mathbf{X}_k) \, \Delta s, \tag{8}\]

\noindent where \(h\) is the Eulerian grid spacing and \(\Delta s\) is the Lagrangian arc length. Fig. 1 illustrates the
computational setup of the four canonical cases used in this study.

\begin{figure}
\centering
\includegraphics{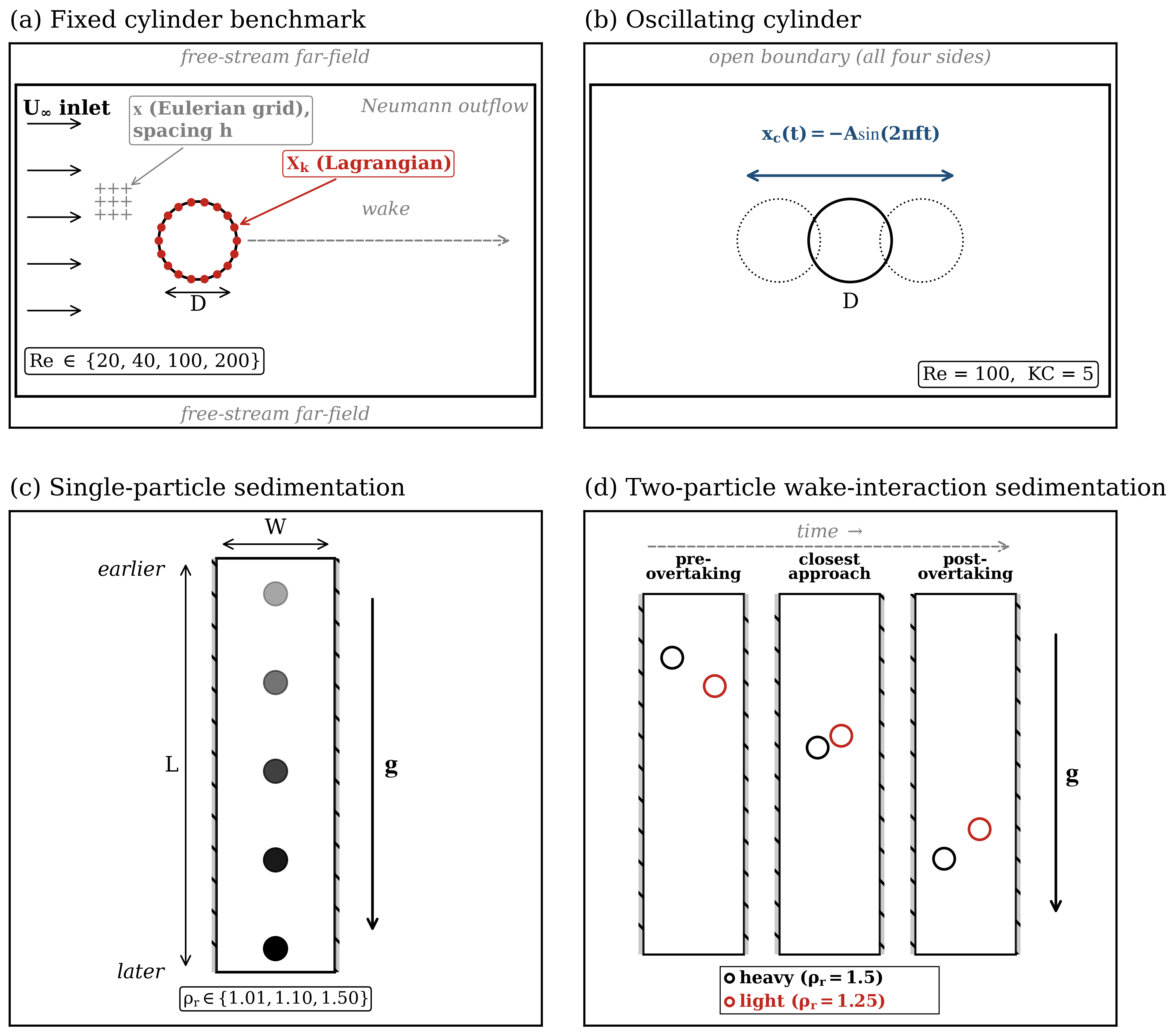}
\caption{\textbf{Fig. 1.} Computational setup of the four canonical cases. (a) Fixed-cylinder benchmark at
\(\mathrm{Re} \in \{20, 40, 100, 200\}\) (uniform inflow / Neumann outflow / far-field Dirichlet free-stream top and
bottom, following the Kang and Hassan {[}20{]} benchmark configuration); Lagrangian marker \(\mathbf{X}_k\) (red) and
Eulerian grid \(\mathbf{x}\) (gray, spacing \(h\)). (b) Oscillating cylinder at \(\mathrm{Re}=100\), \(\mathrm{KC}=5\)
(Dütsch et al.~{[}39{]}). (c) Single-particle sedimentation at the three Wang et al.~{[}16{]} / Glowinski et
al.~{[}40{]} density ratios (\(\rho_s/\rho_f \in \{1.01, 1.1, 1.5\}\)); the \(\rho_s/\rho_f = 1.25\) density-matched
control of Sec. 5.2 reuses this setup. (d) Two-particle wake-interaction (heavy \(\rho_r=1.5\) and light \(\rho_r=1.25\)
particles, gravity along \(+x\); the light particle is initially placed downstream and the heavy particle upstream, so
the upstream/downstream roles exchange during the overtaking / role-exchange sequence (Sec. 5.2.1); Majumder et
al.~{[}23{]}). Coordinate conventions are given in Sec. 2.9; the panel-(d) axis remapping follows the Fig. 7 caption.}
\end{figure}

\subsubsection{2.4 Regularized Delta Functions}\label{regularized-delta-functions}

The present comparison employs two regularized delta functions:

\par\needspace{6\baselineskip}

\emph{Hat function} (2-point support, \(C^0\)):

\nopagebreak[4]\par

\[\phi(r) = \max(1 - |r|, 0). \tag{9}\]

This compact function has a support width \(W = 2\Delta x\). In two dimensions, each Lagrangian point interacts with at
most \(2 \times 2 = 4\) Eulerian nodes with non-zero weights.

\par\needspace{6\baselineskip}

\emph{Peskin 4-point function} (P4) {[}3{]} (4-point support, \(C^1\)):

\nopagebreak[4]\par

\[\phi(r) = \begin{cases}
\tfrac{1}{8}\left(3 - 2|r| + \sqrt{1 + 4|r| - 4r^2}\right), & 0 \leq |r| \leq 1, \\
\tfrac{1}{8}\left(5 - 2|r| - \sqrt{-7 + 12|r| - 4r^2}\right), & 1 \leq |r| \leq 2, \\
0, & |r| > 2.
\end{cases} \tag{10}\]

This function satisfies the partition-of-unity and first-moment conditions {[}3{]}, with a support width
\(W = 4\Delta x\). In two dimensions, each Lagrangian point interacts with at most \(4 \times 4 = 16\) Eulerian nodes
with non-zero weights. The wider support provides smoother interpolation at the cost of a broader effective boundary
thickness.

\subsubsection{2.5 Boundary-Enforcement Schemes}\label{boundary-enforcement-schemes}

\paragraph{2.5.1 Direct Forcing (DF)}\label{direct-forcing-df}

\par\needspace{6\baselineskip}

In the direct forcing approach, the Lagrangian force at each boundary point is computed from a single velocity
interpolation {[}14,37{]}:

\nopagebreak[4]\par

\[\mathbf{F}_L(\mathbf{X}_k) = \frac{2 \rho(\mathbf{X}_k)}{\Delta t} \left[ \mathbf{U}_d(\mathbf{X}_k) - \mathbf{U}(\mathbf{X}_k) \right], \tag{11}\]

\noindent where \(\mathbf{U}_d\) is the desired (wall) velocity and \(\rho(\mathbf{X}_k)\) is the interpolated density. The factor
\(2\) arises because the Guo forcing scheme {[}37{]} incorporates only half of the body force into the macroscopic
velocity definition; the doubled force magnitude compensates, ensuring the full velocity correction within one time
step.

\paragraph{2.5.2 Multi-Direct Forcing (MDF)}\label{multi-direct-forcing-mdf}

\par\needspace{6\baselineskip}

MDF {[}15,16{]} iterates the DF procedure to suppress the residual interface slip that remains after a single DF pass.
Starting from the post-streaming velocity field \(\mathbf{u}^{(0)}\), for \(n = 1, 2, \ldots\):

\nopagebreak[4]\par

\[\mathbf{U}^{(n)}(\mathbf{X}_k) = \sum_{\mathbf{x}} \mathbf{u}^{(n-1)}(\mathbf{x}) \, D_h(\mathbf{x} - \mathbf{X}_k) \, h^2, \tag{12}\]

\[\mathbf{F}_L^{(n)}(\mathbf{X}_k) = \frac{2 \rho(\mathbf{X}_k)}{\Delta t} \left[ \mathbf{U}_d - \mathbf{U}^{(n)} \right], \tag{13}\]

\[\mathbf{u}^{(n)}(\mathbf{x}) = \mathbf{u}^{(n-1)}(\mathbf{x}) + \frac{\omega_{\mathrm{MDF}} \, \Delta t}{2 \rho} \, \mathrm{spread}(\mathbf{F}_L^{(n)})(\mathbf{x}), \tag{14}\]

\noindent with the \emph{MDF spread-relaxation} parameter \(\omega_{\mathrm{MDF}}\) controlling the iteration (this is the
relaxation parameter of the Richardson iteration introduced below, and is a quantity distinct from the LBM
collision-relaxation rate \(\omega = (\nu / c_s^2 + 1/2)^{-1}\) of Eq. (5); the two are kept notationally distinct
throughout). The total IBM force applied over the iteration is the relaxation-weighted accumulation
\(\mathbf{F}_L^{\mathrm{total}} = \sum_{n=1}^N \omega_{\mathrm{MDF}}\,\mathbf{F}_L^{(n)}\), consistent with the
per-iteration velocity increment of Eq. (14). Standard MDF takes \(\omega_{\mathrm{MDF}} = 1\) and a fixed \(N\), but as
Zhang et al.~{[}41{]} showed, the iteration is mathematically a Richardson iteration on a Lagrangian linear system
\(A\mathbf{x} = \mathbf{b}\) (with
\(A_{lm} = \sum_{\mathbf{x}} D_h(\mathbf{x} - \mathbf{X}_l) D_h(\mathbf{x} - \mathbf{X}_m) \, \Delta s \, h^2\)) whose
\emph{sufficient} convergence condition is \(0 < \omega_{\mathrm{MDF}} < 2\,\|A\|_\infty^{-1}\), a conservative row-norm
bound; the spectral-radius condition \(0 < \omega_{\mathrm{MDF}} < 2/\rho(A)\) is the exact Richardson convergence
interval (and, since \(\rho(A) \le \|A\|_\infty\), the wider of the two), but \(\rho(A)\) is not directly accessible at
runtime; increasing \(N\) at fixed \(\omega_{\mathrm{MDF}} = 1\) does \emph{not} monotonically improve accuracy and can
enter the divergence regime once the iteration matrix \(I - \omega_{\mathrm{MDF}} A\) has spectral radius
\(\rho(I - \omega_{\mathrm{MDF}} A) > 1\) --- equivalently, at \(\omega_{\mathrm{MDF}} = 1\), once an eigenvalue of
\(A\) exceeds \(2\) (Zhang et al.~{[}41{]}); the stability of this iteration for freely moving bodies was subsequently
characterized by Suzuki et al.~{[}25{]}. We adopt the \emph{implementable} estimate
\(\omega_{\mathrm{MDF}} \approx \|A\|_\infty^{-1}\) (a row-norm-based estimate, not a spectral optimum) of Zhang et
al.~{[}41{]}, computed once at initialization, together with two implementation-level controls: an \(L_\infty\)
slip-residual stopping criterion
\(r_n = \max_l \| \mathbf{U}_d - \mathbf{U}^{(n)}(\mathbf{X}_l) \|_2 < \mathrm{tol} = 10^{-4}\), and a divergence guard
\(r_n > 1.5\,r_{n-1}\) that breaks the iteration. In the coupled two-particle path the per-particle forces share the
fluid update within each iteration; there the initial relaxation is additionally capped at
\(\omega_{\mathrm{MDF}} \leq 0.8\) and, if the residual grows by more than \(20\%\) between consecutive iterations, that
trial iteration is rejected and the relaxation is halved; the iteration terminates if the halved factor reaches
\(0.125\) of its initial value. The iteration count is bounded by \(N \in [N_{\min}, N_{\max}] = [5,20]\), with the
upper limit recovering the \(N = 20\) benchmark of Kang and Hassan {[}20{]}. Throughout this manuscript ``MDF'' refers
to this adaptive form, denoted \emph{MDF (adaptive \(N \in [5,20]\))} in tables when disambiguation is needed. The
full-grid velocity update at every iteration accounts for the linear \(\mathcal{O}(N)\) scaling of the MDF cost per time
step.

\paragraph{2.5.3 Distribution-Function Correction (DFC)}\label{distribution-function-correction-dfc}

DFC {[}18{]} operates at the distribution-function level rather than the macroscopic velocity level. After the streaming
step yields the post-streaming distributions \(f_i^*(\mathbf{x})\), these are first interpolated to each Lagrangian
marker {[}18{]},

\[f_i^*(\mathbf{X}_k) = \sum_{\mathbf{x}} f_i^*(\mathbf{x}) \, D_h(\mathbf{x} - \mathbf{X}_k) \, h^2, \tag{15}\]

and the desired distribution functions at the marker are constructed from a Lagrangian-level bounce-back relation
{[}18{]},

\[\bar{f}_i(\mathbf{X}_k) = f_{\bar{i}}^*(\mathbf{X}_k) + 2 w_i \rho_f \frac{\mathbf{e}_i \cdot \mathbf{U}_d}{c_s^2}, \tag{16}\]

\noindent where \(\bar{i}\) denotes the opposite direction of \(i\), \(\mathbf{U}_d\) is the desired marker velocity defined
above, and \(\rho_f = \rho_0 = 1\) is the reference fluid density in lattice units. A marker-wise correction scaling
factor \(\lambda_k\) {[}18{]} is then evaluated as

\[\lambda_k = \frac{1}{2 \rho_f \, \Delta s \, W_{\mathrm{sum}}(k)}, \qquad W_{\mathrm{sum}}(k) = \sum_{\mathbf{x}} W_{\mathrm{total}}(\mathbf{x}) \, D_h(\mathbf{x} - \mathbf{X}_k) \, h^2, \tag{17}\]

\noindent with \(W_{\mathrm{total}}(\mathbf{x}) = \sum_{k'} D_h(\mathbf{x} - \mathbf{X}_{k'})\); for stationary boundaries
\(\lambda_k\) is computed once and cached, whereas for moving boundaries it is recomputed at every time step to reflect
the updated marker positions. The resulting distribution-function correction at the marker,

\[\delta f_i(\mathbf{X}_k) = \lambda_k \left[ \bar{f}_i(\mathbf{X}_k) - f_i^*(\mathbf{X}_k) \right], \tag{18}\]

is spread back to the Eulerian lattice as

\[f_i(\mathbf{x}) \leftarrow f_i^*(\mathbf{x}) + \sum_k \delta f_i(\mathbf{X}_k) \, D_h(\mathbf{x} - \mathbf{X}_k) \, \Delta s, \tag{19}\]

and the associated particle-side reaction force follows as {[}18{]}

\[\mathbf{F}_s(\mathbf{X}_k) = -\Delta s \sum_i \mathbf{e}_i \, \lambda_k \left[ \bar{f}_i(\mathbf{X}_k) - f_i^*(\mathbf{X}_k) \right]. \tag{20}\]

In lattice units (\(\Delta x = \Delta t = 1\)) adopted throughout, \(\mathbf{F}_s(\mathbf{X}_k)\) is the per-marker
force on the particle: it is the first velocity moment \(-\Delta s \sum_i \mathbf{e}_i\,\delta f_i(\mathbf{X}_k)\) of
the distribution-function correction, i.e., a momentum increment per unit time step, and the arc-length weight
\(\Delta s\) in Eq. (20) is the single quadrature weight already used in the spreading of Eq. (19), applied once (not a
second volume integration). With \(\Delta t = 1\) the momentum-per-step and the force are numerically equal, so
\(\mathbf{F}_s\) is directly comparable in magnitude and time unit to the DF/MDF Eulerian forcing of Eq. (8) reversed by
Newton's third law.

Unlike DF and MDF, DFC does not use the Guo forcing term; the correction is applied directly to the distribution
functions. The non-iterative DFC therefore requires modest additional cost over DF (1.1--1.3× per step, see Appendix A
Sec.~A.4) while modifying the distribution functions at a more fundamental level. The spatial distribution of the
correction scaling factor \(\lambda_k\) --- which depends on the kernel support and on the relative position of each
marker against the Eulerian grid --- is documented in Sec.~4 as one signature of the kernel-dependent reversal.

\subsubsection{2.6 Reynolds-Number Definitions for Sedimentation
Comparisons}\label{reynolds-number-definitions-for-sedimentation-comparisons}

For a fluid of density \(\rho_f\) and dynamic viscosity \(\mu\), two Reynolds-number bases appear for sedimentation
comparisons in the present manuscript. The fluid-density basis \(\mathrm{Re}_f = \rho_f |U| D / \mu = |U| D / \nu\) is
used by Glowinski et al.~{[}40{]}, Uhlmann {[}14{]}, and Majumder et al.~{[}23{]} (inheriting Uhlmann's definition),
whereas the particle-density basis \(\mathrm{Re}_p = \rho_P |U| D_P / \mu\) is used by Wang et al.~{[}16{]}; in the
latter, \(\rho_P \equiv \rho_s\) is the solid-particle density and \(D_P \equiv D\) the particle diameter, and the
subscript-\(P\) form follows the Wang et al.~{[}16{]} convention adopted in the present manuscript whenever the
cross-source basis is in view, with \(\rho_s\) used otherwise. The two bases are related by
\(\mathrm{Re}_p = \mathrm{Re}_f \cdot (\rho_P / \rho_f)\).

Throughout the manuscript, the density ratio is denoted \(\rho_r \equiv \rho_s / \rho_f\); for the two-particle
wake-interaction case, \(\rho_r^{\mathrm{heavy}} = 1.5\) and \(\rho_r^{\mathrm{light}} = 1.25\) denote the heavy and
light particles. The two particles are labeled by density throughout; their upstream/downstream (leading/trailing) roles
are not fixed but exchange during the overtaking / role-exchange sequence (Sec. 5.2.1), so ``leading'' and ``trailing''
are used only as instantaneous, phase-dependent descriptors.

For sedimentation comparisons, the reported Reynolds number follows the definition used in the corresponding reference
benchmark; the basis adopted by each comparison is stated explicitly in the relevant table caption, and conversion
ratios are provided where multiple bases coexist.

For prescribed bodies (cylinder flows), \(\mathrm{Re} = u_\infty D / \nu\) is used throughout, consistent with the
experimental and numerical reference data of {[}39,42--47{]}.

\subsubsection{2.7 Moving-Body Update and Internal-Mass
Correction}\label{moving-body-update-and-internal-mass-correction}

For the sedimentation benchmark, a rigid circular particle is coupled to the fluid through two-way interaction. The
particle motion is governed by Newton's second law,

\[m_s \frac{d \mathbf{v}}{d t} = \mathbf{F}_{\mathrm{hydro}} + \mathbf{F}_{\mathrm{gravity}}, \qquad I_s \frac{d \Omega}{d t} = T_{\mathrm{hydro}}, \tag{21}\]

\noindent where \(m_s = \rho_s \pi r^2\) is the (two-dimensional) particle mass, \(I_s = \tfrac{1}{2} m_s r^2\) the corresponding
disk moment of inertia, \(\mathbf{F}_{\mathrm{hydro}}\) the hydrodynamic force from the IBM,
\(\mathbf{F}_{\mathrm{gravity}}\) the net gravitational force, \(\Omega\) the angular velocity, and
\(T_{\mathrm{hydro}}\) the hydrodynamic torque about the particle centroid \(\mathbf{X}_c\), evaluated from the
particle-side boundary force with the same sign convention as \(\mathbf{F}_{\mathrm{hydro}}\) (see below): for DF and
MDF, \(T_{\mathrm{hydro}} = -\sum_k (\mathbf{X}_k - \mathbf{X}_c) \times \mathbf{F}_L(\mathbf{X}_k)\,\Delta s\), the
marker arc-length-weighted moment of the Lagrangian fluid-side force reversed by Newton's third law; for DFC,
\(T_{\mathrm{hydro}} = \sum_k (\mathbf{X}_k - \mathbf{X}_c) \times \mathbf{F}_s(\mathbf{X}_k)\), since \(\mathbf{F}_s\)
of Eq. (20) is already the particle-side marker force (the two-dimensional cross product denotes the out-of-plane
scalar). The particle is free to translate and rotate; for a rotating rigid disk the desired marker velocity used in the
boundary-enforcement schemes (Eqs. (11)--(20)) is the rigid-body value
\(\mathbf{U}_d(\mathbf{X}_k) = \mathbf{v} + \Omega \times (\mathbf{X}_k - \mathbf{X}_c)\). The angular velocity is
advanced together with the translation. In the main Velocity-Verlet sedimentation runs the rotational degree of freedom
uses the torque-only Velocity-Verlet update, while the explicit-history internal-mass correction acts on the
translational degree of freedom (Eq. (A3)); in the Euler-explicit wake-interaction cross-check (Sec. 5.2) the rotational
update additionally carries an internal angular-momentum increment
\(\Delta L_{\mathrm{int}}^n = L_{\mathrm{int}}^n - L_{\mathrm{int}}^{n-1}\) recovered from the resolved interior angular
momentum, with \(I_f = \tfrac{1}{2} m_f r^2\) the fictitious-fluid moment of inertia. For the circular-disk geometry the
rotational contribution to the reported terminal and peak Reynolds numbers is negligible, and the internal-mass
correction contrast of Sec. 5.2 --- evaluated with the rotational update held fixed across both ablation arms --- is
governed by the translational correction.

Under a reduced-pressure formulation, the hydrostatic pressure is separated from the Navier--Stokes equations, so that
the LBM solves only the dynamic pressure field without an explicit gravity term in the collision operator. The particle
then experiences only the net gravitational force,

\[\mathbf{F}_{\mathrm{gravity}} = (\rho_s - \rho_f) \pi r^2 \, \mathbf{g}, \tag{22}\]

\noindent where \(\mathbf{g}\) is the gravitational acceleration, directed along \(-\hat{\mathbf{e}}_y\) in the single-particle
sedimentation cases (Sec. 5.1) and along \(+\hat{\mathbf{e}}_x\) in the two-particle wake-interaction configuration
(Sec. 5.2).

The hydrodynamic force on the particle is extracted from the IBM boundary force. For DF and MDF, the Eulerian body force
\(\mathbf{f}(\mathbf{x})\) represents the force exerted \emph{on the fluid} by the boundary; by Newton's third law, the
force on the particle is \(\mathbf{F}_{\mathrm{hydro}} = -\sum_{\mathbf{x}} \mathbf{f}(\mathbf{x}) h^2\). For DFC, the
Lagrangian fluid force \(\mathbf{F}_s(\mathbf{X}_k)\) already represents the force on the particle, so
\(\mathbf{F}_{\mathrm{hydro}} = \sum_k \mathbf{F}_s(\mathbf{X}_k)\) without sign reversal. With this convention, an
upward hydrodynamic drag on a downward-settling particle enters Eq. (21) with the sign opposite to the settling
velocity, as expected physically.

\par\needspace{6\baselineskip}

The particle position and velocity are advanced using the Velocity-Verlet scheme, which is second-order accurate in time
for the resolved hydrodynamic and gravitational forces --- the explicitly lagged internal-mass increment of Appendix A
Sec. A.5 is first-order --- and which serves as the time integrator for the main sedimentation runs of the present
manuscript (the Majumder-aligned wake-interaction cross-check reported in Sec.~5.2 instead uses an Euler-explicit
integrator to match the original Majumder reference setup, as noted there):

\nopagebreak[4]\par

\[\mathbf{v}^{n+1/2} = \mathbf{v}^n + \frac{\Delta t}{2 m_s} \mathbf{F}_{\mathrm{total}}^n, \quad \mathbf{x}^{n+1} = \mathbf{x}^n + \Delta t \, \mathbf{v}^{n+1/2}, \quad \mathbf{v}^{n+1} = \mathbf{v}^{n+1/2} + \frac{\Delta t}{2 m_s} \mathbf{F}_{\mathrm{total}}^{n+1}, \tag{23}\]

\noindent where
\(\mathbf{F}_{\mathrm{total}}^{(\cdot)} = \mathbf{F}_{\mathrm{hydro}}^{(\cdot)} + \mathbf{F}_{\mathrm{gravity}} + \Delta\mathbf{F}_{\mathrm{IM}}^n\),
with the explicitly lagged internal-mass increment \(\Delta\mathbf{F}_{\mathrm{IM}}^n\) of Eq. (A3) evaluated once at
step \(n\) and held fixed across both half-kicks (so only the resolved hydrodynamic part of
\(\mathbf{F}_{\mathrm{total}}\) is updated from \(n\) to \(n+1\)); it is included for the correction-inclusive runs and
set to zero in the without-correction ablation. The half-step velocity update and position advance occur before the
LBM--IBM cycle; the full-step velocity update occurs after the new hydrodynamic force is computed.

The moving-body update uses the Feng and Michaelides {[}48{]} explicit-history internal/fictitious-fluid correction
(Appendix A Eq. (A1); force-equivalent form Eq. (A3)) as the correction-inclusive baseline, with the uncorrected
(boundary-only) configuration used as the sensitivity comparison. The mapping to the Suzuki and Inamuro {[}36{]}
four-scheme taxonomy is detailed in Appendix A Sec. A.5.

Because the terms \emph{internal mass} and \emph{explicit-history} echo the language of physical unsteady forces, three
distinct notions should not be conflated here. First, the explicit internal-mass term used here (Eq. (A3)) is a purely
numerical closure of the boundary-only force evaluation: it compensates the momentum of the fictitious fluid occupying
the solid interior in a regularized-delta IBM, is proportional to the backward difference of the particle velocity, and
vanishes for a stationary body. Second, the inviscid added-mass force of potential-flow theory --- the reaction to
relative acceleration that appears in the equation of motion of Maxey and Riley {[}49{]}, derived for a small rigid
sphere in a three-dimensional flow, and in Auton, Hunt, and Prud'homme {[}50{]}, with an added-mass coefficient set by
the body geometry --- is a physical hydrodynamic response of the flow field to body acceleration; in an
interface-resolved simulation it is represented implicitly in the resolved fluid--body interaction and the recovered
boundary force, to the accuracy of the discretization, not in the correction term. Third, viscous unsteady-inertia and
history contributions --- Basset-history-type forces and finite-Reynolds viscous corrections to the added-mass response
--- are likewise physical effects carried by the resolved flow. Throughout this manuscript, \emph{explicit-history}
names the numerical discretization of Eq. (A3) and is distinct from the Basset history force.

\subsubsection{2.8 Local Diagnostics}\label{local-diagnostics}

Boundary fidelity is quantified by three metrics evaluated on the Eulerian velocity field at the final time step (or an
instantaneous snapshot in the periodic regime). For \(\mathrm{Re}=20\) and 40 this corresponds to a true steady state;
for \(\mathrm{Re}=100\) and 200 it is an instantaneous snapshot of the periodic vortex-shedding regime. The absolute
values may vary with shedding phase; all three schemes are compared at the same stored time step (identical step index
across schemes) of the converged shedding cycle, so the relative ranking and improvement ratios are compared on a common
simulation clock rather than at a phase-aligned instant. The three metrics are the slip error
\(\varepsilon_{\mathrm{slip}}\), the velocity magnitude at the Lagrangian boundary points normalized by \(U_\infty\)
(which should vanish for a stationary cylinder); the leakage flux \(\Phi_{\mathrm{leak}}\), the net mass flux through
the boundary normalized by \(U_\infty \pi D\); and the internal residual velocity \(\|\mathbf{u}\|_{\mathrm{inside}}\),
the mean velocity magnitude inside the solid (with a \(\kappa = 2\) buffer-cell exclusion). Here \(U_\infty\) denotes
the free-stream reference speed, equal to the fixed-cylinder inflow velocity \(u_\infty\).

Two further correction-resolved diagnostics are introduced for Sec.~4: the marker-resolved correction magnitude
\(|\mathbf{F}_{s,k}|\) together with its azimuthal distribution along the cylinder surface, and the near-boundary
pressure / velocity residual in the immediate vicinity of the immersed surface. Here \(|\mathbf{F}_{s,k}|\) denotes the
magnitude of the marker-level distribution-function correction force \(\mathbf{F}_s(\mathbf{X}_k)\) of Eq. (20) --- the
vector moment \(-\Delta s \sum_i \mathbf{e}_i\,\delta f_i(\mathbf{X}_k)\) of the scalar distribution-function correction
\(\delta f_i\) of Eq. (18) --- not the scalar increment \(\delta f_i\) itself.

For moving bodies, two trajectory-resolved diagnostics are used in Sec.~5. The first is the trajectory phase
identification in wake-interaction (pre-overtaking / closest-approach / post-overtaking phases) together with the
corresponding Reynolds and velocity histories. The second is the particle-bound acceleration proxy \(|a^*(t)|\), a
kinematic Lagrangian quantity computed from the particle velocity time series as
\(|a^*| = \sqrt{(d v_x^* / d t^*)^2 + (d v_y^* / d t^*)^2}\).

The internal residual velocity is a known limitation of boundary-only IBM formulations {[}51{]}, which do not include
explicit corrections for the solid interior. Implicit velocity correction (IVC) schemes {[}51{]} provide a complementary
route to suppressing this residual; their integration with the DF / MDF / DFC family compared here is identified as a
future direction (see Sec.~7).

\subsubsection{2.9 Numerical Setup}\label{numerical-setup}

\paragraph{2.9.1 Fixed Cylinder}\label{fixed-cylinder}

A circular cylinder of diameter \(D\) is placed at \((0.4 L_x, 0.5 L_y)\) in a \([0, L_x] \times [0, L_y]\) rectangular
domain with \(L_x = 1.25\), \(L_y = 1.0\) (downstream-extended channel: 0.5 ahead of and 0.75 behind the cylinder
center). The boundary treatment follows the Kang and Hassan {[}20{]} benchmark configuration: a Dirichlet inflow on the
left (\(u_x = u_\infty\), \(u_y = 0\)), a homogeneous Neumann outflow on the right (\(\partial/\partial n = 0\),
recovered from the adjacent interior node), and far-field Dirichlet conditions on the top and bottom that prescribe the
free-stream velocity (\(u_x = u_\infty\), \(u_y = 0\)) rather than a stationary no-slip wall. The inflow velocity is
\(u_\infty = 0.1\) in lattice units. The Reynolds number is \(\mathrm{Re} = u_\infty D / \nu\).

\noindent \textbf{Table I.} Prescribed-body simulation parameters (fixed-cylinder and oscillating-cylinder cases).

\needspace{4\baselineskip}
\vspace{0.8em}
\sbox0{\scriptsize\setlength{\tabcolsep}{3pt}\renewcommand{\arraystretch}{1.2}%
\begin{tabular}{>{\raggedright\arraybackslash}p{0.063\linewidth} >{\raggedright\arraybackslash}p{0.620\linewidth} >{\raggedright\arraybackslash}p{0.268\linewidth}}
\toprule
Parameter & Fixed cylinder & Oscillating cylinder \\
\midrule
Domain & \(L_x \times L_y = 1.25 \times 1.0\) & \(L_x \times L_y = 1.5 \times 1.0\), open (zero-gradient) boundaries on
all four sides \\
Grid & \(2001 \times 1601\) (\(\mathrm{Re}\leq 100\), \(\Delta x = 1/1600\));
\(3001 \times 2401\) (\(\mathrm{Re} = 200\), \(\Delta x = 1/2400\)) & \(2401 \times 1601\) \\
Cylinder diameter & \(D = 40\Delta x\) (\(\mathrm{Re}\leq 100\)); \(D = 60\Delta x\)
(\(\mathrm{Re}=200\)) & \(D = 80\Delta x\) \\
Reynolds number & \(\mathrm{Re} \in \{20, 40, 100, 200\}\) & \(\mathrm{Re} = 100\), \(\mathrm{KC} = 5\) \\
Lagrangian markers & \(\Delta s \approx \pi D / N_b\) (uniform marker spacing on the circle
of circumference \(\pi D\)), equivalently \(N_b = \pi D / \Delta s\) & same \\
Relaxation time & \(\tau = 1/2 + 3\nu\) (\(\tau = 0.59\) at \(\mathrm{Re} = 200\);
cf.~Appendix A Sec. A.1) & \(\tau = 0.74\) (\(\nu = 0.08\)) \\
Sampling interval & 200 time steps & 200 time steps \\
Total steps / averaging & 100,000 (\(\mathrm{Re}\leq 100\), averaged over steps 70,200--100,000,
or the converged tail of an early-converged run); 60,000
(\(\mathrm{Re} = 200\), averaged over steps 30,200--60,000,
\(\approx 9.9\) vortex-shedding cycles at \(\mathrm{St}\approx 0.20\)) & 40,000 (\(T = 4000\) steps × 10 oscillation periods; first 2 periods
(step \(< 8000\)) excluded) \\
\bottomrule
\end{tabular}}
\ifdim\wd0>\textwidth\noindent\resizebox{\textwidth}{!}{\usebox0}\else\noindent\makebox[\textwidth][c]{\usebox0}\fi
\vspace{0.8em}

The blockage ratio is \(D / L_y = 2.5\%\) at \(\mathrm{Re} \leq 100\) and \(2.5\%\) at \(\mathrm{Re} = 200\)
(\(D = 60\Delta x\) with \(\Delta x = 1/2400\)), both well below the 5\% threshold above which blockage effects are
typically considered significant.

\paragraph{2.9.2 Oscillating Cylinder}\label{oscillating-cylinder}

An oscillating-cylinder benchmark following Dütsch et al.~{[}39{]} is configured with
\(\mathrm{Re} = U_{\max} D / \nu = 100\) and Keulegan--Carpenter number \(\mathrm{KC} = U_{\max} / (f_0 D) = 5\). The
cylinder is centered in a \(L_x \times L_y = 1.5 \times 1.0\) domain with diameter \(D = 80\Delta x\)
(\(2401 \times 1601\) grid) and oscillates sinusoidally in a quiescent fluid along the streamwise direction,

\[x_c(t) = -A \sin(2\pi f_0 t),\]

\noindent with displacement amplitude \(A = \mathrm{KC}\,D / (2\pi) \approx 0.796\,D\), frequency
\(f_0 = U_{\max} / (\mathrm{KC}\,D)\), velocity amplitude \(U_{\max} = 0.1\) in lattice units, and oscillation period
\(T = 1/f_0 = \mathrm{KC}\,D / U_{\max} = 4000\) time steps. The kinematic viscosity follows from the target Reynolds
number as \(\nu = U_{\max} D / 100\). Open (zero-gradient) boundary conditions are applied on all four sides, since the
far field is quiescent.

Simulation parameters for the oscillating-cylinder benchmark are reported in Table I (right column).

\paragraph{2.9.3 Single-Particle Sedimentation}\label{single-particle-sedimentation}

A circular particle of diameter \(d\) settles under gravity in a vertical channel, following the canonical configuration
of Glowinski et al.~{[}40{]} and Wang et al.~{[}16{]}. The physical parameters are \(W = 2\) cm (channel width),
\(L = 6\) cm (channel height), \(d = 0.25\) cm (confinement ratio \(W/d = 8\)), \(g = 980\) cm/s\(^2\). The particle is
initially at rest on the channel centerline at \((x, y) = (1, 4)\) cm (laterally centered, released \(8d\) below the top
wall). All four channel walls are no-slip, imposed with the single-node velocity closure of Zou and He {[}52{]}
evaluated at zero wall velocity; the same wall treatment is used for the two-particle channel of Sec. 2.9.4. Four
density ratios are tested: the canonical \(\rho_s / \rho_f \in \{1.01, 1.1, 1.5\}\) at \(\nu = 0.01\) cm\(^2\)/s for the
Sec. 5.1 single-particle baseline, plus \(\rho_s / \rho_f = 1.25\) at \(\nu = 0.1\) cm\(^2\)/s (the Wang et al.~{[}16{]}
low-Reynolds parameter set) as the \emph{density-matched lighter control} used in Sec. 5.2 as a dynamically unmatched
single-particle reference for the wake-interaction light-particle closure sensitivity; the differing viscosity places
the \(\rho_s / \rho_f = 1.25\) control in a distinct kinematic regime (\(\mathrm{Re}_f \approx 13.9\) vs the
wake-interaction light-particle peak \(\approx 238.6\)), as interpreted in Sec. 7.1. Each case uses a cross-channel
(width) resolution \(N_x = 1281\) (\(D = 160\Delta x\), \(\Delta x = W/1280\); the same discretization as the Appendix C
grid analysis), i.e., a full \(1281 \times 3841\) (\(N_x \times N_y\)) grid over the \(W \times L = 2 \times 6\) cm
channel, and the Velocity-Verlet integrator of Sec. 2.7; the relaxation time follows per density ratio from the
Mach-number constraint, with \(\tau = 0.56\), \(0.64\), and \(0.95\) at \(\rho_s/\rho_f = 1.5\), \(1.1\), and \(1.01\),
respectively, and \(\tau = 1.41\) for the \(\rho_s/\rho_f = 1.25\) control.

\paragraph{2.9.4 Two-Particle Wake-Interaction Sedimentation}\label{two-particle-wake-interaction-sedimentation}

The two-particle differential-density wake-interaction configuration follows Majumder et al.~{[}23{]}:
\(\Omega = [0,10] \times [-1,1]\), \(d = 0.2\), heavy particle at \(\mathbf{X}_c^{(1)} = (0.8, -0.13)\) and light
particle at \(\mathbf{X}_c^{(2)} = (1.2, 0.13)\) (with gravity along \(+x\), the light particle thus starts downstream
of the heavy particle), \(\rho_r^{\mathrm{heavy}} = 1.5\) and \(\rho_r^{\mathrm{light}} = 1.25\),
\(\nu = 8 \times 10^{-4}\), lattice \(4001 \times 801\) (\(\Delta x = 0.0025\)), \(\tau = 0.59\),
\(\Delta t = 2.344 \times 10^{-4}\), gravitational acceleration \(g_{\mathrm{lat}} = 2.156 \times 10^{-4}\). The
Reynolds-number basis adopted for the wake-interaction case is the Uhlmann {[}14{]} / Majumder et al.~{[}23{]}
fluid-density form, consistent with the reference-defined reporting principle of Sec. 2.6. The domain, boundary
conditions, grid, and initial placement match the Majumder et al.~{[}23{]} reference configuration, so any finite-domain
(sidewall-mediated) influence is shared with the reference benchmark, and deviations are measured on that common
background.

\paragraph{2.9.5 Taylor--Green Decaying Vortex}\label{taylorgreen-decaying-vortex}

The Taylor--Green vortex {[}53{]} provides an analytical solution for convergence-order verification. The detailed
setup, \(L_2\) error tabulation, and convergence-order estimate are reported in Appendix D.

\subsection{3. Prescribed-Body Regime: Local No-Slip Fidelity}\label{prescribed-body-regime-local-no-slip-fidelity}

For prescribed bodies, the dominant distinction among boundary treatments is not a universal drag ranking but the degree
to which local no-slip and leakage errors are suppressed at the immersed interface. This section organizes the
fixed-cylinder (Sec.~3.1) and oscillating-cylinder (Sec.~3.3) results around three local fidelity diagnostics --- slip
error, leakage flux, and internal residual velocity (Sec.~3.2) --- and treats global integral quantities (\(C_d\),
\(\mathrm{St}\), \(C_l\)) as verification context. The full method × kernel × \(\mathrm{Re}\) tabulation is reported in
Supplementary Material Sec.~S1 (with cross-references in main Appendix B); the grid-sensitivity analysis at
\(\mathrm{Re}=40\) is reported in Supplementary Material Sec.~S2.2 (cross-ref Appendix C Sec.~C.3).

\subsubsection{3.1 Fixed Cylinder}\label{fixed-cylinder-1}

Table II summarizes the time-averaged drag coefficient \(\bar{C}_d\) for all method--delta combinations across four
Reynolds numbers, alongside the IB-LBM and body-fitted reference range compiled from the literature
{[}7,20,42,44--46,51,54,55{]}; complementary impulsively-started wake measurements and high-resolution simulations
{[}56,57{]} support the time-dependent wake structure analyzed in Sec.~B.7. Values are reported as
\(\bar{C}_d = \langle C_d \rangle_{t > t_{\mathrm{conv}}}\), the mean over the converged window (Table II caption).

The matrix reveals systematic trends in the correction direction (Fig. 2). MDF generally reduces \(\bar{C}_d\) relative
to DF on both kernels at \(40 \leq \mathrm{Re} \leq 100\) (largest reduction \(-2.19\%\) for Peskin 4-point at
\(\mathrm{Re} = 100\)); at \(\mathrm{Re} = 20\) MDF and DF coincide within \(0.3\%\) on both kernels. At
\(\mathrm{Re} = 200\), MDF \emph{raises} \(\bar{C}_d\) relative to DF on both kernels (\(+0.82\%\) hat, \(+3.28\%\) P4)
--- the iterative cumulative correction inverts sign as the kernel reversal sets in (cf.~Sec.~4). The DFC correction
direction is itself kernel- and Re-dependent: hat-DFC raises \(\bar{C}_d\) relative to DF at all four Reynolds numbers
(\(+0.32\%\), \(+0.51\%\), \(+1.40\%\), \(+4.94\%\) at \(\mathrm{Re} = 20, 40, 100, 200\)); Peskin 4-point DFC lowers it
at \(\mathrm{Re} \leq 100\) (\(-0.09\%\), \(-0.35\%\), \(-0.89\%\)) and at \(\mathrm{Re} = 200\) raises it
(\(+3.91\%\)). The kernel-dependent direction is one of the principal signatures of the kernel reversal analyzed in
Sec.~4. All \(\mathrm{Re} \leq 100\) values fall within the IB-LBM literature range. At \(\mathrm{Re} = 200\), the
Peskin 4-point cells (1.2569--1.3061, BGK collision; all schemes) underpredict the lower bound 1.349 by
\(3.2\)--\(6.8\%\) --- within \(\pm 7.3\%\) of the most directly comparable IB-LBM target (Wu and Shu {[}51{]}
\(\bar{C}_d = 1.349\), \(\mathrm{St} = 0.193\)) once the full BGK / TRT / CM-MRT P4 range (1.2509--1.3061; Table B.4) is
included. This kernel-dependent low bias is the spreading-kernel smoothness signature analyzed in Sec. 4 (Uhlmann
{[}14{]}, Peskin {[}3{]}). The hat cells (1.3577--1.4247, BGK collision) lie within the literature range.

\noindent \textbf{Table II.} Time-averaged drag coefficient \(\bar{C}_d\) for fixed cylinder
(\(\bar{C}_d = \langle C_d \rangle_{t > t_{\mathrm{conv}}}\), BGK collision). The averaging window is the converged tail
of each run: \(\mathrm{Re} \in \{20, 40, 100\}\) steps 70,200--100,000, except the \(\mathrm{Re} = 20\) hat DF run,
which reached the convergence criterion earlier and is averaged over steps 14,800--21,000; \(\mathrm{Re} = 200\) steps
30,200--60,000 (\(\approx 9.9\) vortex-shedding cycles at \(\mathrm{St} \approx 0.20\)). Literature range compiled from
body-fitted and IB-LBM references; the most directly comparable IB-LBM target at \(\mathrm{Re} = 200\) is Wu and Shu
{[}51{]} \(\bar{C}_d = 1.349\). MDF refers to the adaptive form (Sec.~2.5.2; \(N \in [5,20]\),
\(\omega_{\mathrm{MDF}} \approx \|A\|_\infty^{-1}\), \(L_\infty\) residual early termination).
\(\Delta_{\mathrm{MDF}}\%\) and \(\Delta_{\mathrm{DFC}}\%\) denote the percentage change of the column scheme relative
to DF.

\needspace{4\baselineskip}
\vspace{0.8em}
\sbox0{\scriptsize\setlength{\tabcolsep}{3pt}\renewcommand{\arraystretch}{1.2}%
\begin{tabular}{r l r r r r r l}
\toprule
Re & Delta & DF & MDF & DFC & \(\Delta_\mathrm{MDF}\%\) & \(\Delta_\mathrm{DFC}\%\) & Lit. range \\
\midrule
20 & hat & 2.0777 & 2.0786 & 2.0844 & \(+0.04\%\) & \(+0.32\%\) & 2.076--2.16 \\
20 & P4 & 2.0954 & 2.0893 & 2.0936 & \(-0.29\%\) & \(-0.09\%\) & 2.076--2.16 \\
40 & hat & 1.5625 & 1.5588 & 1.5705 & \(-0.24\%\) & \(+0.51\%\) & 1.555--1.62 \\
40 & P4 & 1.5772 & 1.5644 & 1.5716 & \(-0.81\%\) & \(-0.35\%\) & 1.555--1.62 \\
100 & hat & 1.3965 & 1.3849 & 1.4160 & \(-0.83\%\) & \(+1.40\%\) & 1.364--1.45 \\
100 & P4 & 1.4178 & 1.3867 & 1.4051 & \(-2.19\%\) & \(-0.89\%\) & 1.364--1.45 \\
200 & hat & 1.3577 & 1.3688 & 1.4247 & \(+0.82\%\) & \(+4.94\%\) & 1.349--1.44 \\
200 & P4 & 1.2569 & 1.2982 & 1.3061 & \(+3.28\%\) & \(+3.91\%\) & 1.349--1.44 \\
\bottomrule
\end{tabular}}
\ifdim\wd0>\textwidth\noindent\resizebox{\textwidth}{!}{\usebox0}\else\noindent\makebox[\textwidth][c]{\usebox0}\fi
\vspace{0.8em}

A grid-sensitivity analysis at \(\mathrm{Re}=40\) comparing the baseline resolution (\(N_y = 1601\), \(D = 40\Delta x\))
against a coarser grid (\(N_y = 801\), \(D = 20\Delta x\)) yields a relative \(C_d\) difference that does not exceed
\(1.7\%\) across the five tabulated method--delta combinations (Appendix C Sec. C.3), indicating that the inter-method
ordering in Table II persists across the two tested resolutions and is not reversed by the grid refinement.

\begin{figure}
\centering
\includegraphics{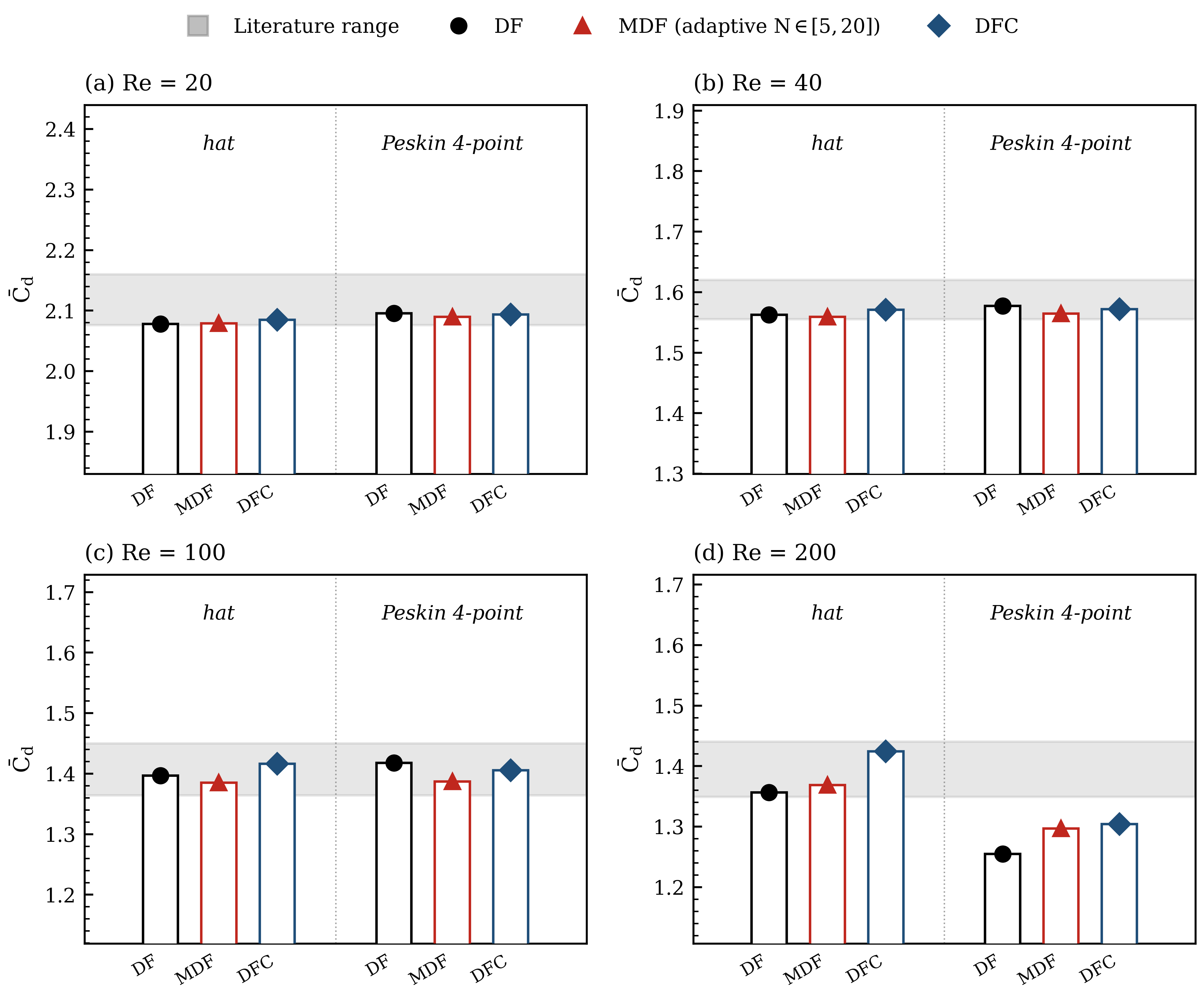}
\caption{\textbf{Fig. 2.} Time-averaged drag coefficient \(\bar{C}_d\) comparison at four Reynolds numbers (BGK
collision): (a) \(\mathrm{Re}=20\), (b) \(\mathrm{Re}=40\), (c) \(\mathrm{Re}=100\), (d) \(\mathrm{Re}=200\). Each panel
shows hat (left) and Peskin 4-point (right) results. Gray bands indicate the literature range. Each panel uses a zoomed
y-axis (non-zero base) to resolve the per-scheme differences. Markers distinguish the three boundary-enforcement schemes
(DF: black circles; MDF (adaptive \(N \in [5,20]\), Sec.~2.5.2): red triangles; DFC: blue diamonds). DFC (blue) shows
opposite correction directions relative to DF for hat versus Peskin 4-point at Re \(\leq 100\) (hat raises
\(\bar{C}_d\), P4 lowers it) but raises \(\bar{C}_d\) on both kernels at \(\mathrm{Re}=200\) --- the
kernel-and-Re-dependent reversal analyzed in Sec.~4.}
\end{figure}

The Strouhal number and lift-coefficient amplitude at \(\mathrm{Re}=100\) and \(\mathrm{Re}=200\) are reported in
Supplementary Material Sec.~S1.2 Tables S2 and S3 (full 36-case matrix; main Appendix B Sec.~B.2 provides the
cross-reference). The unsteady diagnostics bear on local fidelity. Among the tested cases, all six method--delta
combinations at \(\mathrm{Re}=200\) share the uniform FFT-bin Strouhal estimate \(\mathrm{St} = 0.200\) (consistent with
the literature range \(0.193\)--\(0.197\), spanning the IB-LBM value of Wu and Shu {[}51{]} and the experimental value
of Williamson {[}58{]}), at the FFT resolution \(\Delta\mathrm{St} \approx 0.02\) set by the \(\approx 9.9\)-cycle
observation window (Supplementary Material Sec.~S1.2 Table S3); and the \(C_{l,\mathrm{amp}}\) at \(\mathrm{Re}=100\)
exceeds the body-fitted Navier--Stokes literature upper bound (\(C_{l,\mathrm{amp}} = 0.33\); Supplementary Material
Sec. S1.2 Table S2) by \(0.3\)--\(5.2\%\) across all six BGK method--delta combinations, suggesting a solver-level
rather than method-level artifact at this Reynolds number~(the contribution of the SRT-BGK collision operator to this
overprediction is examined under the targeted controls of Sec.~6; any residual gap is hypothesized to stem from LBM
weak-compressibility \(\mathcal{O}(\mathrm{Ma}^2)\) artifacts or two-dimensional assumption limits, with a definitive
attribution requiring direct comparison against an incompressible Navier--Stokes solver --- identified as future work).
At \(\mathrm{Re}=200\), by contrast, the lift response is kernel-dependent: the hat-kernel \(C_{l,\mathrm{rms}}\) is
comparable to the body-fitted reference of Qu et al.~{[}59{]} whereas the Peskin 4-point cases lie systematically below
it (Supplementary Material Sec. S1.2).

\subsubsection{3.2 Boundary Fidelity Metrics}\label{boundary-fidelity-metrics}

Table III reports the mean slip error at the Lagrangian boundary points across method--delta combinations and Reynolds
numbers; the full leakage-flux and internal-residual-velocity matrix is reported in Supplementary Material Sec.~S1.3
(main Appendix B Sec.~B.3 provides the cross-reference). The slip error provides the most discriminating local
diagnostic.

\noindent \textbf{Table III.} Mean slip error \(\varepsilon_{\mathrm{slip,mean}}\), normalized by \(U_\infty\) (selected cases).
The improvement factor is the slip-error reduction ratio
\(\varepsilon_{\mathrm{slip}}^{\mathrm{DF}} / \varepsilon_{\mathrm{slip}}^{\mathrm{MDF}}\) (or
\(\varepsilon_{\mathrm{slip}}^{\mathrm{DF}} / \varepsilon_{\mathrm{slip}}^{\mathrm{DFC}}\)); larger values indicate
stronger slip suppression. MDF refers to the adaptive form (Sec.~2.5.2).

\needspace{4\baselineskip}
\vspace{0.8em}
\sbox0{\scriptsize\setlength{\tabcolsep}{3pt}\renewcommand{\arraystretch}{1.2}%
\begin{tabular}{r l r r r r r}
\toprule
Re & Delta & DF & MDF & DFC & DF→MDF & DF→DFC \\
\midrule
20 & hat & \(6.02\times 10^{-3}\) & \(6.00\times 10^{-4}\) & \(1.99\times 10^{-3}\) & 10.0× & 3.0× \\
20 & P4 & \(1.23\times 10^{-2}\) & \(9.88\times 10^{-5}\) & \(1.39\times 10^{-4}\) & 124.9× & 89.1× \\
40 & hat & \(5.04\times 10^{-3}\) & \(6.17\times 10^{-4}\) & \(1.97\times 10^{-3}\) & 8.2× & 2.6× \\
40 & P4 & \(9.99\times 10^{-3}\) & \(1.25\times 10^{-4}\) & \(1.72\times 10^{-4}\) & 80.1× & 58.1× \\
100 & hat & \(5.29\times 10^{-3}\) & \(7.08\times 10^{-4}\) & \(2.47\times 10^{-3}\) & 7.5× & 2.1× \\
100 & P4 & \(9.73\times 10^{-3}\) & \(1.76\times 10^{-4}\) & \(2.75\times 10^{-4}\) & 55.3× & 35.3× \\
200 & hat & \(5.55\times 10^{-3}\) & \(6.71\times 10^{-4}\) & \(2.32\times 10^{-3}\) & 8.3× & 2.4× \\
200 & P4 & \(1.04\times 10^{-2}\) & \(1.53\times 10^{-4}\) & \(2.55\times 10^{-4}\) & 68.1× & 40.8× \\
\bottomrule
\end{tabular}}
\ifdim\wd0>\textwidth\noindent\resizebox{\textwidth}{!}{\usebox0}\else\noindent\makebox[\textwidth][c]{\usebox0}\fi
\vspace{0.8em}

MDF achieves the strongest slip reduction (\(7.5\)--\(10.0\)× for hat, \(55\)--\(125\)× for P4). The pronounced
improvement under P4 (up to \(125\)×) reflects that P4's wider support generates larger DF slip errors that MDF
iterations effectively correct. DFC's slip improvement is comparable to MDF for P4 (\(35\)--\(89\)× vs \(55\)--\(125\)×)
but markedly weaker for hat (\(2.1\)--\(3.0\)× vs \(7.5\)--\(10.0\)×). This kernel-dependent improvement gap is one of
the local signatures of the DFC kernel reversal analyzed in Sec.~4. Internal residual velocity (1--\(9\%\) of
\(U_\infty\), Supplementary Material Sec.~S1.3) persists across all three schemes. This is a known structural limitation
of boundary-only IBM formulations {[}51{]}, which do not enforce a velocity correction inside the solid. Implicit
velocity correction (IVC) schemes {[}51{]} suppress this residual via interior-node velocity correction; their
integration with the DF / MDF / DFC family compared here is identified as a separate study (Sec.~7).

\begin{figure}
\centering
\includegraphics{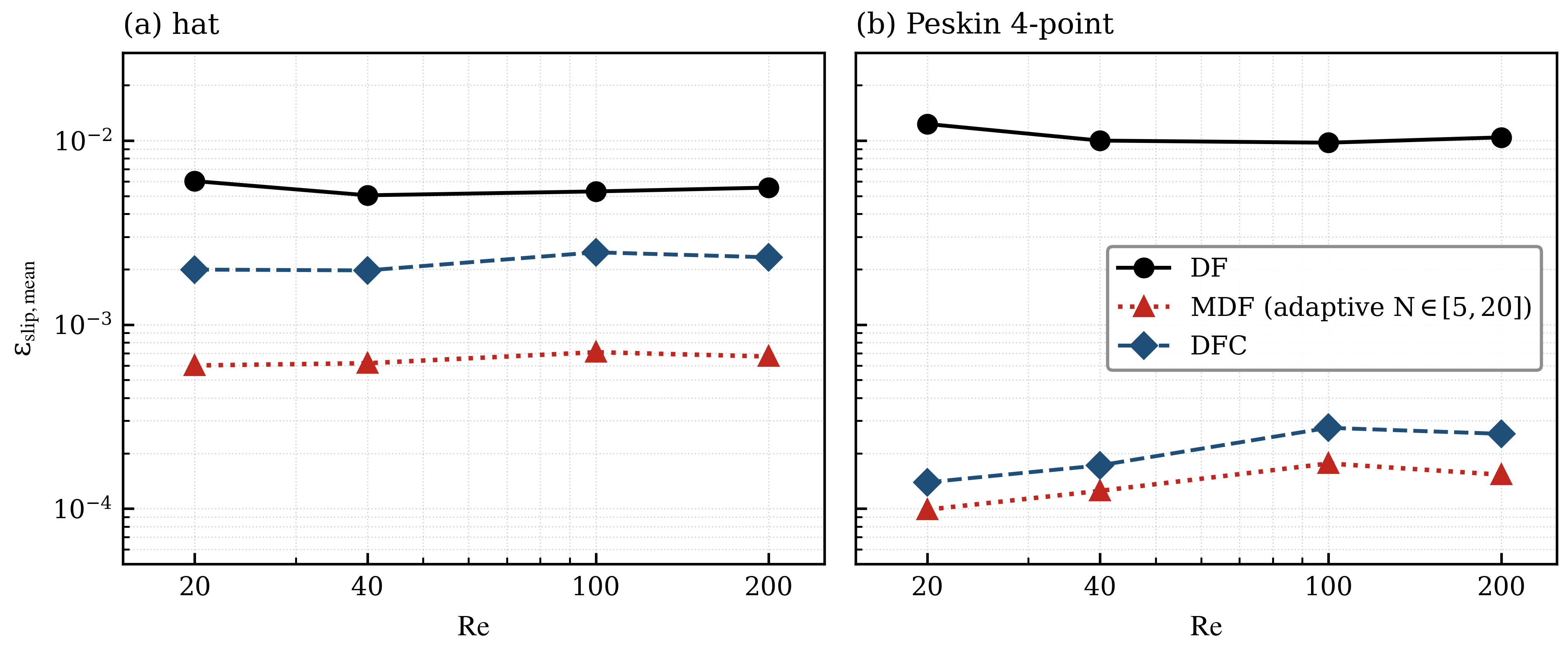}
\caption{\textbf{Fig. 3.} Boundary slip error comparison for (a) hat delta function and (b) Peskin 4-point delta
function at \(\mathrm{Re} \in \{20, 40, 100, 200\}\) on the Table I standard grid (\(D = 40\Delta x\) for
\(\mathrm{Re} \leq 100\) and \(D = 60\Delta x\) for \(\mathrm{Re} = 200\); Table III cases). Note the logarithmic
y-axis. MDF (adaptive \(N \in [5,20]\), Sec.~2.5.2; red) achieves the lowest slip error in all cases; DFC (blue)
approaches MDF performance with P4 but shows weaker improvement with hat --- the kernel-dependent local-fidelity gap
that motivates the correction-redistribution analysis of Sec.~4.}
\end{figure}

\subsubsection{3.3 Oscillating Cylinder}\label{oscillating-cylinder-1}

The oscillating-cylinder case provides a prescribed moving-boundary benchmark. All schemes (DF / MDF / DFC) and both
kernels yield peak drag coefficients close to the Dütsch et al.~{[}39{]} reference value (\(c_d = 2.09\)), which was
obtained by fitting the Morison equation to their in-line force history. The Morison-fitted coefficient and the present
peak coefficient are different observables, so the agreement locates the present results on the reference scale without
constituting a matched-observable comparison; per-case peak drag coefficients are reported in Appendix B Sec. B.8 (Table
B.8).

DFC exhibits markedly elevated delta-function sensitivity in this configuration: the hat--P4 difference is \(2.7\%\)
(\(2.047\) vs \(2.102\)), compared to \(0.6\%\) for DF and \(0.3\%\) for MDF. Under these moving-boundary conditions
\(\lambda_k\) is recomputed at every time step rather than cached, and the analysis of Sec.~4 associates this enhanced
sensitivity with the spatial non-uniformity of \(\lambda_k\) under narrow-support kernels.

In the prescribed-body regime, the principal mechanism is therefore \emph{local no-slip fidelity}: the differences among
DF, MDF, and DFC are most clearly resolved by the slip error and the kernel-dependent slip-improvement gap, rather than
by a universal \(C_d\) ranking.

\subsection{4. Correction-Dominated Regime: DFC Kernel Reversal}\label{correction-dominated-regime-dfc-kernel-reversal}

The DFC reversal points to the spatial distribution of the correction around the interface --- rather than to total
correction magnitude or kernel smoothness alone --- as one signature of the kernel-dependent error. This section reports
marker-resolved correction magnitude, azimuthal slip profiles, and near-boundary pressure / velocity residuals, and
quantifies the spatial non-uniformity of the DFC correction scaling factor \(\lambda_k\) as a kernel-geometric
descriptor (Sec.~4.2). A bounded \(W/\delta\) scale argument (Sec.~4.3) provides a kernel-ordering rationale.

\subsubsection{\texorpdfstring{4.1 Cross-Method \(\Delta C_d\) Sensitivity to Kernel
Choice}{4.1 Cross-Method \textbackslash Delta C\_d Sensitivity to Kernel Choice}}\label{cross-method-delta-c_d-sensitivity-to-kernel-choice}

Fig. 4 reports the \(\Delta C_d\) sensitivity to the kernel choice --- defined as
\((\bar{C}_{d,\,P4} - \bar{C}_{d,\,hat}) / \bar{C}_{d,\,hat} \times 100\%\) on the time-averaged \(\bar{C}_d\) ---
across the three boundary-enforcement schemes (DF, MDF, DFC) and four Reynolds numbers
(\(\mathrm{Re} \in \{20, 40, 100, 200\}\)). The three schemes differ systematically in their kernel sensitivity. MDF
stays within \(|\Delta C_d| \leq 0.51\%\) over \(\mathrm{Re} \leq 100\), consistent with its iterative
cumulative-correction mechanism compensating for kernel-induced inhomogeneity at the boundary; the kernel reversal
extends to MDF only at \(\mathrm{Re} = 200\) (\(-5.16\%\)). DF crosses zero between \(\mathrm{Re} = 100\) (\(+1.53\%\))
and \(\mathrm{Re} = 200\) (\(-7.42\%\)), so \(\bar{C}_{d,\,P4}\) exceeds \(\bar{C}_{d,\,hat}\) at low--moderate
\(\mathrm{Re}\) but is overtaken at \(\mathrm{Re} = 200\). DFC, in turn, crosses zero between \(\mathrm{Re} = 40\)
(\(+0.07\%\)) and \(\mathrm{Re} = 100\) (\(-0.77\%\)), and reaches \(-8.32\%\) at \(\mathrm{Re} = 200\). DFC reverses
earliest (between \(\mathrm{Re}=40\) and \(\mathrm{Re}=100\)) and most strongly (\(-8.32\%\)) at \(\mathrm{Re} = 200\);
DF reverses second (between \(\mathrm{Re}=100\) and \(\mathrm{Re}=200\), \(-7.42\%\)); MDF reverses last (only at
\(\mathrm{Re}=200\), \(-5.16\%\), the smallest of the three at \(\mathrm{Re} = 200\)). This ordering of the
kernel-induced sign-reversal magnitudes is the central signature of Sec.~4.

What controls the DFC reversal is clarified by the correction magnitude and the azimuthal slip profile, considered in
turn.

The total correction magnitude does not explain the reversal. The integrated correction magnitudes
\(\sum_k |\mathbf{F}_{s,k}|\) at \(\mathrm{Re}=100\) are \(0.3926\) for hat and \(0.3911\) for Peskin 4-point ---
comparable across kernels (cf.~Supplementary Material Sec.~S1.3 Table S4). Yet the maximum local slip differs by a
factor of three: \(1.56\%\) of \(U_\infty\) for hat and \(4.72\%\) of \(U_\infty\) for Peskin 4-point. These
marker-sampled maxima are obtained with the bilinear diagnostic probe of Fig. 5 and are distinct from the scheme-level
kernel-interpolated \(\varepsilon_{\mathrm{slip}}\) of Table III, so the kernel ordering of the two slip metrics
differs. The reversal is therefore consistent with the \emph{spatial distribution} of the correction field, of which the
kernel-induced redistribution is one signature, rather than with the total correction.

The azimuthal slip profile shows that slip magnitude and marker-scale enforcement uniformity are independent axes. The
hat profile combines low absolute slip (max \(1.56\%\) of \(U_\infty\) at \(\mathrm{Re} = 100\)) with strong
marker-to-marker variability, with peaks concentrated at locations where \(\lambda_k\) is large (markers positioned
between Eulerian grid nodes). The Peskin 4-point profile combines higher absolute slip (max \(4.72\%\) of \(U_\infty\)
at \(\mathrm{Re} = 100\), \(\sim 3\times\) hat) with a marker-scale-smooth correction scaling (P4 \(\lambda_k\) CV
\(\approx 1.8\%\)), consistent with the wider support averaging out the marker-to-marker variability of the enforcement,
while its slip profile retains a broad angular structure with localized peaks (Sec. 4.2, Fig. 5(c)). The kernel reversal
is therefore associated with the marker-scale non-uniformity of the correction scaling rather than with the absolute
slip magnitude: hat's lower local slip is concentrated at large-\(\lambda_k\) markers, creating a numerically
non-uniform enforcement pattern that becomes significant as the boundary layer thins at high Re. The full
marker-resolved \(\lambda_k\) scaling, correction magnitude, azimuthal slip profile, and near-boundary pressure
deviation diagnostics are reported in main Fig. 5 (Sec.~4.2 below) with the per-case summary statistics in Supplementary
Material Table S4.

\begin{figure}
\centering
\includegraphics{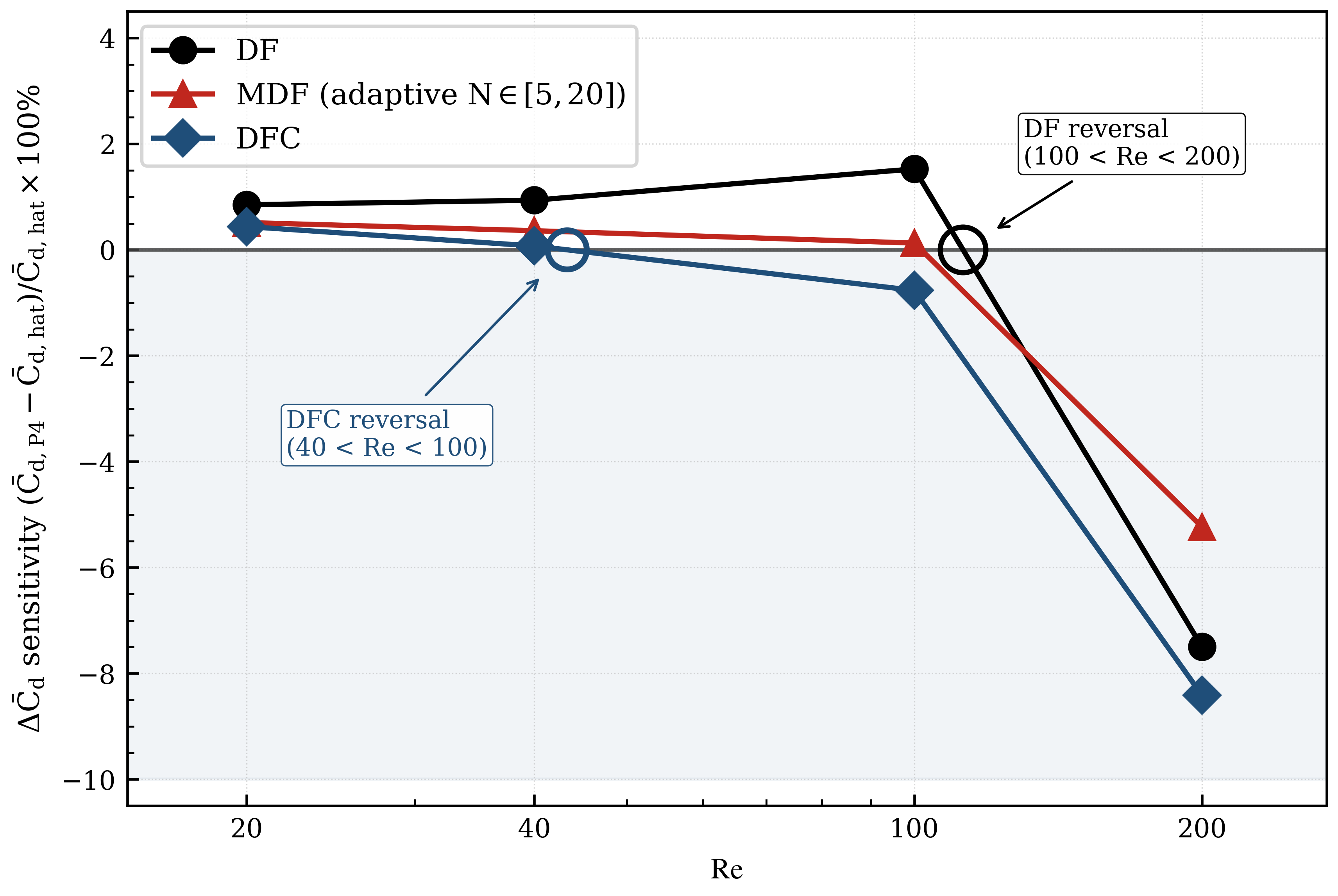}
\caption{\textbf{Fig. 4.} Cross-method \(\Delta C_d\) sensitivity to kernel choice. \(\Delta C_d\) sensitivity
\(= (\bar{C}_{d,\,P4} - \bar{C}_{d,\,hat}) / \bar{C}_{d,\,hat} \times 100\%\) for DF (black circles, solid), MDF (red
triangles, solid), and DFC (blue diamonds, solid) at \(\mathrm{Re} \in \{20, 40, 100, 200\}\), computed from the
converged tail of the time-averaged \(\bar{C}_d\) (Table II caption). The shaded blue band marks
\(\bar{C}_{d,\,P4} < \bar{C}_{d,\,hat}\) (reversed ranking). Reversal positions are bracketing Re intervals between
adjacent measured datapoints. See Sec. 4.1 for the reversal ordering and Sec. 4.2 / Fig. 5 for the marker-resolved
diagnostics.}
\end{figure}

\subsubsection{\texorpdfstring{4.2 Quantitative Analysis of \(\lambda_k\) Spatial
Non-uniformity}{4.2 Quantitative Analysis of \textbackslash lambda\_k Spatial Non-uniformity}}\label{quantitative-analysis-of-lambda_k-spatial-non-uniformity}

The DFC correction scaling factor \(\lambda_k\) depends on the regularized delta function and on the relative position
of each marker against the Eulerian grid. A reduced geometry (\(D = 20\Delta x\), 94 markers) is reported first as the
cleanest single-resolution illustration of the kernel-grid dependence; the primary \(D = 40\Delta x\) benchmark and the
\(D = 160\Delta x\) sedimentation geometry are then reported to bound the kernel-grid contrast against marker-count
effects (the wide-support Peskin 4-point CV scales mildly with marker density, whereas the narrow-support hat CV does
not). On the reduced geometry, the two delta functions produce fundamentally different \(\lambda_k\) distributions. The
hat function gives \(\bar{\lambda} = 0.73\) with a coefficient of variation \(\mathrm{CV} = 18.1\%\) (\(\max/\min\)
ratio \(2.0\)), whereas Peskin 4-point gives \(\bar{\lambda} = 1.32\) with \(\mathrm{CV} = 1.4\%\) (\(\max/\min\) ratio
\(1.09\)).

The two kernels also have different \emph{means} on this geometry (\(\bar{\lambda}_\mathrm{hat} = 0.73\) vs
\(\bar{\lambda}_\mathrm{P4} = 1.32\), a factor of \(\approx 1.82\)); this difference reflects the spreading-kernel sum
normalization and is in itself expected --- the comparison axis of interest here is the \emph{non-uniformity} (CV) on
each kernel, which is the spatial-redistribution metric that connects to the Sec.~4.1 reversal. The absolute standard
deviation \(\sigma_\lambda = \mathrm{CV} \cdot \bar{\lambda}\) recovers the same kernel ordering:
\(\sigma_\mathrm{hat} = 0.131\) vs \(\sigma_\mathrm{P4} = 0.018\), a \(\approx 7\times\) ratio in absolute spread. The
hat function thus produces a substantially larger spatial non-uniformity in the correction scaling than Peskin 4-point
on this reduced geometry, on both the relative (CV) and absolute (\(\sigma\)) axes. To bound the kernel-grid contrast
against marker-count effects, the same diagnostic is computed at the two primary working resolutions: the sedimentation
benchmark (\(D = 160\Delta x\), 503 markers) yields hat \(\mathrm{CV} = 17.9\%\) at the on-node center placement
(\(16.9\%\) averaged over a fixed \(8\times 8\) subgrid of sub-lattice center offsets), Peskin 4-point
\(\mathrm{CV} = 1.1\%\) (unchanged under offset averaging), and the primary fixed-cylinder benchmark on which the
marker-resolved diagnostics of Fig. 5 are computed (\(D = 40\Delta x\), 188 markers) likewise gives hat
\(\mathrm{CV} = 17.3\%\), Peskin 4-point \(\mathrm{CV} = 1.8\%\). Across the three resolutions the hat / Peskin 4-point
CV ratio remains of order \(10\) (\(\approx 10\)--\(16\times\), with the main \(D = 40\Delta x\) case at
\(\sim 10\times\) as the most directly comparable benchmark to Fig. 5), indicating that the qualitative non-uniformity
persists over the tested marker resolutions and is consistent with a delta-function--grid interaction; the Peskin
4-point CV variation (\(1.1\)--\(1.8\%\)) reflects the marker-count sensitivity of the wide-support kernel and is small
relative to the hat baseline of \(\sim 17\)--\(18\%\) in all cases.

This spatial non-uniformity has a direct geometric interpretation. With hat, boundary enforcement alternates between
local over-correction (at markers where \(\lambda_k\) is large, i.e., a marker positioned between grid nodes) and
under-correction (at markers where \(\lambda_k\) is small, i.e., a marker positioned near a grid node), creating a
numerically non-uniform enforcement pattern. At low Re, viscous diffusion smooths out such marker-scale variability. As
Re increases and the boundary layer thins, viscous diffusion is less able to smooth the spatially non-uniform
correction, and the net effect is consistent with a measurably different effective boundary --- in line with the growing
hat--P4 divergence of \(C_d\) in Table II at \(\mathrm{Re}\geq 100\) and the kernel-dependent slip-improvement gap of
Sec.~3.2.

The marker-resolved local-field diagnostics that support this spatial-redistribution interpretation are summarized in
Fig. 5 across three DFC mechanism diagnostic cases (\(\mathrm{Re}=40\) hat, \(\mathrm{Re}=100\) hat, \(\mathrm{Re}=100\)
P4; BGK collision; primary grid resolution). The hat cases exhibit pronounced marker-to-marker non-uniformity in panels
(a)--(b), with peaks colocated at large-\(\lambda_k\) markers (positioned between Eulerian grid nodes), whereas the
Peskin 4-point case shows azimuthally smooth \(\lambda_k\) and \(|\mathbf{F}_{s,k}|\) profiles; panel (c) further shows
that the azimuthal slip residual remains kernel-dependent in magnitude --- Peskin 4-point at \(\mathrm{Re}=100\) retains
a finite slip residual with localized peaks even though its \(\lambda_k\) profile is azimuthally smooth --- a smooth
\(\lambda_k\) profile does not by itself produce a smooth azimuthal slip profile. Fig. 5(c) reports the azimuthal
marker-level slip profile, complementing the marker-averaged slip residual in Fig. 3 (Sec. 3.2). The marker-resolved
local fields are shown at \(\mathrm{Re}\leq 100\), where the redistribution signature is clearest; per-case summary
statistics for the \(\mathrm{Re}=200\) cases, at which the integral kernel reversal is strongest (Fig. 4), are tabulated
in Supplementary Material Table S4.

\begin{figure}
\centering
\includegraphics{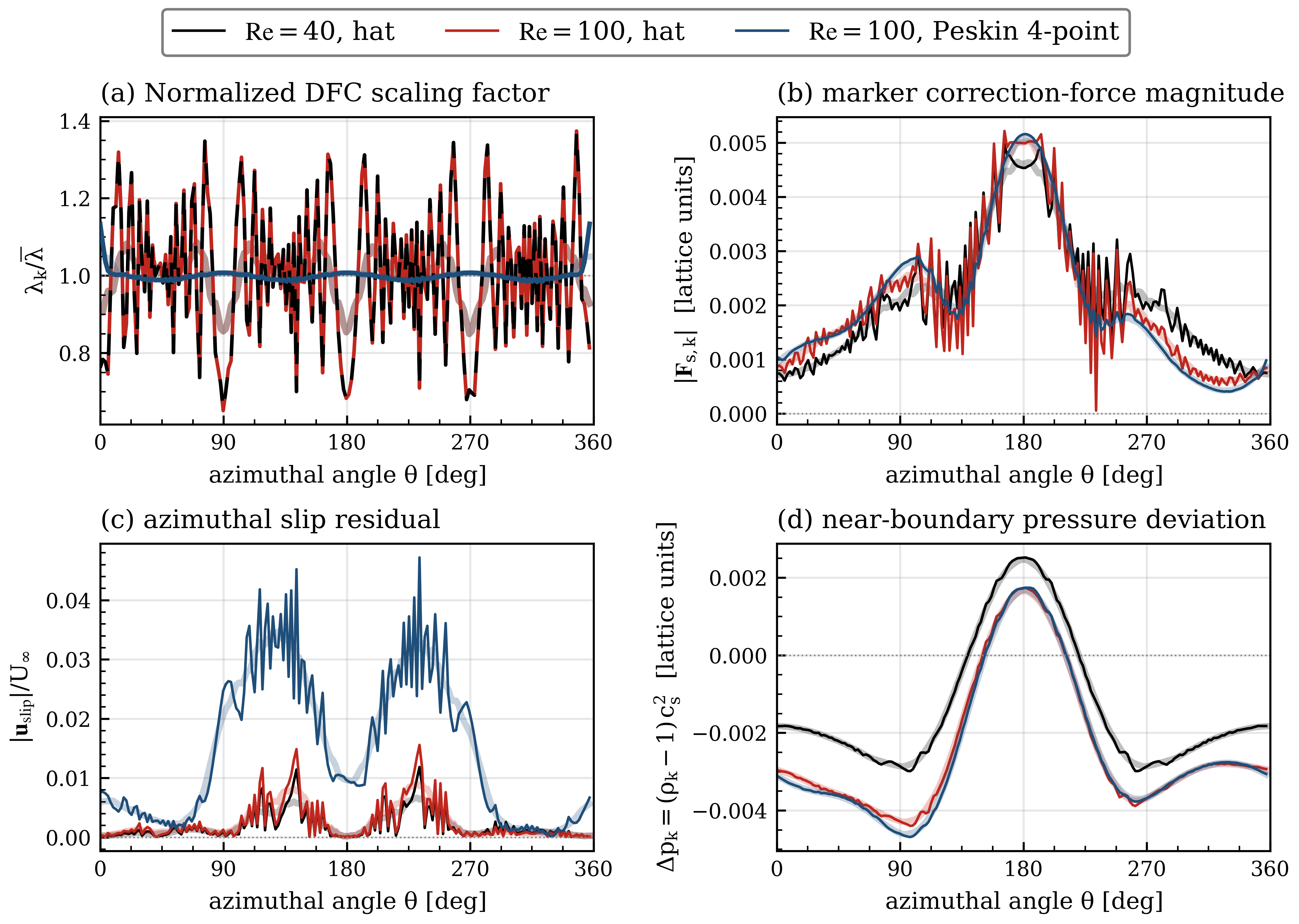}
\caption{\textbf{Fig. 5.} DFC marker-resolved local-field diagnostics across three configurations (\(\mathrm{Re}=40\)
hat, \(\mathrm{Re}=100\) hat, \(\mathrm{Re}=100\) P4; BGK collision; primary fixed-cylinder grid \(D = 40\Delta x\), 188
markers; main Table I). Panels show (a) the marker-resolved DFC scaling factor \(\lambda_k / \overline{\lambda}\) versus
azimuthal angle, (b) the per-marker correction-force magnitude \(|\mathbf{F}_{s,k}|\), (c) the azimuthal slip-velocity
profile \(|\mathbf{u}_{\mathrm{slip}}|/U_\infty\), and (d) the near-boundary pressure deviation
\(\Delta p_k = (\rho_k - 1)\,c_s^2\) at the marker positions. The marker-level slip and pressure fields in panels
(c)--(d) are sampled at the marker positions by bilinear interpolation of the Eulerian field, a diagnostic probe
distinct from the scheme's own hat / Peskin 4-point IBM interpolation of Eq. (7). In every panel, the faint, wider
curves are an 11-point moving-average overlay of the corresponding marker-resolved profile, shown only as a visual guide
to the azimuthal trend and not as separate data.}
\end{figure}

\subsubsection{\texorpdfstring{4.3 Kernel-Support Scale
(\(W/\delta\))}{4.3 Kernel-Support Scale (W/\textbackslash delta)}}\label{kernel-support-scale-wdelta}

A simple scale ratio between the kernel support \(W\) (1D: hat \(W = 2h\), Peskin 4-point \(W = 4h\)) and an estimated
laminar boundary-layer thickness \(\delta \sim D / \sqrt{\mathrm{Re}}\) (Schlichting scaling; the flat-plate Blasius
prefactor \(\delta \approx 5\,D/\sqrt{\mathrm{Re}}\) is used only as an order-of-magnitude estimate) is consistent with
the observed kernel ordering of DFC slip across the available cases. The two-cluster ordering --- hat
\(W/\delta \in [0.06, 0.10]\) with slip \(\sim 1.2\)--\(1.6\%\) \(U_\infty\), P4 \(W/\delta \in [0.19, 0.20]\) with slip
\(\sim 4.7\%\) \(U_\infty\), and a P4/hat slip ratio of \(3.0\)--\(3.4\times\) at fixed Re --- supports a bounded,
kernel-ordering scale rationale. The corresponding rank-correlation supporting diagnostics across the five datapoints
are summarized in Appendix C Sec.~C.6; the detailed tabulation and the small-sample caveat are given in Supplementary
Material Sec. S2.5.

\subsubsection{4.4 Synthesis of the Kernel-Reversal Diagnostics}\label{synthesis-of-the-kernel-reversal-diagnostics}

Across the four DFC mechanism summaries (correction magnitude, azimuthal slip, near-boundary residual, \(\lambda_k\)
non-uniformity) and the \(W/\delta\) scale rationale, the combined evidence is consistent with DFC reversal being one
signature of the spatial distribution of the correction field. The full causal chain --- from non-uniform \(\lambda_k\)
through anisotropic \(|\mathbf{F}_{s,k}|\) corrections to an altered effective boundary shape --- would require
additional global fluid-field reconstruction beyond the boundary-local \(\Delta p_k\) already shown in Fig. 5(d) ---
specifically, the near-wake stress tensor and pressure-gradient field propagated outward from the corrected distribution
functions. This global-field extension is identified as a future direction. The present diagnostics document a
redistribution signature consistent with the observed drag reversal.

In a related diffuse-interface IBM analysis, Peng and Wang {[}60{]} showed analytically that the two-sided force
distribution of the diffuse-interface IBM kernel modifies the momentum equation within the diffused solid--fluid
interface --- where the boundary force is distributed to fluid nodes --- away from the Navier--Stokes equations, and
proposed a single-sided kernel to restore the Navier--Stokes equations in the fluid region. That analysis concerns the
velocity-gradient computation within the diffused interface, whereas the present setting is the kernel-dependent DFC
drag reversal of Sec.~4.1; both observations are nonetheless consistent with the underlying view that \emph{the spatial
distribution of the fluid-side forcing is itself a controlling factor} of the boundary response. The present work
documents this signature in the DF / MDF / DFC family through marker-resolved \(\lambda_k\), slip, and pressure
diagnostics; a unified treatment connecting the two-sided / single-sided kernel choice of {[}60{]} to the
boundary-enforcement schemes compared here is identified as a future direction. More generally, regularized-delta
immersed-boundary formulations are limited to lower-order accuracy at the interface because the spreading kernel smooths
the stress discontinuity that generically occurs there {[}61{]}, so the marker-resolved redistribution diagnosed here is
the DFC-specific signature of a diffuse-interface character shared across the method family. Gruninger and Griffith
{[}62{]} recently introduced composite B-spline delta functions whose polynomial degree differs between the normal and
tangential directions of each velocity component, providing continuously divergence-free velocity interpolation and
gradient-preserving force spreading that eliminate a key source of spurious flows in finite-difference IB computations,
indicating that kernel design remains an active route to controlling the interface fidelity examined here.

\subsection{5. Moving-Particle Regime: Single-Particle Baseline and Wake-Exposed Closure
Sensitivity}\label{moving-particle-regime-single-particle-baseline-and-wake-exposed-closure-sensitivity}

Sec.~5.1 treats the single-particle case as a \emph{reference-defined baseline}: the present manuscript adopts the
Reynolds-number basis defined by the corresponding benchmark and reports each comparison in that basis. The
single-particle cases therefore serve as moving-body baselines for the two-particle analysis. For the targeted control
computations that support the analyses of Secs. 5 and 6 --- the isolated-particle and extended-domain runs, the
marker-retraction and grid-refinement checks, and the cross-scheme closure comparisons --- the analysis windows,
observables, and reporting definitions are stated with each result. Sec.~5.2 then treats the two-particle
differential-density wake-interaction case as the \emph{wake-exposed closure sensitivity} analysis: the wake-exposed
light particle is comparatively more sensitive to explicit internal-mass correction than the isolated single-particle
baseline. The single-particle ablation provides the small-baseline contrast for the wake-interaction comparison.

\subsubsection{5.1 Single-Particle Sedimentation Baseline}\label{single-particle-sedimentation-baseline}

Table IV reports the dimensionless steady-fall maximum settling velocity \(v_y^* = |v_y|_\mathrm{max} / u_g\) for the
three density ratios across all method--delta combinations, with \(u_g = \sqrt{|\rho_s/\rho_f - 1| \, g \, D}\) the
gravitational velocity scale used as the dimensionless unit throughout Sec.~5 and Appendix A Sec.~A.1. The corresponding
settling trajectories and particle-centered streamlines near the terminal plateau are shown in Fig. 6. The reported
\(v_y^*\) value is the maximum of the \(u_g\)-normalized \(|v_y|\) series over the steady-fall segment preceding wall
deceleration (for a single particle in a finite channel this maximum lies within \(\lesssim 0.8\%\) of the terminal
plateau; Supplementary Material Sec. S4.2, Table S10), reported against the corresponding terminal/maximum value of each
reference benchmark (Wang et al.~{[}16{]}, Glowinski et al.~{[}40{]}) rather than a contact-time instantaneous value.
The three-way spread among DF, MDF, and DFC remains within \(0.19\)--\(0.88\%\) across the analyzed density-ratio range
\(\rho_s/\rho_f \in \{1.01, 1.1, 1.5\}\) (Table IV), confirming that --- under a fixed collision model --- the
boundary-enforcement scheme has negligible influence on the settling velocity at this level of resolution; the spread
grows monotonically with \(\rho_s/\rho_f\), suggesting a mild density-ratio dependence that should be re-checked if the
analysis is extended beyond \(\rho_s/\rho_f = 1.5\). Particle-resolved immersed-boundary and lattice-Boltzmann methods
have been applied extensively to settling and particle-laden flows, including interface-resolved single-sphere settling
benchmarks {[}63{]}, two-dimensional lattice-Boltzmann sedimentation of elliptical particles {[}64{]}, and
direct-forcing immersed-boundary suspensions {[}65{]}; the present single-particle cases provide the controlled,
low-sensitivity baseline for the wake-exposed comparison of Sec. 5.2. Channel confinement remains quantitatively
significant in this regime: a recent coupled lattice-Boltzmann--discrete-element study {[}66{]} found channel-width
effects on the terminal settling velocity and on two-particle drafting--kissing--tumbling (DKT) dynamics to persist up
to width-to-diameter ratios of about 15; since every comparison reported here holds the channel geometry fixed across
boundary-enforcement schemes, kernels, and collision models, such confinement effects are controlled by construction.

\noindent \textbf{Table IV.} Dimensionless steady-fall maximum settling velocity \(v_y^* = |v_y|_\mathrm{max} / u_g\) at
\(\rho_s/\rho_f \in \{1.01, 1.1, 1.5\}\) (maximum over the steady-fall segment preceding wall deceleration; the
transient maximum exceeds the terminal plateau by \(\lesssim 0.8\%\), Table S10). Spread
\(= (\max - \min) / \mathrm{mean} \times 100\%\) across the three schemes at each delta function.

\needspace{4\baselineskip}
\vspace{0.8em}
\sbox0{\scriptsize\setlength{\tabcolsep}{3pt}\renewcommand{\arraystretch}{1.2}%
\begin{tabular}{c l r r r r}
\toprule
\(\rho_s/\rho_f\) & Delta & DF & MDF & DFC & Spread \\
\midrule
1.01 & P4 & 0.8791 & 0.8808 & 0.8794 & \(0.19\%\) \\
1.01 & hat & 0.8810 & 0.8818 & 0.8797 & \(0.24\%\) \\
1.1 & P4 & 1.0982 & 1.1045 & 1.1005 & \(0.57\%\) \\
1.1 & hat & 1.1019 & 1.1053 & 1.0983 & \(0.64\%\) \\
1.5 & P4 & 1.2154 & 1.2258 & 1.2203 & \(0.85\%\) \\
1.5 & hat & 1.2204 & 1.2260 & 1.2152 & \(0.88\%\) \\
\bottomrule
\end{tabular}}
\ifdim\wd0>\textwidth\noindent\resizebox{\textwidth}{!}{\usebox0}\else\noindent\makebox[\textwidth][c]{\usebox0}\fi
\vspace{0.8em}

For sedimentation comparisons against the literature, the reported Reynolds number follows the definition used in the
corresponding reference benchmark; the basis is stated in the table caption. Wang et al.~{[}16{]} use a particle-density
basis \(\mathrm{Re}_p = \rho_P |U| D_P / \mu\), whereas Glowinski et al.~{[}40{]} and Uhlmann {[}14{]} use a
fluid-density basis \(\mathrm{Re}_f = |U| D / \nu\). The conversion ratio is
\(\mathrm{Re}_p = \mathrm{Re}_f \cdot (\rho_P / \rho_f)\).

On the common fluid-density basis the present reference-window comparison value (\(\mathrm{Re}_f = 336.30\); the
extended-horizon comparison value is \(349.02\) --- Sec. 6.3) reproduces the Wang et al.~{[}16{]} finite-difference
immersed-boundary family --- where NF denotes the number of direct-forcing iterations in Wang's scheme (NF=1
\(\mathrm{Re}_f = 323.17\), NF=20 \(\mathrm{Re}_f = 335.59\)) --- to within a few percent; this is a reproduction check
within the same diffuse-IBM class. The present value lies about \(23\)--\(28\%\) below the continuous-solver DLM/FEM
range of Glowinski et al.~{[}40{]} (\(\mathrm{Re}_f = 438\)--\(466\)). The offset is not specific to the present
implementation: the Wang finite-difference solver itself lies \(23.4\%\) below the lower Glowinski endpoint on the same
basis. On the common fluid-density basis, the lattice-Boltzmann external-boundary-force method of Parvan et al.~{[}67{]}
and the sharp-interface direct-forcing immersed-boundary method of Badri Ghomizad et al.~{[}68{]} also cluster with the
present and Wang values, below the Glowinski DLM/FEM range (Appendix B Sec. B.5, Table B.6). The fluid solver, interface
treatment, grid and domain, blockage ratio, force-recovery procedure, time integration, and terminal-value criterion are
not uniformly matched across these references, so the cross-study comparison identifies a shared lower cluster among the
compared immersed-boundary implementations rather than a single originating factor. The documented diffuse-kernel
mechanisms --- the regularized delta kernel overestimating the effective hydrodynamic diameter of the body {[}69{]} and
a residual fictitious-fluid/internal flow retained within the body {[}70{]} --- remain the candidate contributors. A
hierarchical sensitivity analysis (Sec. 6.3) bounds each individually tested perturbation well below this offset within
the investigated ranges.

Expressed on the Wang particle-density basis used by that benchmark, the same present result agrees with the Wang NF=20
value to within \(+0.21\%\) (\(\mathrm{Re}_p = 504.45\) vs Wang \(503.38\)). The complete cross-source comparison
(Glowinski, Wang NF=1/NF=20, Parvan, and Badri Ghomizad) and the per-source gap analysis are reported in Appendix B Sec.
B.5 (Table B.6).

A heavy-particle internal-mass correction ablation across the two implemented correction forms (scheme (A) --- a
boundary-only update without an explicit internal-fluid correction --- and the (B-2) Feng--Michaelides explicit-history
correction; the mapping to the Suzuki and Inamuro {[}36{]} four-scheme taxonomy is given in Appendix A Sec.~A.5) yields
a standard-window spread of \(1.27\%\) on the DF baseline (Table B.5); a cross-scheme extension of this ablation gives
spreads of \(1.27\)--\(1.33\%\) across the three boundary-enforcement schemes --- a narrow range with modest
scheme-to-scheme variation, characterized within the tested single-particle configuration (Appendix B Sec. B.5). The DF
single-particle spread is small relative to the light-particle wake-interaction internal-mass sensitivity (\(-16.4\%\)
paired relative to the explicit-history baseline, equivalently a \(13.8\%\) underprediction of the Majumder reference)
of Sec.~5.2 and is consistent with the single-particle case serving as the low-sensitivity reference for the
wake-interaction comparison at the level of resolution achievable in the canonical Glowinski configuration.

\begin{figure}
\centering
\includegraphics{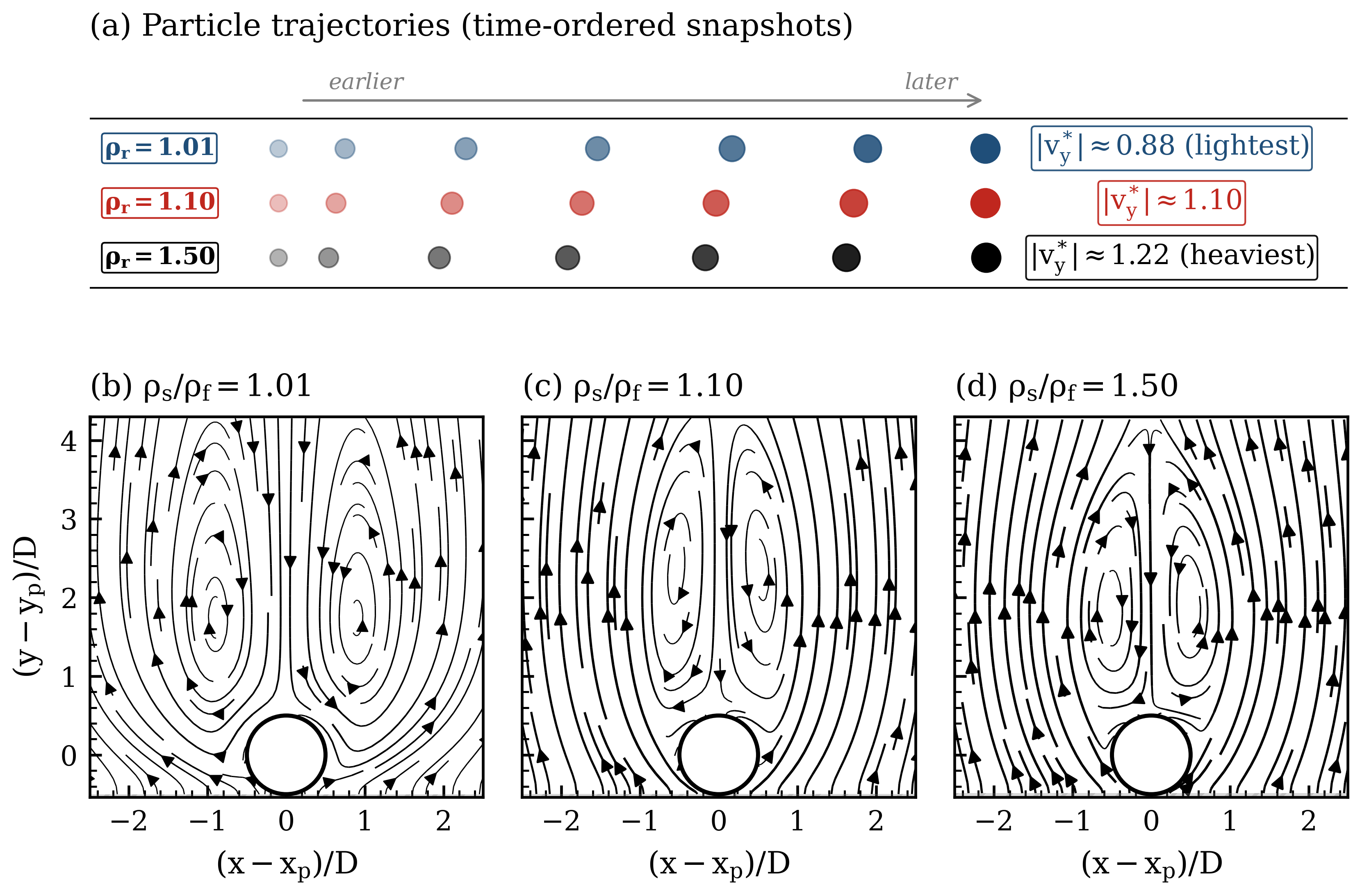}
\caption{\textbf{Fig. 6.} Single-particle sedimentation summary at the three Wang et al.~{[}16{]} / Glowinski et
al.~{[}40{]} density ratios \(\rho_s/\rho_f \in \{1.01, 1.1, 1.5\}\) (Glowinski / Wang vertical-channel setup with
gravity along \(-\hat{\mathbf{e}}_y\) per Eq. (22); panel (a) horizontal direction is time-ordered (earlier to later),
distinct from the Majumder convention of Fig. 7). (a) Seven evenly spaced particle positions in the channel frame
(darker circles later times); the particle remains on the centerline in all three regimes. (b)--(d) Particle-centered
streamlines near the terminal plateau for \(\rho_s/\rho_f = 1.01, 1.1, 1.5\), respectively, computed with DFC and the
Peskin 4-point delta function. The method--delta spread of the dimensionless settling velocity is reported in Table IV.}
\end{figure}

\subsubsection{5.2 Two-Particle Wake-Interaction Sedimentation: Wake-Exposed Closure
Sensitivity}\label{two-particle-wake-interaction-sedimentation-wake-exposed-closure-sensitivity}

The two-particle differential-density wake-interaction configuration of Majumder et al.~{[}23{]} is the moving-body
configuration examined in the present manuscript. This configuration reproduces the \emph{pure wake-interaction} case of
Uhlmann {[}14{]} (Sec. 5.2.2 therein), in which the denser trailing particle overtakes the lighter leading particle
through wake-mediated interaction \emph{without} direct particle--particle contact; it is distinct from the classical
equal-density drafting--kissing--tumbling sequence of Feng, Hu, and Joseph {[}71{]}, which entails near-contact and
tumbling and requires a collision model. That classical DKT configuration continues to serve as a benchmark for
particle-resolved methods: Hui et al.~{[}72{]} recently examined particle pairs settling in non-Newtonian fluids with an
IB-LBM whose hybrid MRT collision scheme is designed in part to reduce the numerical boundary slip at the solid boundary
layer and whose particle equation of motion retains an explicit internal-mass correction {[}36{]}, while Wang, Guo, and
Mi {[}73{]} showed that unequal-sized pairs shorten the contact phase and transition among repeated, one-off, and
suppressed DKT with increasing diameter ratio, explicitly identifying the density difference --- the asymmetry parameter
examined here --- as another significant factor characterizing particle--particle interactions. The Feng and Michaelides
{[}48{]} explicit-history internal-mass correction adopted here has previously been applied to direct-forcing IB-LBM
sedimentation of the equal-density drafting--kissing--tumbling configuration by Eshghinejadfard et al.~{[}74{]}. The
setup here follows Sec. 2.9.4. Reference target values are reported by Majumder on the \emph{fluid-density basis}
\(\mathrm{Re}_f = U D / \nu\), and the DF / Peskin 4-point / Velocity-Verlet / explicit-history reference baseline
matches these target values to within a few percent (Table V).

\paragraph{5.2.1 Trajectory Phase and Reynolds-Number History}\label{trajectory-phase-and-reynolds-number-history}

Fig. 7 shows the trajectories of the two particles during the overtaking event and the corresponding Reynolds-number
history. With gravity along \(+x\), the light particle starts downstream and the heavy particle upstream, and the
upstream/downstream roles exchange once during the sequence (at \(t^* \approx 5.9\) in the reference run). In the
pre-overtaking phase the upstream heavy particle is drawn into the reduced-momentum wake of the leading light particle
and accelerates, since its larger buoyant weight gives it the higher terminal velocity; the closest-approach phase
corresponds to the brief proximity and overtaking event near \(t^* \approx 5.5\), at which the minimum surface gap
between the two particles is \(g_{\min}/D \approx 0.56\), with no geometric contact observed at any stored sample
(consistent with the wake-interaction design of this case; Table C.2); and the post-overtaking phase corresponds to the
subsequent trailing-wake interaction, after which the heavy particle leads and the light particle trails in its
reduced-momentum wake. The light particle becomes the trailing particle in this post-overtaking phase, and the increased
sensitivity to the explicit internal-mass correction coincides with this post-exchange trailing (wake-exposed) phase. In
all recorded two-particle runs at the reference resolution the minimum surface gap is \(g_{\min}/D = 0.52\)--\(0.56\)
(Table C.2), i.e., \(41.8\)--\(45.2\,\Delta x\) at \(D = 80\Delta x\) --- more than ten times the Peskin 4-point support
width (\(4\,\Delta x\)); the spreading and interpolation stencils of the two particles therefore never overlap at any
stored sample, excluding direct kernel-level numerical interference at the evaluated stored states. The grid-refinement
pairs preserve the same non-overlap property (minimum stored-sample surface gaps of \(50\)--\(66\,\Delta x\), well above
the conservative Peskin 4-point support-overlap threshold).

\begin{figure}
\centering
\includegraphics{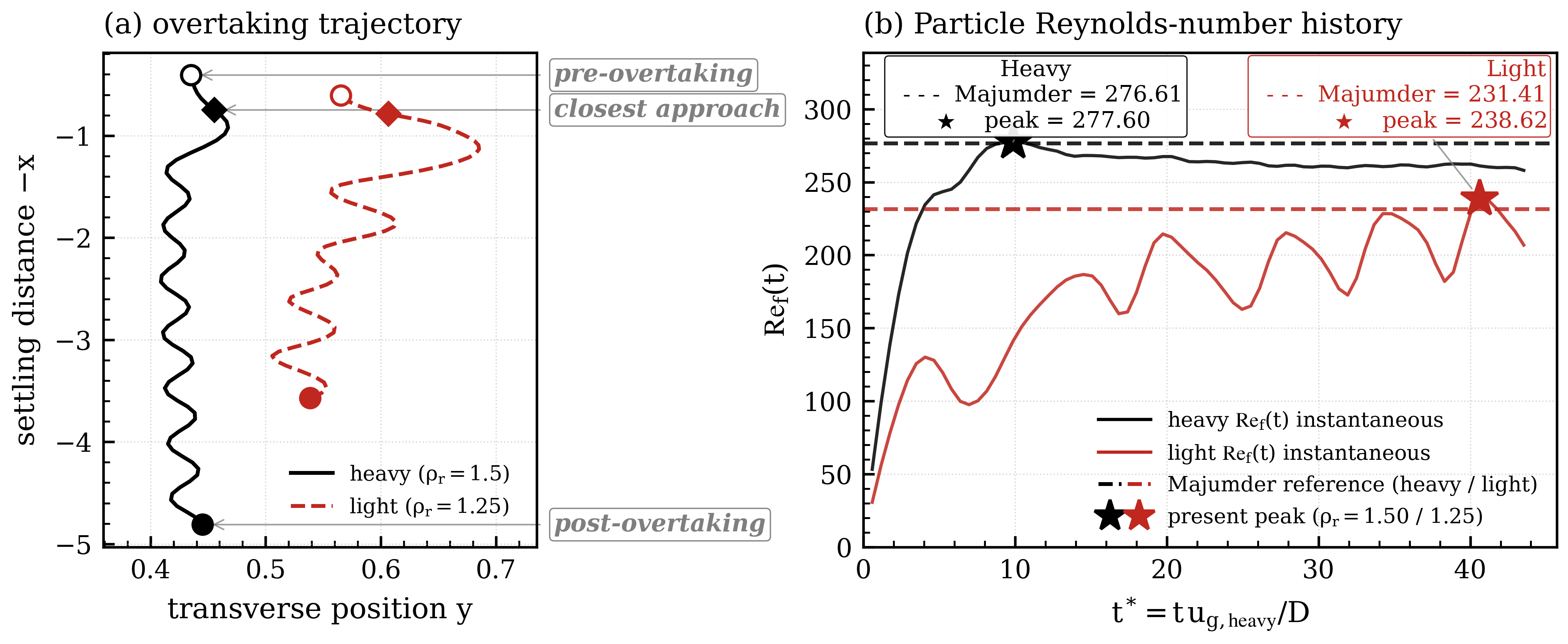}
\caption{\textbf{Fig. 7.} Two-particle wake-interaction trajectory and Reynolds-number history (Majumder et al.~{[}23{]}
setup; gravity along \(+x\), plotted downward as in conventional sedimentation schematics). (a) overtaking trajectory
(pre-overtaking, closest-approach, post-overtaking) showing the light-particle trajectory deflecting around the
heavy-particle wake in the post-overtaking phase; here and throughout, the shared dimensionless time \(t^*\) is
non-dimensionalized with the heavy-particle gravitational velocity \(u_{g,\mathrm{heavy}}\) (the two particles' \(u_g\)
differ by a factor \(\sqrt{2}\)). Closest approach occurs at \(t^* \approx 5.5\) and the upstream/downstream roles
exchange shortly afterward at \(t^* \approx 5.9\) (Sec. 5.2.1). (b) Heavy and light particle Reynolds-number history on
the fluid-density basis \(\mathrm{Re}_f = U D / \nu\) (reference DF / Peskin 4-point / Velocity-Verlet /
explicit-history configuration); annotated baseline values (heavy \(277.60\), light \(238.62\)) are the \(50D\)-window
peaks of the transient curve, with Majumder reference values (276.61 / 231.41) shown as horizontal dashed lines. The
wake-exposed phase corresponds to the post-overtaking segment.}
\end{figure}

\paragraph{5.2.2 Internal-Mass Correction Sensitivity}\label{internal-mass-correction-sensitivity}

Table V reports the internal-mass correction ablation for the two-particle wake-interaction case. The two configurations
correspond to Feng's rigid-body approximation {[}48{]} used here as the baseline \emph{explicit-history internal-mass
correction}, and the \emph{without explicit internal-mass correction} ablation used as a force-evaluation comparison.
The ablation quantifies the sensitivity to whether the internal-mass contribution is \emph{explicitly} corrected in the
particle force evaluation. The ablation targets this numerical closure only --- the only difference between the two arms
(Sec.~2.7); physical added-mass and history-type responses remain part of the resolved hydrodynamic force in both arms,
and because the two arms evolve along different trajectories with different resolved flows, these responses are not
fixed to identical values: the observed contrast is the full dynamical feedback of the numerical intervention. The
mapping to the Suzuki and Inamuro {[}36{]} four-scheme taxonomy is provided in Appendix A Sec.~A.5.

\noindent \textbf{Table V.} Two-particle wake-interaction internal-mass correction comparison, Majumder configuration. Values are
on the \emph{fluid-density basis} \(\mathrm{Re}_f = U D / \nu\) used by Majumder et al.~{[}23{]} and Uhlmann {[}14{]};
percentage deviations are relative to the Majumder reference; the paired baseline-to-ablation shift for the light
particle, relative to the present explicit-history baseline, is \(-16.4\%\), as quoted in Sec. 5.2.
Particle-density-basis values and extended-domain results are reported in Appendix B Table B.7 and Appendix C Table C.3.

\needspace{4\baselineskip}
\vspace{0.8em}
\sbox0{\scriptsize\setlength{\tabcolsep}{3pt}\renewcommand{\arraystretch}{1.2}%
\begin{tabular}{l r r r r}
\toprule
Internal-mass correction & Heavy \(\mathrm{Re}_{f,\max}\) & Light \(\mathrm{Re}_{f,\max}\) & Heavy \(\Delta\%\) & Light \(\Delta\%\) \\
\midrule
\emph{Reference reproduction} &  &  &  &  \\
Reference (Majumder et al.~{[}23{]}) & 276.61 & 231.41 & --- & --- \\
Present, explicit-history (baseline) & 277.60 & 238.62 & \(+0.36\%\) & \(+3.12\%\) \\
\emph{Without-correction ablation} &  &  &  &  \\
Present, without explicit correction (ablation) & 270.62 & 199.52 & \(-2.16\%\) & \(-13.78\%\) \\
\bottomrule
\end{tabular}}
\ifdim\wd0>\textwidth\noindent\resizebox{\textwidth}{!}{\usebox0}\else\noindent\makebox[\textwidth][c]{\usebox0}\fi
\vspace{0.8em}

The observation in this regime is that the wake-exposed light particle is \emph{comparatively} more sensitive to the
explicit internal-mass correction than the heavy particle (Table V): removing the explicit correction yields a
light-particle underprediction relative to the Majumder reference value that is several times the heavy-particle change
on the same ablation; the magnitude of the contrast is window-dependent (Sec. 5.2.4). A within-run comparator is
provided by the heavy particle, which experiences the identical domain, grid, integrator, and numerical treatment within
the same simulation, at a comparable Reynolds-number scale (\(\mathrm{Re}_f \approx 277\) vs \(239\)); its paired
ablation shift remains uniformly \(\leq 3.0\%\) across all four scheme comparisons, clearly separated from the
light-particle band (\(-10.95\%\) to \(-17.71\%\); Appendix C Sec. C.2). The density-axis confound of this comparator
(\(\rho_r = 1.5\) vs \(1.25\)) is complemented by the within-pair phase contrast of Sec. 5.2.3. A matched isolated-light
control --- identical domain, grid, relaxation time, and integrator with the heavy particle removed (explicit-history vs
without-correction) --- provides the most direct isolated-configuration comparison: over its recorded window the signed
contrast is only \(-0.86\%\), i.e., \(0.052\) times the magnitude of the \(50D\) pair contrast. The recorded window does
not establish an eventual isolated-light contrast; the value is used only as a finite-window comparison. Removing the
heavy particle removes not only the wake but also the pair pressure field, event sequence, and momentum redistribution,
so this comparison is associated with the pair-interaction configuration rather than with wake exposure alone.
Single-particle controls at \(\rho_s/\rho_f \in \{1.01, 1.1, 1.25, 1.5\}\) show much smaller shifts in isolated-particle
configurations without two-particle wake interaction (Sec. 5.1, Table B.5), so the wake-exposed response is
regime-specific rather than a generic feature of the explicit correction. The contrast direction is preserved across the
cross-IBM comparison (DF / MDF / DFC) on both the \(50D\) and \(60D\) domains and on a Majumder-aligned configuration
(incompressible-LBGK / Euler-explicit / marker-spacing \(0.83\)); per-case values, denominator definitions, and the four
paired-ablation comparisons are reported in Appendix C Sec. C.2, with the \(50D\)/\(60D\) extended-domain check
tabulated in Table C.3, and the cross-IBM particle-basis matrix is reported in Appendix B Table B.7.

\begin{figure}
\centering
\includegraphics{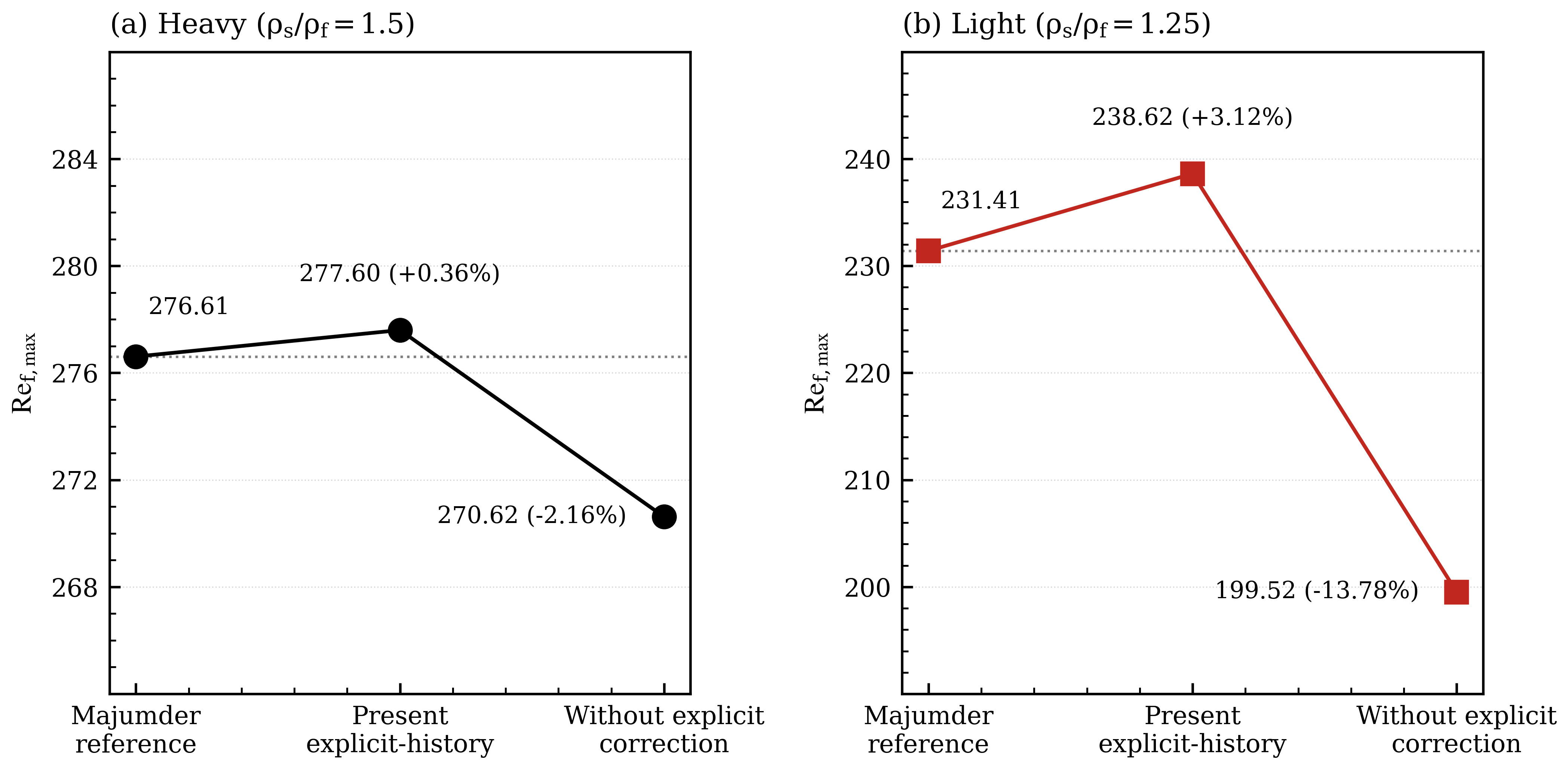}
\caption{\textbf{Fig. 8.} Two-particle wake-interaction internal-mass correction comparison on the Majumder \(50D\)
streamwise domain (fluid-density basis \(\mathrm{Re}_f = U D / \nu\)). Point plot of the observation-window peak
\(\mathrm{Re}_{f,\max}\) for heavy (a) and light (b) particles, comparing the Majumder reference (276.61 / 231.41), the
present explicit-history baseline (277.60 / 238.62), and the without-explicit-correction ablation (270.62 / 199.52).
Each panel uses an independent vertical scale, so the schemes are compared through the relative deviation (Table V)
rather than the absolute \(\mathrm{Re}_{f,\max}\).}
\end{figure}

\paragraph{5.2.3 Particle-Bound Acceleration Diagnostic}\label{particle-bound-acceleration-diagnostic}

Local acceleration diagnostics are consistent with a phase-dependent asymmetry within the wake-interaction pair, in
which the heightened internal-mass-correction sensitivity coincides with the wake-exposed trailing phase. The diagnostic
is computed as a \emph{particle-bound proxy} \(|a^*|\) from the particle velocity time series (the field-level
material-derivative peak is not used because available velocity-field snapshots are single-time). Phase-resolved peak
ratios are tabulated in Appendix C Sec. C.7 (Table C.4).

The light-to-heavy peak acceleration ratio exhibits a \(2.54\times\) phase-dependent contrast between pre-overtaking
(light/heavy \(1.11\times\)) and post-overtaking (light/heavy \(2.81\times\)) on the reference Velocity-Verlet sampling
(\(1.11\times\) / \(2.89\times\), contrast \(2.60\times\), on the Majumder-aligned incompressible-LBGK / Euler
cross-check) (Table C.4), strongest in the post-overtaking phase where the light particle trails the heavy particle and
is wake-exposed --- kinematic support for the wake-exposed geometry as a complement to the Reynolds-number ablation
contrast (Fig. 8). The contrast is a kinematic signature consistent with the wake-exposed geometry rather than a
dynamical measurement: a phase-dependent kinematic synchronization in the post-overtaking phase would produce a similar
signature. Because \(|a^*|\) is normalized by each particle's own gravitational acceleration scale \(u_g^2/D\), the
cross-density ratios (\(1.11\times\), \(2.81\times\)) carry a factor-of-two normalization difference
(\(u_{g,\mathrm{heavy}}^2 / u_{g,\mathrm{light}}^2 = 2\)) and are not raw acceleration ratios; the phase contrast
\(2.54\times\) (\(= 2.81/1.11\)) is scale-invariant for the fixed light/heavy pair and is the normalization-invariant
descriptor. A phase-resolved paired-difference decomposition of the same wake-interaction runs (explicit-history minus
without-correction, Velocity-Verlet pair; Supplementary Material Sec. S4.5) shows the ablation sensitivity itself
developing with the wake-exposed phase: on the event-aligned (primary) axis the peak dimensionless transverse-velocity
difference \(|\Delta v_y^*|\) grows \(6.49\times\) from the pre-closest-approach to the post-exchange phase for the
light particle (\(5.47\times\) for the heavy), and the post-exchange maximum of \(|\Delta v_y^*|\) is \(1.67\times\)
larger for the wake-exposed light particle (each particle's difference is normalized by its own \(u_g\), so, as with the
acceleration proxy, cross-particle ratios are not raw velocity ratios). The phase maxima are taken over unequal recorded
exposures (\(9\) pre- vs \(65\) post-exchange stored samples); phase means on the common-time axis, which reduce this
sensitivity, show that the post-exchange amplification persists. Per-phase values and the common-time (secondary)
alignment are tabulated in Supplementary Material Sec. S4.5.

\paragraph{5.2.4 Domain and Window Sensitivity}\label{domain-and-window-sensitivity}

The three scheme pairs have additionally been re-run on a \(60D\)-extended streamwise domain at matched grid resolution
(\(N_y = 801\)); the extended-domain check (Appendix C Sec. C.2, Table C.3) preserves the direction of the
closure-sensitivity ablation contrast at an attenuated magnitude on every scheme. The signed contrast has the same
direction in the DF / MDF / DFC comparisons on both recorded domains (Sec. C.2) --- the \emph{magnitude varies with the
domain and observation window for the case without explicit internal-mass correction}. Per-domain magnitudes,
denominator definitions, and saturation diagnostics for both \(50D\) and \(60D\) values are tabulated in Appendix C
Sec.~C.2 (Table C.3).

The harmonized \(50D\) MDF and DFC values characterize the cross-method spread at that domain length; the
domain-extension analysis is reported for all three scheme pairs (Sec.~C.2; the extension-run inventory is documented in
Supplementary Material Sec.~S2.6).

\subsubsection{5.3 Discussion of the Moving-Particle Mechanism}\label{discussion-of-the-moving-particle-mechanism}

In the moving-particle regime, the single-particle controls show only small internal-mass shifts in isolated-particle
configurations without two-particle wake interaction (heavy- and lighter-density, dynamically unmatched; Sec. 5.1, Table
B.5), whereas the two-particle wake-interaction case shows a \emph{heightened internal-mass-closure sensitivity that
coincides with the post-exchange trailing (wake-exposed) phase} in the present dataset: the wake-exposed light particle
exhibits this heightened sensitivity (Table V), with the domain and window dependence of the magnitude characterized in
Sec. 5.2.4. Trajectory phase diagnostics, the particle-bound acceleration proxy, the density-matched lighter
single-particle control, and a Majumder-aligned wake-interaction cross-check constrain alternative explanations, and the
additional phase-resolved, isolated-particle, and coupled-refinement controls strengthen the association between the
large recorded finite-window contrast and the post-exchange two-particle configuration. The isolated high-Galileo-number
(high-Ga) control reaches the peak-Reynolds range of the two-particle light response yet records a contrast of only
\(-1.06\%\) (Appendix B Sec. B.5). The resulting interpretation is a configuration-dependent finite-window sensitivity
(Sec. 5.2.4). The complementary collision-model controls (Sec. 6) bound the relaxation-side response within this regime.

\subsection{6. Collision-Model and Sedimentation-Domain
Controls}\label{collision-model-and-sedimentation-domain-controls}

The TRT and CM-MRT calculations are used as \emph{collision-model controls}. Within the tested regimes and
reference-defined peak metrics --- prescribed-body comparisons at \(\mathrm{Re} \leq 200\) (two-dimensional) and the
reported moving-particle cases --- collision-relaxation changes are secondary to the boundary and closure mechanisms
probed in Sec.~3--Sec.~5. No conclusion is drawn for higher-Reynolds-number regimes, where collision-model differences
become more prominent {[}28{]}. The controlled collision spread for the fixed-cylinder and two-particle wake-interaction
reference cases is reported in Sec. 6.1. Sec. 6.3 consolidates a hierarchical sensitivity analysis of the
single-particle cross-formulation gap across the tested numerical factors. The method × kernel × collision case matrices
are compiled in main Appendix B (collision spread Sec.~B.4, single-particle Sec.~B.5, wake-interaction case Sec.~B.6)
with the corresponding \(\bar{C}_d\) / \(\mathrm{St}\) / \(C_l\) full matrices in Supplementary Material Sec.~S1.

\subsubsection{6.1 Collision-Model Control}\label{collision-model-control}

The fixed-cylinder \(\bar{C}_d\) collision spread and the two-particle wake-interaction peak-Reynolds collision spread
are reported together as the collision-model control diagnostic. For the fixed cylinder, the BGK / TRT / CM-MRT spread
reaches a maximum of \(0.94\%\) at \(\mathrm{Re} = 100\) across the twelve-combination matrix (Appendix B Table B.4);
for the two-particle wake-interaction reference baseline (DF / Peskin 4-point / Velocity-Verlet / explicit-history), the
heavy- and light-particle spreads are \(0.21\%\) and \(0.69\%\) on the fluid-density basis. Representative cases are
summarized in Table VI and Fig. 9. Across the prescribed-body regime at the tested Reynolds numbers
(\(\mathrm{Re} \leq 200\)), the collision-only contribution is therefore \emph{bounded} relative to the boundary and
kernel effects (DF→MDF changes reach approximately \(3.3\%\) at \(\mathrm{Re} = 200\) on P4 and the DFC kernel-dependent
shifts reach approximately \(-8.3\%\) at \(\mathrm{Re} = 200\)); in the wake-interaction case, the collision spread is
an order of magnitude smaller than the light-particle internal-mass correction contrast (Sec. 5.2 / Table V). The CM-MRT
operator is implemented in the D2Q9 central-moment form, adapted from {[}38, Appendix B{]}. The two-particle
wake-interaction case matrix at the reference baseline configuration is reported in Appendix B Sec. B.6 (Table B.7).

\noindent \textbf{Table VI.} Collision-model control summary across BGK / TRT / CM-MRT. (a) Fixed-cylinder \(\bar{C}_d\) for
selected cases that span the spread range of the twelve-combination matrix (full matrix in Appendix B Table B.4). (b)
Two-particle wake-interaction reference baseline (DF / Peskin 4-point / Velocity-Verlet / explicit-history; heavy
\(\rho_r = 1.5\) / light \(\rho_r = 1.25\)) peak \(\mathrm{Re}_{f,\max}\) on the \emph{fluid-density basis}. MDF refers
to the adaptive form (Sec. 2.5.2).

\needspace{4\baselineskip}
\vspace{0.8em}
\sbox0{\scriptsize\setlength{\tabcolsep}{3pt}\renewcommand{\arraystretch}{1.2}%
\begin{tabular}{l l r r r r}
\toprule
Block & Case / Quantity & BGK & TRT & CM-MRT & Spread \\
\midrule
(a) Fixed cylinder (\(\bar{C}_d\)) & Re=100, DF + hat & 1.3965 & 1.3834 & 1.3947 & \(0.94\%\) (max) \\
(a) Fixed cylinder (\(\bar{C}_d\)) & Re=100, MDF + hat & 1.3849 & 1.3720 & 1.3829 & \(0.93\%\) \\
(a) Fixed cylinder (\(\bar{C}_d\)) & Re=100, DFC + P4 & 1.4051 & 1.4021 & 1.4046 & \(0.21\%\) \\
(a) Fixed cylinder (\(\bar{C}_d\)) & Re=200, MDF + P4 & 1.2982 & 1.2971 & 1.2984 & \(0.10\%\) (min) \\
(b) wake-interaction reference (\(\mathrm{Re}_{f,\max}\)) & Heavy (\(\rho_r = 1.5\)) & 277.60 & 278.19 & 277.62 & \(0.21\%\) \\
(b) wake-interaction reference (\(\mathrm{Re}_{f,\max}\)) & Light (\(\rho_r = 1.25\)) & 238.62 & 238.24 & 239.89 & \(0.69\%\) \\
\bottomrule
\end{tabular}}
\ifdim\wd0>\textwidth\noindent\resizebox{\textwidth}{!}{\usebox0}\else\noindent\makebox[\textwidth][c]{\usebox0}\fi
\vspace{0.8em}

\subsubsection{\texorpdfstring{6.2 Sedimentation Domain and Observation-Window Sensitivity (Standard Channel vs \(60D\)
Extended
Horizon)}{6.2 Sedimentation Domain and Observation-Window Sensitivity (Standard Channel vs 60D Extended Horizon)}}\label{sedimentation-domain-and-observation-window-sensitivity-standard-channel-vs-60d-extended-horizon}

The single-particle case of Sec. 5.1 is run in the canonical Glowinski et al.~{[}40{]} / Wang et al.~{[}16{]}
configuration (\(W = 2\) cm \(\times\) \(L = 6\) cm physical channel, \(d = 0.25\) cm, \(24D\) streamwise extent
(\(L/D = 24\)), particle initialized on the channel centerline at \((x, y) = (1, 4)\) cm); the two-particle
wake-interaction case of Sec. 5.2 inherits the same Glowinski / Majumder family configuration on the Majumder \(50D\)
streamwise domain. Throughout this section, ``standard-channel baseline'' denotes the full contact-safe recorded horizon
of this \(24D\) standard channel (\(y^* \leq 15.5\) in fall units of \(D\)). The domain dependence of the Sec. 5.1
single-particle body-local peak Reynolds number and the Sec. 5.2 closure-sensitivity ablation contrast is assessed in
two separate analyses: an extended-channel analysis of the heavy single-particle case (\(\rho_s/\rho_f = 1.5\)), and the
\(60D\) two-particle re-runs. In the single-particle extension the body-local \(|v_y^*|\) peak is reached at
\(y^* \approx 25\)--\(33\) (in \(D\) units; \(y^*\) dimensionless throughout) in the extended runs (DF earliest, DFC
latest), and the body-local peak Reynolds number is consistent with near-saturation over the tested extended horizons.
The ``\(60D\) extended'' label denotes a tall channel of height \(H/D = 60\), with \(52D\) of fall clearance below the
release position and a contact-safe recorded horizon of \(y^* \leq 51.5\); the wake-interaction body-local consistency
check and the \(60D\)-extended domain check are summarized below.

On the single-particle side, the analysis is reported in Appendix C Sec. C.2; the body-local peak Reynolds number
changes uniformly by \(+3.78\%\) to \(+4.63\%\) between the standard channel and the full \(60D\) horizon across all six
tested configurations (DF / MDF / DFC × BGK / TRT × Verlet × explicit-history) --- comparable to roughly one third of
the light-particle ablation contrast (Sec. 5.2 / Table V) that underpins the wake-exposed closure sensitivity (the
largest \(60D\) shift, \(4.63\%\), is \(\approx 1/3\) of the contrast), and the within-IBM BGK / TRT collision-spread
component of the \(60D\) Reynolds number remains below \(0.08\%\) on each IBM family (Appendix C Sec. C.2), consistent
with the bounded collision-spread reported in Sec.~6 (Table VI). The complete six-case \(60D\) analysis therefore
extends the body-local single-particle result across the tested schemes and collision operators.

For the two-particle wake-interaction case, the sedimentation of Sec. 5.2 inherits the same body-local Glowinski /
Majumder configuration on a \(50D\) streamwise domain, and the channel-length adequacy on the body-local metric
(Sec.~5.2 Table V peak \(\mathrm{Re}_{f,\max}\) at the wake-exposed particle) is supported by a body-local consistency
check: the light-particle \(\mathrm{Re}_{f,\max}\) peak is reached at \(27.1D\) downstream of the light initial position
on the explicit-history baseline case and at \(26.0D\) on the without-explicit-correction ablation case, leaving a
\(16.9D\) / \(18.0D\) streamwise center-to-outlet clearance (Table C.2). On the same from-initial-displacement basis
(\(y^*\) denotes the streamwise displacement from the particle's initial position in \(D\) units, as adopted throughout
Sec.~C.2; here \(y^*\to 0\) at \(t=0\)), the light-particle peak distance (\(26\)--\(27D\)) sits inside the body-local
peak-location band of the Sec.~C.2 single-particle analysis (\(25D\)--\(33D\) across DF / MDF / DFC × BGK / TRT cases).

The body-local consistency check is consistent with near-saturation of the explicit-history peak over the tested
\(50D\)/\(60D\) windows (\(-0.59\%\) to \(-1.02\%\) shifts between \(50D\) and \(60D\) across the three schemes),
whereas the without-explicit-correction case continues to develop at \(50D\). The ablation direction --- wake-exposed
light particle slowed under without-explicit-correction --- is shared by the DF/MDF/DFC comparisons on both the \(50D\)
and the \(60D\)-extended domains (Sec. 5.2.4). The \(60D\) without-correction maxima occur at or within one stored
sample of the run end, so the comparison is reported over the recorded windows. The magnitude of the without-correction
case attenuates between the \(50D\) Majumder window and the \(60D\)-extended window, each on its own without-correction
denominator (per-window magnitudes tabulated in Appendix C Sec.~C.2 Table C.3; the \(13.8\%\) value of Sec.~5.2 uses the
Majumder-reference denominator and is not directly equated to those window-specific values). The fixed-cylinder (Sec.~3)
and oscillating-cylinder (Sec.~3.3) benchmarks are not subject to the finite settling-window constraints of the
sedimentation cases and their domain adequacy is supported by the grid-sensitivity analysis of Appendix C Sec.~C.3 and
C.4. Far-wake quantities and three-dimensional wake-transition claims are outside the scope of the present analysis; the
\(60D\) analysis is positioned as a sedimentation-domain adequacy control.

\begin{figure}
\centering
\includegraphics{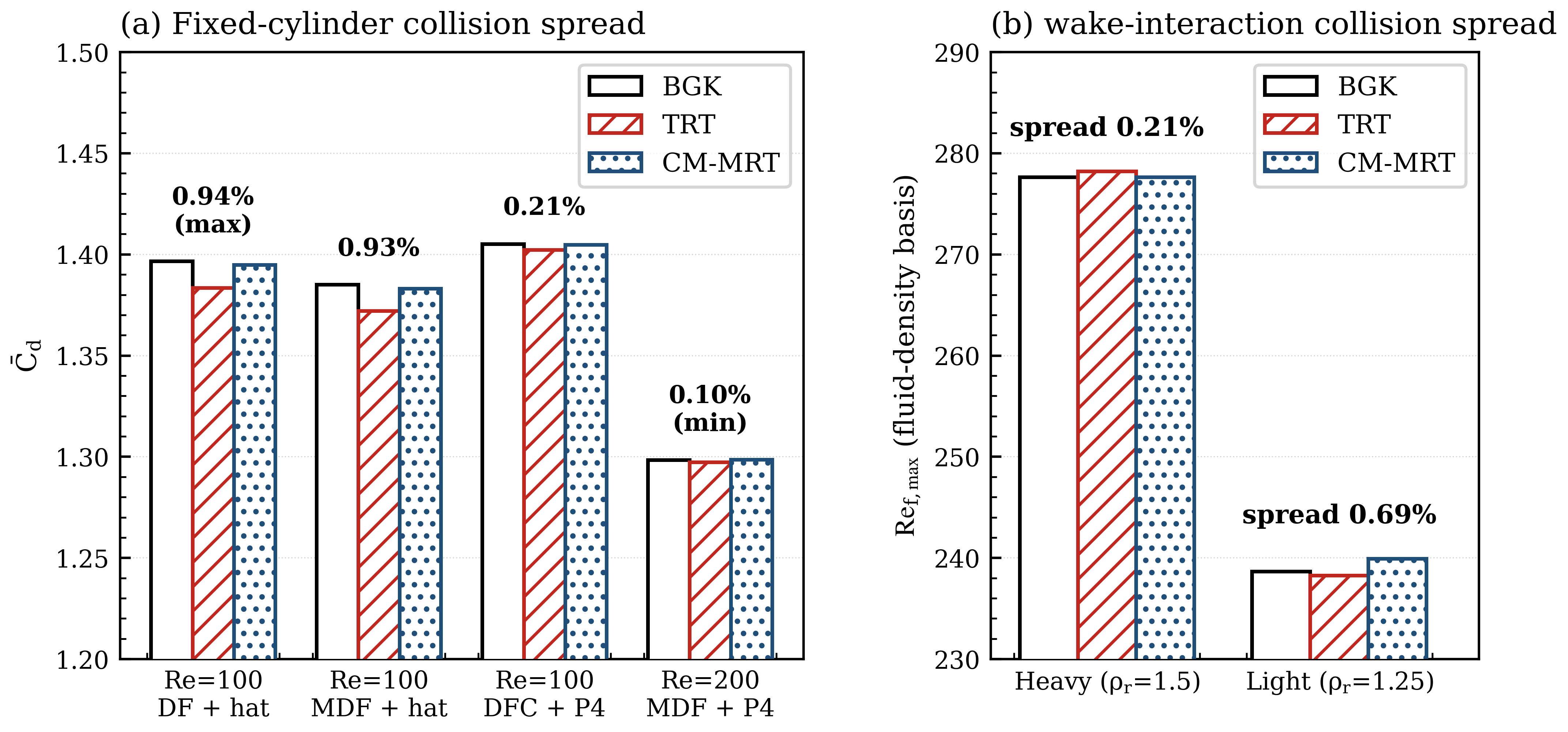}
\caption{\textbf{Fig. 9.} Collision-model summary across the two reference cases. (a) Fixed-cylinder \(\bar{C}_d\)
collision spread (BGK / TRT / CM-MRT) for selected cases that span the full twelve-combination matrix in Appendix B
Table B.4; the maximum across the twelve-combination matrix is \(0.94\%\). (b) Two-particle wake-interaction peak
\(\mathrm{Re}_{f,\max}\) collision spread for the reference DF / Peskin 4-point / Velocity-Verlet / explicit-history
case (heavy \(\rho_r = 1.5\) / light \(\rho_r = 1.25\)) on the \emph{fluid-density basis} \(\mathrm{Re}_f = U D / \nu\)
used in Sec.~5.2 / Table V: \(0.21\%\) (heavy) and \(0.69\%\) (light). Absolute values are not compared across panels;
only the collision spread is compared. Each panel uses a zoomed y-axis (non-zero base) to resolve the per-case spread,
and the spread value is annotated above each bar.}
\end{figure}

\subsubsection{6.3 Hierarchical Sensitivity Analysis of the Single-Particle Cross-Formulation
Gap}\label{hierarchical-sensitivity-analysis-of-the-single-particle-cross-formulation-gap}

This section consolidates the single-particle cross-formulation gap diagnosis of Sec. 5.1. The reference-window
discrepancy is \(23.22\)--\(27.83\%\) for the original reference-window value \(\mathrm{Re}_f = 336.30\) against the
DLM/FEM range \(438\)--\(466\) (absolute gap \(\Delta\mathrm{Re}_f = 101.70\)--\(129.70\)). Extending the observation
horizon raises the present extended-horizon comparison value to \(349.02\) and reduces the comparison gap to
\(20.32\)--\(25.10\%\) (\(\Delta\mathrm{Re}_f = 88.98\)--\(116.98\)); the longer observation horizon therefore accounts
for part of the difference, and the residual difference is bounded factor-by-factor below. Table VII bounds each
numerically testable factor by its measured span; per-row data sources are the existing tabulations (Table C.1, the Sec.
C.5 grid checks, Tables B.4, VI, and B.5) and the marker-retraction test, cross-referenced to avoid duplication. Because
the relative shifts use different denominators, row magnitudes are compared through the absolute
\(|\Delta\mathrm{Re}_f|\) column only, and gap fractions are quoted against the matching layer (standard-window rows
against the reference-window gap; extended-window rows against the extended-horizon residual). The two-particle
grid-refinement check is a two-particle observable and is reported in Sec. C.5, not in this table.

\begin{landscape}

\noindent \textbf{Table VII.} Hierarchical sensitivity analysis of the single-particle cross-formulation gap: tested perturbations
with absolute and relative shifts. Percentages were computed from unrounded values. For the internal-fluid closure row
the absolute shift is converted from the particle-density basis as
\(|\Delta\mathrm{Re}_f| = |\Delta\mathrm{Re}_p| / 1.5\). The boundary-scheme relative shifts are normalized by the DF
value in the same window.

\nopagebreak[4]\par\vspace{-0.2em}

\needspace{4\baselineskip}
\vspace{0.8em}
\sbox0{\scriptsize\setlength{\tabcolsep}{3pt}\renewcommand{\arraystretch}{1.2}%
\begin{tabular}{>{\raggedright\arraybackslash}p{0.189\linewidth} >{\raggedright\arraybackslash}p{0.180\linewidth} >{\raggedright\arraybackslash}p{0.050\linewidth} >{\raggedright\arraybackslash}p{0.371\linewidth} >{\raggedright\arraybackslash}p{0.159\linewidth}}
\toprule
Perturbation & Observable / window / shift denominator & \(|\Delta\mathrm{Re}_f|\) & Relative shift & Interpretation \\
\midrule
Grid resolution (\(D = 160 \to 240\Delta x\); DF, Euler pair; Verlet
bridge \(0.0002\%\)) & peak \(\mathrm{Re}_f\) / standard window / self-baseline & \(0.37\) & \(+0.109\%\) & minor contribution \\
Channel height (\(24D \to 60D\); six configurations) & \(|v_y^*|\)-peak \(\mathrm{Re}_f\) / matched \(y^* \leq 15.5\) /
self-baseline & \(4.52\)--\(4.82\) & \(+1.33\)--\(+1.42\%\) & minor contribution \\
Observation horizon (\(y^*\!: 15.5 \to 51.5\); six configurations) & \(|v_y^*|\)-peak \(\mathrm{Re}_f\) / extended \(60D\) window /
self-baseline & \(8.02\)--\(10.93\) & \(+2.35\)--\(+3.19\%\) & largest single tested factor; narrows the gap \\
Additional single-particle \(60D \to 80D\) increment (DF, two
configurations) & \(|v_y^*|\)-peak \(\mathrm{Re}_f\) / extended horizon / self-baseline & \(\approx 0.6\) & \(\approx +0.18\%\) & small additional increment \\
Collision operator (DF, BGK vs TRT, tall-\(60D\)) & \(|v_y^*|\)-peak \(\mathrm{Re}_f\) / extended \(60D\) window / relative
to \(\mathrm{Re}_f = 349.02\) & \(0.26\) & \(0.075\%\) & bounded within the tested two-operator comparison \\
Boundary-enforcement scheme, standard window (max--min across DF / MDF /
DFC) & \(|v_y^*|\)-peak \(\mathrm{Re}_f\) / standard window / self-baseline & \(2.88\) (BGK); \(3.12\) (TRT) & \(0.86\%\); \(0.93\%\) & modest, window-specific \\
Boundary-enforcement scheme, extended window† & \(|v_y^*|\)-peak \(\mathrm{Re}_f\) / extended \(60D\) window /
self-baseline & \(5.73\) (BGK); \(5.71\) (TRT) & \(1.64\%\) & larger, window-dependent spread \\
Internal-fluid closure, (A) vs (B-2) (DF; MDF; DFC) & peak \(\mathrm{Re}_p\) / standard window / Wang same-source \(503.38\) & \(4.25\); \(4.46\); \(4.37\) & \(1.27\%\); \(1.33\%\); \(1.30\%\) & modest scheme-to-scheme variation within the tested configuration \\
Interface regularization (inward marker retraction
\(0.3 / 0.5 / 1.0\,\Delta x\)) & peak \(\mathrm{Re}_f\) / standard window / retraction-\(0\) baseline & \(0.83\); \(1.39\); \(2.79\) & \(+0.247\%\); \(+0.413\%\); \(+0.830\%\) (monotone) & bounded one-sided shift \\
Cross-study cluster & terminal / peak \(\mathrm{Re}_f\) / per-reference conventions & --- & IBM cluster \(327\)--\(336\); present reference-window \(336.30\);
extended-horizon comparison \(349.02\) (reported separately); DLM/FEM
\(438\)--\(466\); Wang FD self-reported \(-23.4\%\) & shared lower cluster of the compared immersed-boundary implementations \\
\bottomrule
\end{tabular}}
\ifdim\wd0>\textwidth\noindent\resizebox{\textwidth}{!}{\usebox0}\else\noindent\makebox[\textwidth][c]{\usebox0}\fi
\vspace{0.8em}

† The extended-window boundary-scheme spread is an arithmetic comparison derived from the six-configuration matrix of
Table C.1. Per-scheme values for both windows are reported in the Supplementary Material; the scheme ordering DF
\textless{} DFC \textless{} MDF is unchanged between the two windows and the two collision operators.

\end{landscape}

Within the tested inward marker-retraction range of \(0\)--\(1\Delta x\), the standard-window peak \(\mathrm{Re}_f\)
increases monotonically by \(0.247\)--\(0.830\%\) (\(\Delta\mathrm{Re}_f = 0.83\)--\(2.79\)), toward the DLM/FEM range;
this bounds the tested interface-location sensitivity at \(2.2\)--\(2.7\%\) of the reference-window gap. The
standard-window boundary-scheme spread corresponds to \(2.2\)--\(3.1\%\) of the reference-window gap, and the larger
extended-window boundary-scheme spread to approximately \(4.9\)--\(6.4\%\) of the extended-horizon residual;
finite-window dependence is observed in both the single-particle closure comparison (Sec. C.2) and this boundary-scheme
spread.

The hierarchical tests quantitatively bound the candidate factors accessible within the present framework: each
individually tested perturbation is bounded well below the remaining cross-formulation offset over the investigated
ranges, with the observation-horizon extension as the largest single contributor, narrowing the comparison to a residual
difference of \(20.32\)--\(25.10\%\). The internal-fluid closure spread lies in the narrow range \(1.27\)--\(1.33\%\)
with modest scheme-to-scheme variation (Sec.~B.5), and the tested marker retraction produces a bounded, monotone
one-sided shift. The four immersed-boundary implementations represented in the comparison --- including a
sharp-interface method --- form a cluster below the reported DLM/FEM values, so this lower-cluster observation is not
specific to the present scheme; identifying its controlling element calls for a matched cross-code comparison in which
the remaining factors are held fixed or decomposed in a designed factorial study.

\subsection{7. Discussion: Relative Roles of the Mechanisms and
Limitations}\label{discussion-relative-roles-of-the-mechanisms-and-limitations}

\subsubsection{7.1 Relative Roles of the Interface Mechanisms}\label{relative-roles-of-the-interface-mechanisms}

The mechanisms identified in Sec.~3--Sec.~6 are compared by their relative roles and the strength of their supporting
evidence. Each probes a distinct interface momentum-transfer mechanism, and each is supported by the local diagnostics
named in Sec.~2.8.

Local no-slip fidelity (Sec. 3) distinguishes DF, MDF, and DFC in the prescribed-body regime. Global integral quantities
(\(C_d\), \(\mathrm{St}\), \(C_l\)) reach the literature range for most method--delta combinations, but the local slip
error and the kernel-dependent slip-improvement gap reveal differences that are masked by the integrals. In particular,
MDF achieves an order-of-magnitude slip reduction under both kernels, whereas DFC under hat shows a kernel-dependent
fidelity gap that prefigures the kernel reversal of Sec.~4.

Correction redistribution (Sec. 4) is a signature of the DFC kernel reversal. The total correction magnitude alone does
not distinguish hat from Peskin 4-point at \(\mathrm{Re} = 100\); the difference lies in the \emph{spatial distribution}
of the correction field, quantified by an order-of-magnitude larger coefficient of variation of \(\lambda_k\) under hat
(\(\approx 10\)--\(16\times\) across the analyzed resolutions; \(\sim 10\times\) on the primary \(D=40\Delta x\) Fig. 5
case). As the boundary layer thins at higher Re, this anisotropic enforcement is associated with a measurably different
effective boundary, and the kernel-dependent \(\bar{C}_d\) ranking inverts. The marker-resolved \(\lambda_k\), slip, and
pressure diagnostics (Fig. 5) are reported together as a redistribution signature; they support, but do not close, the
causal chain to the Sec.~4.1 reversal.

Wake-exposed closure sensitivity (Sec. 5) is probed by the freely sedimenting cases. The single-particle baseline
reproduces the Wang finite-difference immersed-boundary family on the common fluid-density basis to within a few percent
(the residual gap from the continuous-solver DLM/FEM range is discussed in Sec. 5.1) and shows a \(1.27\%\) DF
heavy-reference ablation spread (spreads of \(1.27\)--\(1.33\%\) with modest scheme-to-scheme variation; Appendix B Sec.
B.5); the two-particle wake-interaction case shows an increased sensitivity to explicit internal-mass correction in the
wake-exposed light particle (\(13.8\%\) underprediction relative to the Majumder reference value on the \(50D\)
measurement window). The present dataset shows that this heightened internal-mass-closure sensitivity coincides with the
post-exchange trailing (wake-exposed) phase. Several controls constrain the alternative explanations (scheme artifact,
integrator/compressibility artifact, isolated-particle baseline). The density-matched lighter single-particle ablation
(\(\rho_s/\rho_f = 1.25\), DF / Peskin 4-point / Velocity-Verlet, Sec.~5.1) is retained only as a dynamically unmatched
single-particle comparison without two-particle wake interaction; its small paired shift is tabulated in Appendix B Sec.
B.5 and is not used as a wake-exposure control. The Majumder-aligned wake-interaction cross-check (incompressible-LBGK /
Euler-explicit / marker-spacing \(0.83\)) reproduces the contrast with a \(-13.91\%\) light-particle deficit relative to
the Majumder reference, supporting that it is not explained solely by an integrator or compressibility artifact of the
reference configuration. Trajectory phase identification and particle-bound acceleration diagnostics (phase-dependent
light-to-heavy peak-ratio contrast, Table C.4) are consistent with the wake-exposed geometry as a kinematic signature.
The heavy particle serves as a within-run comparator for the wake-exposed light-particle response (Sec. 5.2.2). A
matched isolated-light configuration and an approximately peak-\(\mathrm{Re}_f\)-matched isolated control extend the
isolated-particle comparisons (Sec. 5.2.2, Appendix B Sec. B.5); both record only small contrasts within their recorded
windows, consistent with the configuration-dependent finite-window sensitivity of Sec. 5.3. Four paired-ablation
comparisons (cross-IBM, spanning DF / MDF / DFC; Appendix C Sec. C.2) bound the heavy-particle peak-Reynolds ablation
spread uniformly at \(\leq 3.0\%\) on every tested boundary-enforcement scheme while keeping the light-particle
without-explicit-correction deficit in a single \(-10\%\) to \(-18\%\) band, supporting separation of the light-particle
wake-exposed response from a boundary-enforcement scheme artifact. The signed direction is shared by the DF / MDF / DFC
comparisons on both the \(50D\) and \(60D\)-extended domains, while the magnitude remains dependent on the domain and
reporting window (Appendix C Sec. C.2).

Collision-model controls (Sec. 6) bound the relaxation-side response. The fixed-cylinder collision-only spread is at
most \(0.94\%\) across the twelve-combination matrix (Table B.4; the maximum is attained at \(\mathrm{Re} = 100\) for DF
+ hat); the reference wake-interaction case collision spread is \(0.21\%\) (heavy) and \(0.69\%\) (light) --- both an
order of magnitude smaller than the light-particle ablation contrast (Sec. 5.2 / Table V). Within the tested regimes
(prescribed-body comparisons at \(\mathrm{Re} \leq 200\) and the reported moving-particle cases), the relaxation-side
response remains secondary to the boundary, correction, and closure mechanisms.

Across the four mechanisms, the relative performance of DF, MDF, and DFC is regime-dependent: each canonical benchmark
is most diagnostic of the specific interface momentum-transfer mechanism it activates, so the informative comparison
proceeds mechanism by mechanism rather than as a single cross-benchmark ordering.

The preferential sensitivity of the wake-exposed light particle to the explicit internal-mass correction is consistent
with the discrete force closure. In the wake-interaction post-overtaking phase the trailing light particle traverses the
leading particle's reduced-momentum wake, where the local relative velocity and its rate of change are expected to be
larger and more unsteady than in the isolated-particle baseline (the particle-bound acceleration proxy of Sec. 5.2.3 is
consistent with this expectation). The Feng and Michaelides {[}48{]} explicit-history term, as discretized in our Eq.
(A3), supplies a \emph{particle-velocity-history-based} internal-mass contribution
\(\Delta\mathbf{F}_{\mathrm{IM}}^n = (m_f / \Delta t)(\mathbf{v}_s^n - \mathbf{v}_s^{n-1})\) that the boundary-only IBM
force omits. The treatment of this fictitious interior fluid has a long history in particle-resolved methods: in
lattice-Boltzmann moving-boundary schemes the interior fluid is treated as an internal mass carried by the particle
{[}75{]}, an alternative formulation repopulates fluid nodes so that no interior fluid is retained at any solid-to-fluid
density ratio {[}76{]}, and direct-forcing methods that integrate the interior momentum directly over the particle
volume remove the density-ratio singularity carried by an effective-mass denominator as \(\rho_s \to \rho_f\) {[}70{]}.
Because the discrete term scales with the step-to-step change of the particle velocity, it is largest when that velocity
changes most rapidly, as in the wake-exposed post-overtaking phase; when \(\rho_s\) approaches \(\rho_f\) the same
increment becomes a larger fraction of the small net buoyancy \((\rho_s - \rho_f) V_p g\) (with \(V_p = \pi r^2\) the
two-dimensional particle volume), so the wake-exposed light particle is the configuration in which omitting the
correction produces the largest change in the force evaluation. This is a property of the discrete moving-body closure
under the elevated, more unsteady relative velocity of the post-exchange trailing phase: Eq. (A3) is a backward
difference of the particle velocity, a force-evaluation term distinct from a field-level fluid-acceleration (added-mass)
response and from the viscous unsteady-history (Basset-type) contributions carried by the resolved flow (Sec. 2.7). The
\(\rho_s/\rho_f = 1.25\) single-particle control is interpreted as a density-matched but dynamically unmatched
single-particle reference rather than a same-\(\mathrm{Re}\) isolated-mechanism comparison (the two configurations
sample different kinematic regimes, \(\mathrm{Re}_f \approx 13.9\) vs \(\approx 238.6\); see the Fig.~7 caption and
Appendix~B Sec.~B.5): in this reference the explicit-history versus without-explicit-correction shift is small at
\(\rho_s/\rho_f = 1.25\) (tabulated in Appendix B Sec. B.5; Sec.~5.1), a bounded low-Reynolds single-case observation
that is not used as a wake-exposure control; we cannot, however, exclude that the bounded heavy-particle ablation spread
(\(\leq 3.0\%\) across DF / MDF / DFC) and the case-spread of the light-particle deficit (\(-10.95\%\) DFC to
\(-17.71\%\) MDF, a \(\sim 7\)-percentage-point range across boundary-enforcement schemes; Sec.~C.2) carry an additional
sub-mechanism-specific component superimposed on the wake-exposed response.

The DFC kernel reversal connects to two geometric properties of the spreading kernel. The marker-resolved correction
non-uniformity is set by how each kernel samples the underlying Eulerian grid: the narrow hat support
(\(W = 2\Delta x\)) is strongly sensitive to whether a marker falls between or near grid nodes, producing the large
\(\lambda_k\) coefficient of variation reported in Sec. 4.2, whereas the wider Peskin 4-point support
(\(W = 4\Delta x\)) averages over this positioning and stays nearly uniform. The Reynolds-number dependence enters
through the boundary layer: its thickness on a stationary cylinder scales as \(\delta \sim D / \sqrt{\mathrm{Re}}\)
(Schlichting-type boundary-layer scaling; the flat-plate Blasius prefactor \(\delta \approx 5\,D/\sqrt{\mathrm{Re}}\)
serves only as an order-of-magnitude estimate), so the kernel-support-to-layer ratio \(W/\delta\) grows with Reynolds
number and orders the absolute slip across kernels (Sec. 4.3). As the boundary layer thins, viscous diffusion is less
able to smooth the non-uniform correction of the narrow kernel; the measured enforcement anisotropy (Sec. 4.2) and the
kernel-dependent \(\bar{C}_d\) ranking inversion co-occur in this regime --- a chain supported, but not closed, by the
marker-resolved diagnostics. The diffuse-interface kernel analysis of Peng and Wang {[}60{]}, which shows that a
regularized spreading kernel alters the local velocity gradient and the effective no-slip enforcement at the interface,
is the continuum counterpart of this marker-resolved redistribution.

Collision-model effects are secondary in the tested regimes for a structural reason. The lattice Boltzmann collision
operator controls the relaxation rates of hydrodynamic and non-hydrodynamic moments separately; Lallemand and Luo
{[}33{]} demonstrated that BGK couples these rates and produces viscosity-dependent dispersion errors, while TRT
{[}31{]} and central-moment MRT {[}38{]} decouple them. In the present analysis, the boundary-enforcement scheme
(Sec.~3) and the kernel choice (Sec.~4) determine the effective interface, while the collision operator chiefly modifies
how non-hydrodynamic moments dissipate inside the bulk fluid (for DF and MDF the boundary forcing enters collision
through the Guo source term of Sec. 2.2, while DFC applies its correction at the distribution level, Sec. 2.5.3; in
either case the scheme-to-scheme variation is held fixed here); under the bounded Reynolds-number range and
reference-defined peak metrics tested here, the bulk dissipation difference is small compared with the boundary-induced
response, and the BGK / TRT / CM-MRT spread is therefore \(\leq 0.94\%\) on the fixed cylinder and \(\leq 0.69\%\) on
the reference wake-interaction case. Coreixas et al.~{[}28{]} document that collision-model differences become more
prominent at higher Reynolds numbers and in three-dimensional configurations --- outside the present analysis scope.

MDF achieves an order-of-magnitude slip reduction through iteration. DF computes the boundary force from a single
residual velocity at the marker and spreads it once per time step; any residual that the spread misses survives to the
next step. MDF iteratively re-evaluates the marker velocity after each spread and accumulates successive corrections
{[}41{]}. Under the adaptive \(N \in [5,20]\) stopping rule and the \(L_\infty\) slip-residual early-termination
criterion of Sec.~2.5.2, the observed slip residual decreases with iteration count and reaches the order-of-magnitude
reduction reported in Table III within the adaptive window. The iteration is the algorithmic reason MDF tracks the
boundary more tightly than a single-shot DF for the same kernel.

\subsubsection{7.2 Limitations}\label{limitations}

The conclusions reached here are bounded by the regimes, metrics, and discretizations of the present analysis. Detailed
denominator definitions, domain analyses, and per-case ablation tables are provided in Appendices B and C and in the
supplementary material.

All cases are two-dimensional, and the resolved range is finite. The fixed cylinder spans
\(\mathrm{Re} \in \{20, 40, 100, 200\}\), the oscillating cylinder is at \(\mathrm{Re} = 100\), and sedimentation
density ratios cover \(\rho_s/\rho_f \in \{1.01, 1.1, 1.25, 1.5\}\) (single-particle, where \(\rho_s/\rho_f = 1.25\) is
the Sec.~5.2 density-matched lighter control) with \(\rho_r = 1.5\) heavy / \(1.25\) light for the wake-interaction
pair. Three-dimensional wake-transition claims, far-wake claims, and behavior at higher Re or more extreme density
ratios are outside the present scope.

The moving-particle conclusions are reported on defined domains and windows. The two-particle wake-interaction case uses
the Majumder \(50D\) streamwise domain; the sign of the ablation contrast is shared by DF/MDF/DFC on both the \(50D\)
domain and the \(60D\) extension, while the magnitude varies with the domain and window in the
without-explicit-correction case (Appendix C Sec.~C.2, Table~C.3). A harmonized-window coupled space--time refinement
comparison and a \(60D\) without-explicit-correction single-particle control are reported in Secs.~C.5 and C.2, and the
recorded contrasts are finite-window quantities. Reference comparisons (Wang, Majumder, Uhlmann) adopt each source as a
reporting procedure; the internal-mass correction ablation is a force-evaluation comparison of the discrete closure.

The DFC mechanism is supported by local diagnostics rather than by a closed stress-budget proof. The DFC kernel reversal
interpretation in Sec. 4 rests on four mutually consistent local diagnostics (\(\lambda_k\) scaling, correction
magnitude, azimuthal slip, near-boundary pressure deviation) and a \(W/\delta\) scale ordering across the available
cases (with a single minor inversion at the highest \(W/\delta\); supporting rank-correlation diagnostics in Appendix C
Sec.~C.6). The full causal chain from non-uniform \(\lambda_k\) through anisotropic correction to near-wake stress and
pressure-gradient field is identified as a future direction; a closed angular pressure-drag decomposition is not
constructed in the present analysis. The Sec.~5.2.3 particle-bound acceleration proxy is reported as a kinematic
signature of the wake-exposed geometry.

Implicit velocity correction, cumulant collision, and full collision-operator studies remain outside the present scope.
The boundary-only direct-forcing framework enforces no-slip through a Lagrangian surface force and treats the internal
fluid as a fictitious extension; \emph{implicit velocity correction (IVC) schemes} {[}51{]} that explicitly correct the
internal velocity field and their integration with the DF / MDF / DFC family compared here are identified as a separate
study. Cumulant collision models (e.g., the parameterized cumulant LBM applied by Xing et al.~{[}77{]} to subcritical-Re
cylinder flow) and a full BGK / TRT / MRT / cumulant collision-operator study across the full kernel and case matrix
exceed the scope of the present three-way IBM benchmark.

\subsubsection{7.3 Outlook}\label{outlook}

The mechanism-resolved framing suggests several extensions.

Integration of IVC schemes {[}51{]} with the DF / MDF / DFC family compared here would clarify whether the boundary-only
enforcement limitation reported in Sec.~3.2 (internal residual velocity 1--9\% of \(U_\infty\)) is suppressed by
interior-node correction without affecting the kernel-reversal mechanism of Sec.~4 or the wake-exposed closure
sensitivity of Sec.~5. Such an integration would close the boundary-only gap left by the present comparison.

In three dimensions, the DFC memory overhead increases (\(D3Q19\) correction storage vs \(D2Q9\)), and the wake-exposed
closure sensitivity may be altered by three-dimensional wake instabilities absent in 2D. Whether the light-particle
closure sensitivity (\(-16.4\%\) paired ablation contrast on the \(50D\) window) persists, amplifies, or saturates in 3D
is an open question. The computational route to such three-dimensional extensions is already established:
block-structured adaptive-mesh IB-LBM solvers demonstrate efficient large-scale GPU performance, with validations
extending to drafting--kissing--tumbling sphere pairs and hundreds of freely settling spheres {[}78{]}, while
heterogeneous CPU--GPU frameworks scale fully resolved particle-laden flows to tens of billions of lattice cells
{[}79{]}.

A full extension of the targeted controls of Sec. 6 to cumulant operators would test whether the bounded \(\leq 0.94\%\)
fixed-cylinder spread and the \(0.21\%/0.69\%\) wake-interaction spread persist or grow when relaxation rates are
allowed to differ across higher-order moments.

Beyond the boundary-local \(\Delta p_k\) diagnostic of Fig. 5(d), a global reconstruction of the near-wake stress tensor
and pressure-gradient field propagated from the corrected distribution functions would establish the full causal chain
from non-uniform \(\lambda_k\) to altered effective boundary shape.

\subsection{8. Conclusions}\label{conclusions}

A controlled three-way benchmark of direct forcing (DF), multi-direct forcing (MDF), and distribution-function
correction (DFC) for the immersed-boundary lattice Boltzmann method has been conducted with matched solver settings
across each benchmark configuration and analyzed within a mechanism-resolved framing of interface momentum transfer.
Rather than ranking the three schemes, the analysis identifies which interface momentum-transfer mechanism each
canonical benchmark activates, supported by case-resolved local diagnostics.

For prescribed bodies, the differences among the three schemes are most clearly discriminated by local no-slip fidelity
rather than by a universal drag ranking. MDF achieves an order-of-magnitude slip reduction under both kernels
(\(7.5\)--\(10.0\times\) for hat, \(55\)--\(125\times\) for Peskin 4-point) at the tested Reynolds numbers; global drag
alone is insufficient as a discriminator. DFC under hat shows a kernel-dependent local fidelity gap that prefigures the
Sec.~4 kernel reversal.

The DFC \(\bar{C}_d\) ranking reversal between the hat and Peskin 4-point kernels at \(\mathrm{Re} \geq 100\) is
consistent with the spatial redistribution of the marker-resolved correction field, supported by four mutually
consistent local diagnostics (\(\lambda_k\) scaling --- coefficient of variation differs by \(\approx 10\)--\(16\times\)
across analyzed resolutions; correction magnitude; azimuthal slip; near-boundary pressure deviation) and a bounded
\(W/\delta\) scale ordering. The diagnostics do not by themselves constitute a closed causal proof.

For freely moving particles, the two-particle wake-interaction case shows the wake-exposed light particle to be
comparatively more sensitive to the explicit internal-mass correction in the force evaluation: a \(-16.4\%\) paired
shift relative to the explicit-history baseline (equivalently a \(13.8\%\) underprediction of the Majumder reference
value) on the \(50D\) measurement window. The signed ablation contrast has the same direction in the DF / MDF / DFC
comparisons on both the \(50D\) and \(60D\)-extended domains, and the magnitude varies with the domain and window in the
without-explicit-correction case, attenuating on the longer domain for every scheme (per-domain values in Appendix C
Sec. C.2 Table C.3). The within-run heavy-particle comparator remains at or below \(3.0\%\) on every tested
boundary-enforcement scheme, and the isolated-particle controls, both without two-particle wake interaction, record only
small finite-window contrasts (\(-1.06\%\) for the approximately peak-\(\mathrm{Re}_f\)-matched high-Ga control and
\(-0.86\%\) for the matched isolated-light configuration). The heightened internal-mass-closure sensitivity coincides
with the post-exchange trailing (wake-exposed) phase. Trajectory phase diagnostics, a particle-bound acceleration proxy,
the isolated-particle controls, and a Majumder-aligned wake-interaction cross-check (\(-13.91\%\) light-particle
deficit) narrow the plausible alternatives and associate the heightened sensitivity with the post-exchange two-particle
configuration, a configuration-dependent finite-window sensitivity.

Targeted collision-model controls remain secondary: within the tested regimes and reference-defined peak metrics
(prescribed-body comparisons at \(\mathrm{Re} \leq 200\), two-dimensional, and the reported moving-particle cases), the
BGK / TRT / CM-MRT spread is bounded --- at most \(0.94\%\) across the twelve-combination fixed-cylinder matrix (Table
B.4) and \(0.21\%\) (heavy) / \(0.69\%\) (light) for the reference correction-inclusive wake-interaction baseline. The
supporting case matrices are reported in Appendix B.

These results replace a universal ranking of DF, MDF, and DFC with a regime-dependent interpretation of how IB-LBM
simulations transfer momentum across the Eulerian--Lagrangian interface: different benchmarks activate different
mechanisms, and the appropriate question is which mechanism each benchmark probes. Conclusions are restricted to
body-localized metrics within the tested regimes (prescribed-body comparisons at \(\mathrm{Re} \leq 200\),
two-dimensional; moving-particle cases as reported); far-wake quantities, three-dimensional wake-transition claims, full
cumulant collision-operator studies, IVC integration with the DF / MDF / DFC family, and configurations outside the
analyzed Reynolds-number / density-ratio / discretization range are outside scope (cf.~Sec.~7.2).

\subsection{Supplementary Material}\label{supplementary-material}

See the supplementary material accompanying this article for the full BGK / TRT / CM-MRT × DF / MDF / DFC × hat / Peskin
4-point fixed-cylinder \(\bar{C}_d\), \(\mathrm{St}\), and \(C_l\) matrices and the BGK boundary-fidelity matrix (Sec.
S1), the extended grid / marker / boundary-condition analysis (Sec. S2), the qualitative wake-vorticity comparison (Sec.
S3), and the raw single-particle and two-particle wake-interaction sedimentation trajectories and time series that
underlie the Sec.~5 mechanism analyses (Sec.~S4).

\subsection{Acknowledgments}\label{acknowledgments}

The author acknowledges no external funding for this work.

\subsection{Author Declarations}\label{author-declarations}

\subsubsection{Conflict of Interest}\label{conflict-of-interest}

The author has no conflicts to disclose.

\subsubsection{Author Contributions}\label{author-contributions}

\textbf{Hongju Jo}: Conceptualization (lead); Methodology (lead); Software (lead); Validation (lead); Formal analysis
(lead); Investigation (lead); Data curation (lead); Visualization (lead); Writing --- original draft (lead); Writing ---
review and editing (lead).

\subsection{Data Availability Statement}\label{data-availability-statement}

Processed benchmark data (status summaries, single- and two-particle sedimentation histories, reference data, and
convergence arrays), analysis scripts, simulation configuration files, and source code are openly available in
\mbox{Zenodo} at \mbox{https://doi.org/10.5281/zenodo.22232642} and in the public GitHub repository
(\mbox{https://github.com/mulgae-life/iblbm-mechanism-resolved}, release v1.0.0), including the
local-diagnostic, internal-mass correction ablation, targeted collision-control, and grid/domain analysis datasets
compiled for the present study. These reproduce the reported tables and the figures derived from the processed data;
figures that require large raw velocity fields, marker-resolved DFC diagnostic arrays, or raw two-particle
velocity-field snapshots rely on an optional raw-field package that is not bundled in the repository and is available
from the corresponding author upon reasonable request.

\subsection{Appendix A --- Numerical Formulation (Details)}\label{appendix-a-numerical-formulation-details}

\subsubsection{A.1 Grid, Time-Step, and Stability Constraints}\label{a.1-grid-time-step-and-stability-constraints}

The lattice spacing \(\Delta x\) and time step \(\Delta t\) are set in lattice units throughout. The kinematic viscosity
in lattice units is \(\nu_\mathrm{lat} = c_s^2 (\tau - 1/2) \Delta t\) with \(c_s^2 = 1/3\). For the fixed cylinder,
\(\tau = 1/2 + 3 \nu_\mathrm{lat}\) with \(\nu_\mathrm{lat}\) chosen so that
\(\mathrm{Re} = u_\infty D / \nu_\mathrm{lat}\) matches the target Reynolds number (see Table I). Stability requires
\(\tau > 1/2\); the empirical stability bound for SRT-BGK lattice operators near \(\tau = 1/2\) is documented by
Sterling and Chen {[}80{]}. In the present BGK runs, we keep \(\tau\) above an empirical safety threshold of
\(\tau \approx 0.55\) to avoid numerical instability near the \(\tau \to 1/2\) relaxation limit at
\(\mathrm{Re} = 200\). The \(\mathrm{Re} = 200\) fixed-cylinder runs therefore use \(D = 60\Delta x\) on the
\(3001 \times 2401\) grid (Table I), which gives \(\tau = 0.59\) (margin \(0.04\) above the safety bound), rather than
the coarser \(N_y = 1601\) (\(D = 40\Delta x\)) configuration (\(\tau = 0.56\), marginal). A finer \(N_y = 3201\)
(\(D = 80\Delta x\)) configuration (\(\tau = 0.62\)) was tested but is not adopted as the production grid for the
\(\mathrm{Re} = 200\) comparison; the \(3001 \times 2401\) baseline of Table I is adopted instead, and the converged
tail of the \(\mathrm{Re} = 200\) runs (steps 30,200--60,000, approximately \(9.9\) vortex-shedding cycles at
\(\mathrm{St} \approx 0.20\)) is used for the time-averaged \(\bar{C}_d\) in Table II, so that the present
\(\mathrm{Re} = 200\) comparison is reference-defined with the IB-LBM target of Wu and Shu {[}51{]}.

For sedimentation, the lattice parameters are determined by a Mach-constrained strategy: given the Archimedes number
\(\mathrm{Ar} = (\rho_s/\rho_f - 1) g d^3 / \nu^2\), the minimum lattice diameter \(D_\mathrm{lat}\) is computed from
the requirement \(\mathrm{Ma} = \sqrt{3 \, \mathrm{Ar}} \cdot \nu_\mathrm{lat} / D_\mathrm{lat} < 0.08\), with a safety
factor of 4 to ensure adequate \(\tau\) margin above the BGK stability limit and sufficient IBM resolution
(\(D_\mathrm{lat} \geq 40\)). The gravitational velocity scale \(u_g = \sqrt{|\Delta\rho/\rho_f| g D}\) and the
dimensionless time \(t^* = t \, u_g / D\) are used for non-dimensionalization.

\subsubsection{A.2 MDF Convergence Monitor}\label{a.2-mdf-convergence-monitor}

All MDF runs in this manuscript use the adaptive form defined in Sec.~2.5.2:
\(\omega_{\mathrm{MDF}} \approx \|A\|_\infty^{-1}\) (Zhang et al.~{[}41{]}) and an \(L_\infty\) slip-residual stopping
criterion (tolerance \(\mathrm{tol} = 10^{-4}\)) with a divergence guard \(r_n > 1.5\,r_{n-1}\) that breaks the
iteration. Iteration is bounded by \(N \in [N_{\min}, N_{\max}] = [5,20]\). The unrelaxed MDF Richardson iteration can
diverge when the spectral radius of the iteration matrix \(I - A\) exceeds unity --- equivalently, when an eigenvalue of
the symmetric positive-definite interpolation--spreading matrix \(A\) exceeds \(2\) --- under certain delta-function and
grid configurations (Zhang et al.~{[}41{]}); for freely moving bodies the coupled-iteration stability additionally
depends on the particle-to-fluid density ratio, as characterized by Suzuki et al.~{[}25{]}. The \(r_n > 1.5\,r_{n-1}\)
guard is the single-particle safeguard; the coupled two-particle path uses the \(20\%\) residual-growth guard with
relaxation halving (Sec. 2.5.2). For the single-particle sedimentation runs, the stored MDF diagnostics are
predominantly tolerance-terminated, typically in \(5\)--\(7\) iterations; in the coupled two-particle runs the MDF
iteration operates at the \(N_{\max} = 20\) ceiling with end-of-iteration \(L_\infty\) residuals of
\(1.2\)--\(1.4\times10^{-4}\), and is reported as an \(N_{\max}\)-capped evaluation.

\subsubsection{A.3 Two-Particle Wake-Interaction (No Direct
Contact)}\label{a.3-two-particle-wake-interaction-no-direct-contact}

For the two-particle wake-interaction configuration the inter-particle interaction is hydrodynamic by construction ---
no collision or contact model is active, so the only coupling pathway in the numerical model is the resolved fluid field
--- consistent with the Majumder et al.~{[}23{]} reference setup, which reproduces the pure wake-interaction case of
Uhlmann {[}14{]} (Sec. 5.2.2 therein): the denser trailing particle overtakes the lighter leading particle with no
geometric contact observed at any stored sample. At every stored state in the analysis window the conservative surface
clearance at closest approach (minimum surface gap \(g_{\min}/D \approx 0.56\); Table C.2) exceeds the Peskin 4-point
support-overlap threshold, so direct stencil overlap is excluded at the evaluated stored states. This is consistent with
Uhlmann's report that this configuration requires no collision model.

\subsubsection{A.4 IBM FLOP Analysis}\label{a.4-ibm-flop-analysis}

Table A.1 reports the per-marker, per-step IBM FLOP counts. The breakdown distinguishes the per-stencil delta-function
evaluation overhead \(C_\delta\) (hat \(C_\delta = 13\), Peskin 4-point \(C_\delta = 31\)) from the field-pass cost. DFC
processes 9 distribution functions (vs 3 macroscopic fields for DF), so the relative DFC/DF ratio depends on the kernel
--- a heavier stencil overhead dilutes the cost of additional fields.

\noindent \textbf{Table A.1.} IBM FLOP count per Lagrangian marker per time step. The counts enumerate the full delta-function
stencil (\(S = 3\) for hat, \(S = 5\) for Peskin 4-point; \(S^2 = 9\) and \(25\)); the outermost nodes carry zero
weight, so the non-zero support is \(2 \times 2\) and \(4 \times 4\) as in Sec. 2.4.

\needspace{4\baselineskip}
\vspace{0.8em}
\sbox0{\scriptsize\setlength{\tabcolsep}{3pt}\renewcommand{\arraystretch}{1.2}%
\begin{tabular}{c l r r}
\toprule
Method & Operation & Hat (\(S^2 = 9\)) & P4 (\(S^2 = 25\)) \\
\midrule
DF & Interpolation (3 fields) & 198 & 1,000 \\
DF & Force computation & 8 & 8 \\
DF & Spreading (2 fields) & 162 & 900 \\
DF & Total & 368 & 1,908 \\
DFC & \(f_i\) interpolation (9 fields) & 360 & 1,450 \\
DFC & Bounce-back + deviation & 81 & 81 \\
DFC & \(\delta f\) spreading (9 fields) & 288 & 1,250 \\
DFC & Fluid force & 54 & 54 \\
DFC & Total (stationary) & 783 & 2,835 \\
\bottomrule
\end{tabular}}
\ifdim\wd0>\textwidth\noindent\resizebox{\textwidth}{!}{\usebox0}\else\noindent\makebox[\textwidth][c]{\usebox0}\fi
\vspace{0.8em}

The total per-step cost (LBM + IBM) is approximately \(1.0\times\) DF for DF, \(2\)--\(3.3\times\) DF for MDF (the
spread reflects the adaptive \(N \in [5,20]\) window of Sec. 2.5.2; on the linear \(0.74 + 0.26\,N\) per-step cost model
the lower bound corresponds to \(N \approx N_{\min}\) and the upper \(3.3\times\) to \(N \approx 10\); the
single-particle sedimentation diagnostics are predominantly tolerance-terminated at typical \(N \approx 5\)--\(7\),
while the coupled two-particle runs operate at the \(N_{\max} = 20\) ceiling, corresponding to \(\approx 5.9\times\)),
and \(1.1\)--\(1.3\times\) DF for stationary DFC (\(1.4\)--\(1.5\times\) for moving DFC, where \(\lambda_k\) must be
recomputed every step). The IBM operations constitute approximately \(26\%\) of the DF per-step cost, and the relative
IBM cost ratios among the three schemes remain similar at higher resolutions.

\subsubsection{A.5 Internal-Mass Correction: Suzuki--Inamuro Taxonomy and Force-Equivalent
Form}\label{a.5-internal-mass-correction-suzukiinamuro-taxonomy-and-force-equivalent-form}

The boundary-only hydrodynamic force \(\mathbf{F}_{\mathrm{hydro}}\) recovered from the IBM forcing does not include the
contribution of the fictitious internal fluid that occupies the solid region in a regularized-delta IBM formulation. The
explicit-history form of Feng and Michaelides {[}48{]} supplies this contribution as an algebraic correction to the
particle force evaluation,

\[\mathbf{v}^{n+1}_s = \left(1 + \frac{\rho_f}{\rho_s}\right) \mathbf{v}^n_s - \frac{\rho_f}{\rho_s} \mathbf{v}^{n-1}_s + \frac{\Delta t}{m_s} \left[ \mathbf{F}_{\mathrm{hydro}}^n + \mathbf{F}_{\mathrm{gravity}} \right]. \tag{A1}\]

Here \(\mathbf{F}_{\mathrm{hydro}}^n\) is the particle-side hydrodynamic force of Sec. 2.7 --- for DF/MDF,
\(\mathbf{F}_{\mathrm{hydro}} = -\sum_{\mathbf{x}} \mathbf{f}(\mathbf{x})\,h^2\), the fluid-side Eulerian forcing
\(\mathbf{f}\) of Sec. 2.3 reversed by Newton's third law --- corresponding to the surface-traction term written as
\(\sum_l \mathbf{f}_l\,dV_l\) in the original Feng and Michaelides {[}48{]} equation;
\(\mathbf{F}_{\mathrm{gravity}} = (m_s - m_f)\,\mathbf{g}\) is the reduced-gravity term of Eq. (22). The denominator is
\(m_s\), the rigid-body particle mass (Feng and Michaelides {[}48{]} convention; the alternative
\(m_{\mathrm{eff}} = m_s - m_f\) denominator of Uhlmann {[}14{]} is a different scheme --- cataloged as scheme (B-1) of
{[}36{]} --- that is not used in the present manuscript). The previous-step particle-velocity term
\((\rho_f / \rho_s)(\mathbf{v}^n_s - \mathbf{v}^{n-1}_s)\) is the quantity referred to by the numerical
``explicit-history'' label of Feng and Michaelides {[}48{]}; it is not a Basset history force.

Eq. (A1) is \emph{not} used as a stand-alone time integrator here; the time integration of the particle position and
velocity is performed by the Velocity-Verlet scheme of Sec.~2.7. Eq. (A1) is used as an \emph{internal-mass
force-evaluation correction}: the difference between Eq. (A1) and the corresponding \emph{no-internal-mass} form (scheme
(A) of Suzuki and Inamuro {[}36{]}) is interpreted as the internal-fluid / fictitious-mass contribution that the IBM
boundary force omits, and is added to \(\mathbf{F}_{\mathrm{hydro}}\) before the Velocity-Verlet update.

The four-scheme taxonomy of Suzuki and Inamuro {[}36{]} catalogs the available internal-mass treatments: (A) no
internal-mass effect (boundary-only force used as is); (B-1) Uhlmann denominator substitution
\(m_\mathrm{eff} = m_s - m_f\) {[}14{]} that lumps the fictitious-fluid mass into the effective particle mass; (B-2)
Feng--Michaelides explicit-history correction {[}48{]} adding an explicit previous-step particle-velocity term (the
numerical ``explicit-history'' discretization of Sec. 2.7, distinct from the physical Basset history force); and (C)
Lagrangian-points approximation distributing the internal-mass effect over the immersed markers. The present three-way
IB-LBM benchmark implements (A) and (B-2) --- with (B-2) Feng explicit-history as the reference baseline matching the
standard direct-forcing IB-LBM particle-flow literature {[}74{]} --- and treats (B-1) and (C) as outside the present
scope. Within this taxonomy, the correction-inclusive baseline of Sec.~5.1 and Sec.~5.2 corresponds to scheme (B-2); the
corresponding \emph{no-explicit-correction} ablation corresponds to scheme (A) of {[}36{]}, which sets the internal
fluid contribution to zero and uses the boundary-only update

\[\mathbf{v}^{n+1}_{s,\,\mathrm{scheme\,(A)}} = \mathbf{v}^n_s + \frac{\Delta t}{m_s} \left[ \mathbf{F}_{\mathrm{hydro}}^n + \mathbf{F}_{\mathrm{gravity}} \right]. \tag{A2}\]

Concretely, the internal-mass force correction extracted from Eq. (A1) relative to the scheme (A) update Eq. (A2) is

\[\Delta \mathbf{F}_{\mathrm{IM}}^n = \frac{m_s}{\Delta t} \left[ \mathbf{v}^{n+1}_{s,\,\mathrm{Eq.}(A1)} - \mathbf{v}^{n+1}_{s,\,\mathrm{scheme\,(A)}} \right] = \frac{m_f}{\Delta t} \left( \mathbf{v}^n_s - \mathbf{v}^{n-1}_s \right), \tag{A3}\]

which is a force-equivalent form of the explicit-history velocity increment of Feng and Michaelides {[}48{]} (a backward
first-order finite-difference of the previous-step velocity) and depends only on \(m_f = \rho_f V_p\) and the
previous-step velocity history. This \(\Delta \mathbf{F}_{\mathrm{IM}}^n\) is added to \(\mathbf{F}_{\mathrm{hydro}}\)
before the Velocity-Verlet update of Eq. (23), so the Velocity-Verlet integrator receives a corrected total force at
each step rather than being replaced by Eq. (A1). Because \(\Delta \mathbf{F}_{\mathrm{IM}}^n\) enters the total force
\(\mathbf{F}_{\mathrm{total}}\) of the Velocity-Verlet update (Eq. (23)), each of the two half-kicks receives half of
the internal-mass increment, \(\tfrac{1}{2}\,\Delta t\,\Delta \mathbf{F}_{\mathrm{IM}}^n / m_s\); the history velocities
\(\mathbf{v}_s^n\) and \(\mathbf{v}_s^{n-1}\) are evaluated at the current step and held fixed across the two
half-updates. The internal-mass term is therefore treated explicitly (first-order in time, lagging the resolved force),
while the resolved hydrodynamic and gravitational terms retain the second-order Velocity-Verlet accuracy. In a fully
resolved immersed-boundary formulation, the resolved unsteady fluid-inertia response {[}49,50{]} remains in the flow
solution and the boundary force (see the three-notion distinction of Sec. 2.7); the present comparison quantifies
sensitivity to whether the fictitious-fluid / internal-mass contribution is \emph{explicitly} corrected in the particle
force evaluation. Eshghinejadfard et al.~{[}74{]} adopt the same Feng--Michaelides explicit-history correction in their
direct-forcing IB-LBM particle-flow simulations, including 2D wake-interaction sedimentation and 3D single-sphere
settling.

\subsection{Appendix B --- Method × Kernel × Collision Case
Matrices}\label{appendix-b-method-kernel-collision-case-matrices}

The targeted-control summary statistics of Sec.~6 are supported by the underlying method × kernel × collision case
matrices. The collision-spread tabulation (Sec.~B.4), the single-particle internal-mass ablation matrix (Sec.~B.5), and
the two-particle wake-interaction case matrix (Sec.~B.6) --- all of which underpin the mechanism analyses of Sec.~5 and
Sec.~6 --- are reported here. The full \(\bar{C}_d\) and \(\mathrm{St} / C_l\) matrices, the BGK boundary-fidelity
matrix, and the qualitative vorticity comparison are reported in Supplementary Material Sec.~S1.

\subsubsection{\texorpdfstring{B.1 Fixed-Cylinder \(\bar{C}_d\) Full
Matrix}{B.1 Fixed-Cylinder \textbackslash bar\{C\}\_d Full Matrix}}\label{b.1-fixed-cylinder-barc_d-full-matrix}

The complete \(\bar{C}_d = \langle C_d \rangle_{t > t_{\mathrm{conv}}}\) matrix across BGK / TRT / CM-MRT × DF / MDF /
DFC × hat / Peskin 4-point × \(\mathrm{Re} \in \{20, 40, 100, 200\}\) (72 cases) is reported in Supplementary Material
Sec.~S1.1 Tables S1(a)--(d). The 24-case BGK subset reproduces the main text Sec.~3.1 Table II; the collision spread
across the twelve-combination matrix is summarized in Sec.~B.4 below.

\subsubsection{B.2 Strouhal Number and Lift Amplitude}\label{b.2-strouhal-number-and-lift-amplitude}

The full \(\mathrm{St}\) and lift-amplitude matrix at \(\mathrm{Re} = 100\) (\(C_{l,\mathrm{amp}}\), peak-to-mean) and
\(\mathrm{Re} = 200\) (\(C_{l,\mathrm{rms}}\), converged tail; Table II caption) across all 36 BGK / TRT / CM-MRT × DF /
MDF / DFC × hat / Peskin 4-point × \(\mathrm{Re} \in \{100, 200\}\) cases is reported in Supplementary Material
Sec.~S1.2 Tables S2 and S3. Two results from the main text Sec.~3.1 are: all \(\mathrm{Re} = 200\) cases share the
uniform FFT-bin Strouhal estimate \(\mathrm{St} = 0.200\), consistent with the literature range {[}51,58{]} at the FFT
resolution (\(\Delta\mathrm{St} \approx 0.02\)); the \(C_{l,\mathrm{rms}}\) at \(\mathrm{Re} = 200\) for the hat kernel
(\(0.44\)--\(0.49\) across the nine collision × method cases) is comparable to the Qu et al.~{[}59{]} body-fitted
reference \(C_{l,\mathrm{rms}} = 0.4678\) (relative ratio \(0.93\)--\(1.06\times\)), whereas the P4 cases
(\(0.33\)--\(0.41\)) lie systematically below it (\(0.70\)--\(0.87\times\)). The kernel-dependent gap in this 2D
vortex-shedding regime is consistent with the kernel-dependent boundary thickness analyzed in Sec.~4: the wider P4
support produces a thicker effective immersed boundary that attenuates the in-plane \(C_l\) amplitude. The relevant
three-dimensional transition, the Mode-A Floquet instability, sets in at \(\mathrm{Re}_c \approx 188.5\) {[}43{]}; the
present simulations are two-dimensional and 3D wake-transition effects are explicitly outside scope.

\subsubsection{B.3 Boundary Fidelity (Slip / Leakage / Internal
Residual)}\label{b.3-boundary-fidelity-slip-leakage-internal-residual}

The representative slip-error matrix is reported in main Sec.~3.2 Table III. The full BGK collision × DF / MDF / DFC ×
hat / Peskin 4-point × Re matrix of leakage flux \(\Phi_\mathrm{leak}\) and internal residual velocity
\(\|\mathbf{u}\|_\mathrm{inside}\) is summarized in Supplementary Material Sec.~S1.3. Across all 24 BGK cases,
\(|\Phi_\mathrm{leak}| < 4.1 \times 10^{-4}\) with sign alternation (near-zero net flux) and
\(\|\mathbf{u}\|_\mathrm{inside} / U_\infty \in [0.01, 0.09]\) --- the boundary-only IBM internal-residual band
identified in Sec.~3.2 and Sec.~7.2.

\subsubsection{B.4 Collision-Model Spread Across Twelve Method--Kernel--Reynolds
Cases}\label{b.4-collision-model-spread-across-twelve-methodkernelreynolds-cases}

\noindent \textbf{Table B.4.} Complete BGK / TRT / CM-MRT collision spread of \(\bar{C}_d\) across the twelve fixed-cylinder cases
(3 boundary-enforcement schemes × 2 kernels × \(\mathrm{Re} \in \{100, 200\}\) = 12 cases; \(\mathrm{Re}=20\) / 40
entries appear in Sec.~S1.1 only as \(\bar{C}_d\) values, not as a collision-spread analysis). Spread is computed as
\((\max - \min) / \mathrm{mean} \times 100\%\) across the three collision operators (BGK, TRT, CM-MRT) at fixed method,
kernel, and Re.

\needspace{4\baselineskip}
\vspace{0.8em}
\sbox0{\scriptsize\setlength{\tabcolsep}{3pt}\renewcommand{\arraystretch}{1.2}%
\begin{tabular}{r l l r r r r}
\toprule
Re & Method & Kernel & BGK \(\bar{C}_d\) & TRT \(\bar{C}_d\) & CM-MRT \(\bar{C}_d\) & Spread (\%) \\
\midrule
100 & DF & hat & 1.3965 & 1.3834 & 1.3947 & 0.94 (max) \\
100 & DF & P4 & 1.4178 & 1.4095 & 1.4166 & 0.59 \\
100 & MDF & hat & 1.3849 & 1.3720 & 1.3829 & 0.93 \\
100 & MDF & P4 & 1.3867 & 1.3777 & 1.3853 & 0.65 \\
100 & DFC & hat & 1.4160 & 1.4126 & 1.4156 & 0.24 \\
100 & DFC & P4 & 1.4051 & 1.4021 & 1.4046 & 0.21 \\
200 & DF & hat & 1.3577 & 1.3663 & 1.3539 & 0.91 \\
200 & DF & P4 & 1.2569 & 1.2509 & 1.2584 & 0.60 \\
200 & MDF & hat & 1.3688 & 1.3643 & 1.3656 & 0.33 \\
200 & MDF & P4 & 1.2982 & 1.2971 & 1.2984 & 0.10 (min) \\
200 & DFC & hat & 1.4247 & 1.4231 & 1.4233 & 0.11 \\
200 & DFC & P4 & 1.3061 & 1.3026 & 1.3059 & 0.27 \\
\bottomrule
\end{tabular}}
\ifdim\wd0>\textwidth\noindent\resizebox{\textwidth}{!}{\usebox0}\else\noindent\makebox[\textwidth][c]{\usebox0}\fi
\vspace{0.8em}

Across the tested twelve-combination matrix, the maximum collision-only spread is \(0.94\%\) (attained at
\(\mathrm{Re} = 100\) for DF + hat); the minimum is \(0.10\%\) at \(\mathrm{Re} = 200\) MDF + Peskin 4-point. The
\(\mathrm{Re} = 200\) rows are computed on the converged tail (steps 30,200--60,000, \(\approx 9.9\) vortex-shedding
cycles at \(\mathrm{St} \approx 0.20\)) of the \(3001 \times 2401\) baseline grid (cf.~main Sec.~2.9.1 Table I; Appendix
A Sec.~A.1). The corresponding \(\mathrm{St}\) and \(C_{l,\mathrm{rms}}\) spreads at \(\mathrm{Re} = 200\) (reported in
Supplementary Material Sec.~S1.2 Table S3; the \(\mathrm{Re} = 100\) counterpart is reported as \(\mathrm{St}\) and
\(C_{l,\mathrm{amp}}\) in Table S2) are bounded by the same order of magnitude in the tested regime.

\subsubsection{B.5 Single-Particle Sedimentation Matrix}\label{b.5-single-particle-sedimentation-matrix}

\noindent \textbf{Table B.5.} Heavy-reference single-particle ablation spread (\(\rho_s/\rho_f = 1.5\), DF / BGK / Peskin 4-point
/ Velocity-Verlet) across the two implemented internal-mass schemes (A) and (B-2) (catalog of Suzuki and Inamuro
{[}36{]}). \(\mathrm{Re}_p\) is the particle-density basis \(\rho_P |U| D_P / \mu\) used by Wang et al.~{[}16{]}.

\needspace{4\baselineskip}
\vspace{0.8em}
\sbox0{\scriptsize\setlength{\tabcolsep}{3pt}\renewcommand{\arraystretch}{1.2}%
\begin{tabular}{l r r}
\toprule
Scheme & Heavy \(\mathrm{Re}_p\) & \(\Delta\%\) vs Wang value, same-source basis (503.38) \\
\midrule
(A) no-internal-mass & 498.08 & \(-1.05\%\) \\
(B-2) Feng explicit-history & 504.45 & \(+0.21\%\) \\
Spread (max \(-\) min) over (A) / (B-2) & \(6.37\) & \(1.27\%\) \\
\bottomrule
\end{tabular}}
\ifdim\wd0>\textwidth\noindent\resizebox{\textwidth}{!}{\usebox0}\else\noindent\makebox[\textwidth][c]{\usebox0}\fi
\vspace{0.8em}

The (B-1) Uhlmann denominator substitution and the (C) Lagrangian-points approximation of Suzuki and Inamuro {[}36{]}
are outside the present scope. The two implemented schemes yield a DF heavy-reference spread of \(1.27\%\), small
relative to the \(-16.4\%\) paired light-particle wake-interaction ablation contrast (Sec.~5.2). The corresponding
isolated-particle ablations at the lighter density ratios (DF / Peskin 4-point / Velocity-Verlet, \(N_x = 1281\)) yield
(B-2) Feng explicit-history \(\to\) (A) without-explicit-correction terminal-\(\mathrm{Re}_f\) shifts uniformly bounded
at \(\leq 0.7\%\): \(-0.32\%\) at \(\rho_s/\rho_f = 1.01\) (\(34.40 \to 34.29\)), \(-0.70\%\) at \(\rho_s/\rho_f = 1.1\)
(\(135.90 \to 134.95\)), and \(-0.14\%\) at \(\rho_s/\rho_f = 1.25\) (\(13.934 \to 13.915\)), supporting the inference
that the heavy single-particle bound extends across the analyzed light density-ratio range
(\(\rho_s/\rho_f \in \{1.01, 1.1, 1.25\}\)) for isolated-particle configurations without two-particle wake interaction.
The full per-density-ratio single-particle tables (BGK collision operator; DF / MDF / DFC boundary schemes with hat /
Peskin 4-point kernels, together with the explicit-history / without-correction ablation pairs) are provided in the
public repository, alongside the BGK / TRT / CM-MRT collision-operator comparison at \(\rho_s/\rho_f = 1.5\) (see Data
Availability Statement).

Extending this single-particle ablation to MDF and DFC (\(\rho_s/\rho_f = 1.5\), BGK / Peskin 4-point / Velocity-Verlet,
standard window, Wang same-source denominator \(503.38\)) gives Wang-normalized paired spreads, computed from unrounded
values, of \(1.27\%\) for DF, \(1.33\%\) for MDF, and \(1.30\%\) for DFC. The three spreads lie in the narrow range
\(1.27\)--\(1.33\%\), with the MDF and DFC values marginally above the DF value --- modest scheme-to-scheme variation;
the observed \(1.27\)--\(1.33\%\) range applies to the three schemes in this single-particle configuration (per-arm
values in the Supplementary Material).

A separate high-Ga isolated control at \(\rho_s/\rho_f = 1.25\) with the standard kinematic viscosity (\(\nu = 0.01\))
raises the isolated-particle Reynolds number into the band of the two-particle light response: it reaches peak Reynolds
numbers of \(228.70\) and \(226.26\) in the explicit-history and without-correction arms, respectively, both within the
prespecified matching band \([215, 263]\) (\(\pm 10\%\) of the \(50D\) reference light peak \(238.62\)). Its recorded
signed finite-window contrast is \(-1.06\%\), corresponding to \(0.065\) times the magnitude of the \(50D\) pair-case
light contrast. Although both isolated-control peaks fall within this matching band, the isolated configuration records
only a small fraction of the pair-case sensitivity over the recorded window.

On the common fluid-density basis for the single-particle Wang case, the compared implementations cluster consistently
below the continuous-solver DLM/FEM range: the DLM/FEM result of Glowinski et al.~{[}40{]}
(\(\mathrm{Re}_f = 438\)--\(466\)) lies above the immersed-boundary cluster at \(\mathrm{Re}_f = 327\)--\(336\). Wang's
own NF=20 finite-difference result is \(\mathrm{Re}_f = 335.59\) on this basis, i.e., \(-23.4\%\) relative to the lower
Glowinski endpoint. The present lattice-Boltzmann baseline (\(\mathrm{Re}_f = 336.30\)) lies \(-23.2\%\) to \(-27.8\%\)
relative to the Glowinski range. The present method, Wang et al.~{[}16{]}, the lattice-Boltzmann external-boundary-force
method of Parvan et al.~{[}67{]}, and the sharp-interface direct-forcing immersed-boundary method of Badri Ghomizad et
al.~{[}68{]} all exhibit a comparable offset on this benchmark, with reported terminal/peak Reynolds-number values on
the fluid basis clustering at \(327\)--\(336\) (NF=20 converged values; the under-converged Wang NF=1 value of
\(323.17\) is excluded) across two fluid discretizations (finite difference, lattice Boltzmann) and both diffuse and
sharp-interface treatments, whereas the fictitious-domain DLM/FEM result of Glowinski et al.~{[}40{]} lies at
\(438\)--\(466\). The fluid solver, interface treatment, grid and domain, blockage ratio, force-recovery procedure, time
integration, and terminal-value criterion are not uniformly matched across these references; resolving the controlling
element of this difference would require matching the solver, kernel, time integrator, grid, and terminal-value
convention across the references --- a matched cross-code comparison outside the present three-way IB-LBM benchmark
scope. On the particle-density basis the present baseline reproduces the Wang NF=20 value within \(+0.21\%\) (Table
B.6), reported as a reproduction check within the same diffuse-IBM class.

\noindent \textbf{Table B.6.} Single-particle reported terminal/peak Reynolds-number cross-source comparison
(\(\rho_s/\rho_f = 1.5\) Wang case; each source's value follows its own reporting criterion --- terminal or peak, and
the associated window; cf.~Sec. 5.1). The basis column states the Reynolds-number definition adopted by each source. The
\(\mathrm{Re}_f\) equiv. column converts each source's reported value to the fluid-density Reynolds-number basis
\(\mathrm{Re}_f = |U| D / \nu = \mathrm{Re}_p / 1.5\) for comparison across sources that use different bases. The match
against the Wang et al.~{[}16{]} particle-density basis is reported as a reproduction check within the same diffuse-IBM
class. NF denotes the number of direct-forcing iterations in Wang's {[}16{]} scheme. The fluid-basis values cluster
across diffuse and sharp-interface immersed-boundary methods on two fluid discretizations (Sec. 5.1).

\needspace{4\baselineskip}
\vspace{0.8em}
\sbox0{\scriptsize\setlength{\tabcolsep}{3pt}\renewcommand{\arraystretch}{1.2}%
\begin{tabular}{>{\raggedright\arraybackslash}p{0.269\linewidth} >{\raggedright\arraybackslash}p{0.261\linewidth} >{\raggedright\arraybackslash}p{0.166\linewidth} >{\raggedleft\arraybackslash}p{0.135\linewidth} >{\raggedleft\arraybackslash}p{0.119\linewidth}}
\toprule
Source & Method & Basis & \(\mathrm{Re}\) (source basis) & \(\mathrm{Re}_f\) equiv. \\
\midrule
Glowinski et al.~{[}40{]} & DLM/FEM & fluid: \(|U| D / \nu\) & 438--466 & 438--466 \\
Wang et al.~{[}16{]}, NF=1 & DF-IBM + 4th FD & particle: \(\rho_P U D / \mu\) & 484.75 & 323.17 \\
Wang et al.~{[}16{]}, NF=20 & MDF-IBM + 4th FD & particle: \(\rho_P U D / \mu\) & 503.38 & 335.59 \\
Parvan et al.~{[}67{]} & LBM + external boundary force & particle: \(\rho_P U D / \mu\) & 491.56--497.18 & 327.71--331.45 \\
Badri Ghomizad et al.~{[}68{]} & sharp-interface DF-IBM (MLS) + FD & particle: \(\rho_P U D / \mu\) & 498.56--499.94 & 332.37--333.29 \\
Present (DF, P4, explicit-history) & DF-IBM + LBM (BGK) & particle: \(\rho_P U D / \mu\) & 504.45 & 336.30 \\
Present \(-\) Wang NF=20 & --- & --- & \(+0.21\%\) & \(+0.21\%\) \\
\bottomrule
\end{tabular}}
\ifdim\wd0>\textwidth\noindent\resizebox{\textwidth}{!}{\usebox0}\else\noindent\makebox[\textwidth][c]{\usebox0}\fi
\vspace{0.8em}

\subsubsection{B.6 Two-Particle Wake-Interaction Case Matrix}\label{b.6-two-particle-wake-interaction-case-matrix}

\noindent \textbf{Table B.7.} Observation-window peak \(\mathrm{Re}_{p,\max}\) (particle-density basis
\(\mathrm{Re}_p = \rho_P |U| D_P / \mu\)) for the wake-interaction heavy and light particles across the analyzed
configurations. The reference baseline reported in Table VI is the same reference realization
\(\mathit{DF}\,/\,P4\,/\,\mathrm{Velocity}\text{-}\mathrm{Verlet}\,/\,\mathrm{explicit}\text{-}\mathrm{history}\), heavy
\(\rho_r = 1.5\) / light \(\rho_r = 1.25\) that supplies the Sec.~5.2 closure-sensitivity baseline (Table V,
fluid-density basis \(\mathrm{Re}_{f,\max} = 277.60 / 238.62\)); Table V (fluid-density basis) and this Table B.7
(particle-density basis) report the same run on the two reference bases (the bases differ by \emph{each particle's own}
density ratio --- heavy \(\rho_r = 1.5\), light \(\rho_r = 1.25\)).

\needspace{4\baselineskip}
\vspace{0.8em}
\sbox0{\scriptsize\setlength{\tabcolsep}{3pt}\renewcommand{\arraystretch}{1.2}%
\begin{tabular}{>{\raggedright\arraybackslash}p{0.392\linewidth} >{\raggedleft\arraybackslash}p{0.087\linewidth} >{\raggedleft\arraybackslash}p{0.087\linewidth} >{\raggedright\arraybackslash}p{0.384\linewidth}}
\toprule
Case & Heavy \(\mathrm{Re}_p\) & Light \(\mathrm{Re}_p\) & Note \\
\midrule
DF / BGK / Verlet / explicit-history (B-2) & 416.41 & 298.27 & reference baseline \\
DF / TRT / Verlet / explicit-history & 417.29 & 297.80 & collision-only variant \\
DF / CM-MRT / Verlet / explicit-history & 416.43 & 299.86 & collision-only variant \\
Heavy collision spread & \(0.21\%\) & --- & --- \\
Light collision spread & --- & \(0.69\%\) & --- \\
DFC / BGK / Verlet / explicit-history & 418.68 & 293.64 & DFC method-axis \\
DFC / TRT / Verlet / explicit-history & 419.56 & 290.10 & DFC + TRT \\
MDF / TRT / Verlet / explicit-history & 424.77 & 296.67 & MDF + TRT \\
DF / BGK / Verlet / scheme A (no explicit corr.) & 405.94 & 249.40 & ablation; light \(-13.78\%\) vs Majumder reference value \\
DF / BGK / Verlet / scheme B-2 (Feng explicit-history) & 416.41 & 298.27 & identical to reference baseline by construction \\
\bottomrule
\end{tabular}}
\ifdim\wd0>\textwidth\noindent\resizebox{\textwidth}{!}{\usebox0}\else\noindent\makebox[\textwidth][c]{\usebox0}\fi
\vspace{0.8em}

The collision-only sub-matrix (reference DF / Peskin 4-point / Velocity-Verlet / explicit-history case, heavy
\(\rho_r = 1.5\) / light \(\rho_r = 1.25\), × BGK / TRT / CM-MRT) yields the \(0.21\%\) (heavy) / \(0.69\%\) (light)
spread reported in the main text Table VI. Across method-axis variations (DFC, MDF) on the same explicit-history
correction, the heavy / light \(\mathrm{Re}_p\) remain within approximately \(\pm 2.0\%\) / \(\pm 2.7\%\) of the
reference baseline.~The time-resolved trajectories and Reynolds-number histories are available in the public repository
(see Data Availability Statement).

\subsubsection{B.7 Vorticity Field Comparison (Qualitative)}\label{b.7-vorticity-field-comparison-qualitative}

A qualitative wake-structure cross-check at \(\mathrm{Re} = 100\) and \(\mathrm{Re} = 200\) shows the wake morphology
(shoulder-vortex extent, flow separation, and vortex-shedding pattern) to be consistent across methods; no recirculation
length or separation angle is quantitatively extracted from these contours. The narrative summary, consistency with
Bouard and Coutanceau {[}56{]} and Koumoutsakos and Leonard {[}57{]}, and the per-method vorticity snapshots are
reported in Supplementary Material Sec.~S1.4; the field-level wake-vorticity figure for the DF / MDF / DFC × Peskin
4-point cases at \(\mathrm{Re} \in \{100, 200\}\) is reported in Supplementary Material Sec.~S3 Fig. S1.

\subsubsection{B.8 Oscillating-Cylinder Peak Drag}\label{b.8-oscillating-cylinder-peak-drag}

\noindent \textbf{Table B.8.} Oscillating cylinder peak drag coefficient (\(\mathrm{Re}=100\), \(\mathrm{KC}=5\); BGK collision;
cf.~Sec. 3.3). MDF refers to the adaptive form (Sec. 2.5.2). The Dütsch et al.~{[}39{]} reference value (\(c_d = 2.09\))
is a Morison-fit coefficient and a different observable from the present peak drag; the tabulated peak coefficients
indicate proximity to the reference scale and are not a quantitative validation metric.

\needspace{4\baselineskip}
\vspace{0.8em}
\sbox0{\scriptsize\setlength{\tabcolsep}{3pt}\renewcommand{\arraystretch}{1.2}%
\begin{tabular}{l r r r r}
\toprule
Delta & DF & MDF & DFC & Dütsch {[}39{]} \\
\midrule
hat & \(2.094\) & \(2.084\) & \(2.047\) & 2.09 \\
P4 & \(2.081\) & \(2.078\) & \(2.102\) & 2.09 \\
\bottomrule
\end{tabular}}
\ifdim\wd0>\textwidth\noindent\resizebox{\textwidth}{!}{\usebox0}\else\noindent\makebox[\textwidth][c]{\usebox0}\fi
\vspace{0.8em}

\subsection{Appendix C --- Grid, Domain, Marker, and Boundary-Condition
Analysis}\label{appendix-c-grid-domain-marker-and-boundary-condition-analysis}

\subsubsection{C.1 Domain Adequacy (Body-Local Analysis Windows)}\label{c.1-domain-adequacy-body-local-analysis-windows}

The body-local quantities used to support the mechanism-resolved claims are evaluated within the configurations
summarized in main Table I. The fixed- and oscillating-cylinder benchmarks adopt the standard streamwise / transverse
extents of their respective references. Domain adequacy is \emph{observable-dependent}: body-localized metrics (\(C_d\),
\(\mathrm{St}\), \(C_l\), slip / leakage / internal residual) are stable within the analyzed band, whereas far-wake
metrics depend on the streamwise extent and lie outside the present analysis. The detailed observable-dependent
rationale is reported in Supplementary Material Sec.~S2.1; the sedimentation standard-channel vs \(60D\)
extended-horizon analysis that supports the Sec.~5.2 wake-exposed closure sensitivity is provided in Sec. C.2 below.

\subsubsection{C.2 Sedimentation Channel 60D Extended-Domain
Analysis}\label{c.2-sedimentation-channel-60d-extended-domain-analysis}

The domain dependence of the Sec. 5.1 single-particle body-local peak Reynolds number and the Sec. 5.2
closure-sensitivity ablation contrast is assessed in two separate analyses: the single-particle question through an
extended-channel re-run on the heavy single-particle case (\(\rho_s/\rho_f = 1.5\)) in the Glowinski / Majumder
tall-channel at matching grid resolution \(N_x = 1281\) (Table C.1), and the ablation-contrast question through the
\(60D\) two-particle re-runs reported later in this section (Table C.3). Six configurations were tested for the
single-particle extension --- DF / MDF / DFC × BGK / TRT, all with Velocity-Verlet integration and the explicit-history
(B-2) internal-mass correction. The body-local \(|v_y^*|\) peak occurs at \(y^* \approx 25\)--\(33\) (\(y^*\) is
dimensionless, in \(D\) units; DF earliest, DFC latest), and the body-local peak Reynolds number is consistent with
near-saturation within the tested \(60D\) streamwise extent. Values are reported on the standard fluid-density basis
\(\mathrm{Re}_f = U D / \nu\) (Sec. 2.6), with the standard-channel baseline sourced from the Sec. 5.1 single-particle
reference runs at matched discretization.

\noindent \textbf{Table C.1.} Sedimentation single-particle (\(\rho_s/\rho_f = 1.5\), \(N_x = 1281\)) body-local peak Reynolds
number, standard fluid-density basis (cf.~Sec.~2.6), evaluated with the observable defined in Table IV (unsmoothed
\(|v_y^*|\) peak; the velocity-magnitude variant \(|\mathbf{v}^*|\) differs by \(0.0041\)--\(0.0572\%\) across the six
configurations on the full extended horizon). Three windows are reported per configuration: the standard \(24D\) channel
on its full contact-safe horizon (\(y^* \leq 15.5\)), the \(60D\) tall channel evaluated on the matched
\(y^* \leq 15.5\) window (isolating the channel-height effect), and the \(60D\) tall channel on its full contact-safe
horizon (\(y^* \leq 51.5\), adding the observation-horizon effect); the decomposition columns give the channel-height
component, the observation-horizon component, and their combination. Six configurations (DF / MDF / DFC × BGK / TRT, all
with Velocity-Verlet integration and explicit-history (B-2) internal-mass correction) are tested; records are restricted
to the pre-contact horizon with finiteness, in-domain, and step-monotonicity checks (one non-physical terminal record
removed from the MDF / TRT standard history). The standard-channel baseline values are sourced from the Sec.~5.1
single-particle reference runs at matched discretization.

\needspace{4\baselineskip}
\vspace{0.8em}
\sbox0{\scriptsize\setlength{\tabcolsep}{3pt}\renewcommand{\arraystretch}{1.2}%
\begin{tabular}{>{\raggedright\arraybackslash}p{0.074\linewidth} >{\raggedleft\arraybackslash}p{0.270\linewidth} >{\raggedleft\arraybackslash}p{0.270\linewidth} >{\raggedleft\arraybackslash}p{0.156\linewidth} >{\raggedleft\arraybackslash}p{0.057\linewidth} >{\raggedleft\arraybackslash}p{0.057\linewidth} >{\raggedleft\arraybackslash}p{0.066\linewidth}}
\toprule
Case & \(\mathrm{Re}_f^{\mathrm{std}}\) (\(y^* \leq 15.5\)) & \(\mathrm{Re}_f^{60D}\) (\(y^* \leq 15.5\)) & \(\mathrm{Re}_f^{60D}\) (full) & Channel & Horizon & Combined \\
\midrule
DF / BGK & \(336.30\) & \(341.00\) & \(349.02\) & \(+1.40\%\) & \(+2.35\%\) & \(+3.78\%\) \\
DF / TRT & \(336.54\) & \(341.24\) & \(349.28\) & \(+1.40\%\) & \(+2.36\%\) & \(+3.78\%\) \\
MDF / BGK & \(339.18\) & \(344.00\) & \(354.75\) & \(+1.42\%\) & \(+3.12\%\) & \(+4.59\%\) \\
MDF / TRT & \(339.66\) & \(344.18\) & \(354.99\) & \(+1.33\%\) & \(+3.14\%\) & \(+4.51\%\) \\
DFC / BGK & \(337.64\) & \(342.33\) & \(353.26\) & \(+1.39\%\) & \(+3.19\%\) & \(+4.63\%\) \\
DFC / TRT & \(337.66\) & \(342.35\) & \(353.24\) & \(+1.39\%\) & \(+3.18\%\) & \(+4.61\%\) \\
\bottomrule
\end{tabular}}
\ifdim\wd0>\textwidth\noindent\resizebox{\textwidth}{!}{\usebox0}\else\noindent\makebox[\textwidth][c]{\usebox0}\fi
\vspace{0.8em}

For the single-particle case (Table C.1), the body-local peak Reynolds number shifts by \(+3.78\%\) to \(+4.63\%\)
between the standard channel and the full \(60D\) horizon across all six configurations (max \(-\) min \(= 0.85\)
percentage points) --- well within the \(13.8\%\) light-particle Majumder-referenced deficit of Sec. 5.2 (ratio
\(\approx 3\times\)). The within-IBM BGK / TRT spread of the \(60D\) Reynolds number remains below \(0.08\%\) on each
IBM family (DF \(0.075\%\), MDF \(0.070\%\), DFC \(0.007\%\); computed from unrounded values), consistent with Sec. 6
(Table VI). The DF pair re-run on an \(80D\) tall channel (contact-safe horizon \(y^* \leq 71.5\), same \(|v_y^*|\)
observable definition) shifts the full-horizon peak Reynolds number by only about \(+0.18\%\) relative to \(60D\) (BGK
\(+0.179\%\), TRT \(+0.176\%\)), confirming a small additional \(60D\)-to-\(80D\) increment on the two tested DF
configurations. For the tested DF / BGK / Peskin 4-point / Velocity-Verlet single-particle configuration, the
Wang-normalized paired ablation spread (explicit-history vs without-correction, same-source denominator \(503.38\))
decreases from \(1.092\%\) on the matched \(y^* \leq 15.5\) window of the tall-\(60D\) channel to \(0.104\%\) on its
full contact-safe \(y^* \leq 51.5\) horizon; both values lie below the \(1.27\)--\(1.33\%\) cross-scheme spread range of
Sec. B.5, with the spread decreasing substantially over the longer window. The standard-channel analysis window used in
Sec. 5.1 is therefore body-locally adequate for the \emph{single-particle explicit-history peak-Re domain-shift}
analysis; the closure-sensitivity ablation contrast is a two-particle wake-interaction quantity addressed below (Table
C.3), while far-wake quantities (e.g., wake length, vortex shedding frequency at \(30D\) or \(60D\) downstream of the
particle) and three-dimensional wake-transition claims are \emph{outside} the scope of this analysis.

The reference wake-interaction runs bound the body-local peak adequacy. The two-particle wake-interaction case uses the
Majumder \(50D\) streamwise extent (the Majumder domain \(\Omega = [0,10] \times [-1,1]\) with \(d = 0.2\), the Sec.
2.9.4 configuration), with the light-particle initial position \(6D\) downstream of the gravity-direction inflow,
leaving \(44D\) of streamwise extent between the initial light-particle center and the outlet boundary. The
light-particle peak occurs \(26\)--\(27D\) downstream of its initial position (equivalent to \(y^* = 26\)--\(27\) on the
from-initial-displacement basis), inside the single-particle peak-location band \(y^* = 25\)--\(33\) observed in Table
C.1 over the tested configurations, with a streamwise center-to-outlet distance at peak of \(16\)--\(18D\). The Majumder
\(50D\) wake-interaction domain is therefore body-locally adequate for the \emph{explicit-history baseline case only}
(light-particle trail-peak shifts of \(-0.59\%\) to \(-1.02\%\) between \(50D\) and \(60D\) across the three schemes,
Table C.3); the \emph{without explicit correction} case continues to develop beyond \(50D\) (\(+5.12\%\) to \(+10.41\%\)
shifts, Table C.3), so its magnitude varies with the domain and is reported with the \(50D / 60D\) values side-by-side
(\(19.60\% / 7.45\%\) for DF, \(21.51\% / 10.52\%\) for MDF, \(12.30\% / 5.73\%\) for DFC; Table C.3), while the
qualitative direction is shared by the DF/MDF/DFC comparisons on both domains (Sec. 5.2.4). The direct extraction of the
body-local peak from the existing wake-interaction reference runs is given in Table C.2, and the same cases were re-run
per scheme on a \(60D\)-extended domain at matched grid resolution (\(N_y = 801\)), reported in Table C.3.

The Sec. 5.2 light-particle closure-sensitivity contrast (\(-16.4\%\) paired; \(13.8\%\) against the Majumder reference)
is measured on the reference DF / Peskin 4-point / Velocity-Verlet / explicit-history configuration; four
paired-ablation comparisons (DF / MDF / DFC, both Velocity-Verlet and Euler-explicit integrators, compressible and
incompressible LBGK) were performed under the same \(50D\) wake-interaction configuration. The heavy-particle ablation
spread (computed as \(|\mathrm{Re}_{\mathrm{eh}} - \mathrm{Re}_{\mathrm{none}}|/\mathrm{mean}\) on each scheme) is
uniformly \(\leq 3.0\%\) --- DF \(2.55\%\) (\(277.60 \to 270.62\)), MDF \(2.99\%\) (\(282.52 \to 274.21\)), DFC
\(2.87\%\) (\(279.12 \to 271.23\)), and Majumder-aligned DF \(2.32\%\) (\(274.50 \to 268.21\)) --- small relative to the
\(-16.4\%\) paired light-particle contrast. The light-particle without-correction deficit (relative to the
explicit-history baseline, \((\mathrm{Re}_{\mathrm{none}} - \mathrm{Re}_{\mathrm{eh}})/\mathrm{Re}_{\mathrm{eh}}\))
falls in a single \(-10\%\) to \(-18\%\) band --- DFC \(-10.95\%\) (\(234.91 \to 209.19\)), MDF \(-17.71\%\)
(\(237.95 \to 195.82\)), Majumder-aligned DF \(-13.43\%\) (\(230.14 \to 199.22\)), DF \(-16.39\%\)
(\(238.62 \to 199.52\)) --- cleanly separated from the heavy \(\leq 3.0\%\) spread. The wake-exposed response therefore
appears across the DF / MDF / DFC schemes and both integrators; the case-specific variability (\(-10.95\%\) to
\(-17.71\%\)) may reflect scheme- or integrator-dependent timing or numerical details, which the present comparisons do
not separate; across all four schemes the contrast retains its sign and magnitude band, so the contrast separates the
light-particle wake-exposed response from a boundary-enforcement scheme artifact in the present dataset.

\noindent \textbf{Table C.2.} Two-particle wake-interaction body-local peak position and streamwise center-to-outlet distance
(measured from the particle center; the particle-surface clearance is \(0.5D\) smaller), Majumder configuration (xmax
\(= 50D\), \(N_y = 801\), gravity-direction \(+x\)). Peak Re is on the standard fluid-density basis
(\(\mathrm{Re}_f = U D / \nu\)). The minimum center-to-center distance over each trajectory corresponds to a surface gap
of \(g_{\min}/D = 0.52\)--\(0.56\) across all eighteen recorded runs (method \(\times\) collision \(\times\)
correction), no geometric contact is observed at any stored sample, consistent with the wake-interaction design in the
sense of Uhlmann {[}14{]} (Sec. 5.2.2 therein).

\needspace{4\baselineskip}
\vspace{0.8em}
\sbox0{\scriptsize\setlength{\tabcolsep}{3pt}\renewcommand{\arraystretch}{1.2}%
\begin{tabular}{>{\raggedright\arraybackslash}p{0.206\linewidth} >{\raggedright\arraybackslash}p{0.166\linewidth} >{\raggedleft\arraybackslash}p{0.073\linewidth} >{\raggedleft\arraybackslash}p{0.060\linewidth} >{\raggedleft\arraybackslash}p{0.199\linewidth} >{\raggedleft\arraybackslash}p{0.246\linewidth}}
\toprule
Case & Particle & Peak \(\mathrm{Re}_f\) & Peak step & Peak distance from initial {[}D{]} & Center-to-outlet distance at peak {[}D{]} \\
\midrule
explicit-history (B-2) & Heavy (\(\rho_r = 1.5\)) & \(277.60\) & \(8500\) & \(8.4\) & \(37.6\) \\
explicit-history (B-2) & Light (\(\rho_r = 1.25\)) & \(238.62\) & \(35000\) & \(27.1\) & \(16.9\) \\
without explicit correction (A) & Heavy (\(\rho_r = 1.5\)) & \(270.62\) & \(9500\) & \(8.8\) & \(37.2\) \\
without explicit correction (A) & Light (\(\rho_r = 1.25\)) & \(199.52\) & \(36500\) & \(26.0\) & \(18.0\) \\
\bottomrule
\end{tabular}}
\ifdim\wd0>\textwidth\noindent\resizebox{\textwidth}{!}{\usebox0}\else\noindent\makebox[\textwidth][c]{\usebox0}\fi
\vspace{0.8em}

A \(60D\)-extended domain check completes the wake-interaction domain analysis. The three wake-interaction scheme pairs
of Sec. 5.2 (DF / MDF / DFC, each Peskin 4-point / Velocity-Verlet / BGK \(\times\) explicit-history /
without-correction) were re-run on a \(60D\)-extended streamwise domain at matched grid resolution (\(N_y = 801\),
gravity-direction \(+x\), \(\rho_s/\rho_f \in \{1.5, 1.25\}\)), each run terminating when the leading particle reaches
the \(60D\) streamwise offset stop (steps \(45\,000\)--\(47\,000\) across the six runs). For the explicit-history
baselines the trail-peak Reynolds number shifts only marginally between \(50D\) and \(60D\) (\(-0.80\%\) DF, \(-0.59\%\)
MDF, \(-1.02\%\) DFC), supporting that the Majumder \(50D\) domain is body-locally adequate for the \emph{baseline}
case, whereas the without-correction trail peaks shift substantially (\(+10.41\%\) DF, \(+9.30\%\) MDF, \(+5.12\%\)
DFC), each of those trajectories continuing to increase until near the end of its \(60D\) run. The wake-exposed
trail-peak ablation contrast is accordingly \(19.60\%\) / \(21.51\%\) / \(12.30\%\) (DF / MDF / DFC) on the \(50D\)
Majumder domain (intra-domain measurements with each scheme's without-correction case as denominator) and \(7.45\%\) /
\(10.52\%\) / \(5.73\%\) on the \(60D\)-extended domain (each on that domain's own without-correction denominator); the
\(13.8\%\) value of Sec. 5.2 uses the Majumder reference \(231.41\) as denominator on the same \(50D\) measurement
window. The \(60D\) without-correction values are end-of-window estimates (each trail peak falls at or within one stored
sample of the run end); the qualitative direction is preserved on every scheme, bracketed against the single-particle
\(+3.78\%\)--\(+4.63\%\) band and the heavy-particle ablation spread \(\leq 3.0\%\).

\noindent \textbf{Table C.3.} Two-particle wake-interaction body-local peak Reynolds: \(50D\) Majumder vs \(60D\)-extended domain
check across the three boundary-enforcement schemes. Peak Re is on the standard fluid-density basis
(\(\mathrm{Re}_f = U D / \nu\)); the light particle (\(\rho_r = 1.25\)) is tabulated. Schemes B-2 and A in the
Suzuki--Inamuro taxonomy denote the explicit-history baseline and the without-explicit-correction ablation,
respectively. The heavy-particle peaks are identical at the reported precision between the two domains on every scheme
(DF \(277.60\) / \(270.62\), MDF \(282.52\) / \(274.21\), DFC \(279.12\) / \(271.23\) for the two arms), occurring at
steps \(8\,500\)--\(9\,500\), substantially earlier than the light-particle trail peaks. The \(60D\) without-correction
values are evaluated at the end of the recorded window (each trail peak falls at or within one stored sample of the run
end).

\needspace{4\baselineskip}
\vspace{0.8em}
\sbox0{\scriptsize\setlength{\tabcolsep}{3pt}\renewcommand{\arraystretch}{1.2}%
\begin{tabular}{l l r r r}
\toprule
Scheme & Case (light particle) & \(50D\) Peak \(\mathrm{Re}_f\) & \(60D\) Peak \(\mathrm{Re}_f\) & \(\Delta\%\) \\
\midrule
DF & explicit-history (B-2) & \(238.62\) & \(236.72\) & \(-0.80\%\) \\
DF & without explicit correction (A) & \(199.52\) & \(220.30\) & \(+10.41\%\) \\
MDF & explicit-history (B-2) & \(237.95\) & \(236.55\) & \(-0.59\%\) \\
MDF & without explicit correction (A) & \(195.82\) & \(214.03\) & \(+9.30\%\) \\
DFC & explicit-history (B-2) & \(234.91\) & \(232.51\) & \(-1.02\%\) \\
DFC & without explicit correction (A) & \(209.19\) & \(219.91\) & \(+5.12\%\) \\
DF & wake-exposed gap (eh \(-\) none) & \(+19.60\%\) & \(+7.45\%\) & --- \\
MDF & wake-exposed gap (eh \(-\) none) & \(+21.51\%\) & \(+10.52\%\) & --- \\
DFC & wake-exposed gap (eh \(-\) none) & \(+12.30\%\) & \(+5.73\%\) & --- \\
\bottomrule
\end{tabular}}
\ifdim\wd0>\textwidth\noindent\resizebox{\textwidth}{!}{\usebox0}\else\noindent\makebox[\textwidth][c]{\usebox0}\fi
\vspace{0.8em}

\subsubsection{C.3 Fixed-Cylinder Grid Sensitivity}\label{c.3-fixed-cylinder-grid-sensitivity}

A grid sensitivity analysis at \(\mathrm{Re} = 40\) comparing the baseline resolution (\(N_y = 1601\),
\(D = 40\Delta x\)) against a coarser grid (\(N_y = 801\), \(D = 20\Delta x\)) yields a relative \(C_d\) difference that
does not exceed \(1.7\%\) across the five tabulated method--delta combinations, consistent with the inter-method
differences in Sec. 3.1 Table II persisting across the two tested resolutions and not reversed by grid refinement at
this level. The full \(\mathrm{Re} = 40\) table is reported in Supplementary Material Sec.~S2.2 Table S5; the
corresponding complementary DF + Peskin 4-point grid sweeps at \(\mathrm{Re} = 100\)
(\(N_y \in \{641, 961, 1281, 1601\}\)) and \(\mathrm{Re} = 200\) (\(N_y \in \{961, 1281, 1601, 1921\}\), complementary
to the \(2401\) production grid of main Table I) are \emph{available in the public repository} and not tabulated in
Table S5.

\subsubsection{C.4 Oscillating-Cylinder Grid}\label{c.4-oscillating-cylinder-grid}

For the oscillating-cylinder benchmark at \(\mathrm{KC} = 5\), \(\mathrm{Re} = 100\), a DF / BGK / Peskin 4-point grid
series at \(N_y \in \{801, 1281, 1601\}\) (\(D = 40, 64, 80\Delta x\); \(N_y = 1601\) is the \(2401 \times 1601\)
production grid of Table~I) yields converged-cycle peak drag coefficients of \(2.140\), \(2.093\), and \(2.081\): the
peak drag changes by \(2.9\%\) from the coarsest grid to the production grid and by \(0.59\%\) between the two finest
grids (relative changes computed from the unrounded peak values with the production-grid value as denominator). The
per-grid outputs are reported in Supplementary Material Sec.~S2.3 and are available in the public repository; no
oscillating-cylinder grid finer than the production grid was run, and no cross-method oscillating-cylinder grid
comparison is made.

\subsubsection{C.5 Sedimentation Grid}\label{c.5-sedimentation-grid}

The sedimentation grid robustness on the body-local peak-Reynolds metric is supported by three complementary checks.

For the Wang single-particle heavy case (\(\rho_s/\rho_f = 1.5\)), a grid cross-check between the production resolution
(\(D = 160\Delta x\)) and a \(1.5\times\) refined grid (\(D = 240\Delta x\); Euler-integrator pair) gives a peak
\(\mathrm{Re}_f\) shift of \(+0.109\%\) (\(336.30\) vs \(336.67\)); the integrator contrast on the production grid
(Verlet vs Euler) is \(0.0002\%\) on the same metric. The single-particle and wake-interaction cases share the same
boundary-enforcement scheme (DF / Peskin 4-point), the same gravitational driver, and the same body-local peak-Reynolds
metric, so the single-particle grid check provides a supporting consistency check for the use of the same body-local
peak-Reynolds metric in the wake-interaction case on the same boundary-enforcement scheme; the wake-interaction grid
sensitivity itself is assessed by the coupled space--time refinement series reported below.

The wake-interaction body-local peak position is also internally consistent: the wake-interaction light-particle
\(\mathrm{Re}_{f,\max}\) peak is reached at \(26\)--\(27D\) downstream of the light-particle initial position (Table
C.2), which sits \emph{inside} the body-local peak-location band \(25\)--\(33D\) observed in the single-particle \(60D\)
extended-channel analysis across six tested configurations (Sec.~C.2, \(+3.78\%\)--\(+4.63\%\) shift band). This
body-local agreement supports the \emph{explicit-history} wake-interaction baseline; the
\emph{without-explicit-correction} magnitude remains domain- and window-sensitive, as summarized in Table C.3 and the
direct re-run described next.

A direct \(60D\)-extended wake-interaction re-run on the three scheme pairs (DF / MDF / DFC, Peskin 4-point /
Velocity-Verlet × explicit-history / without explicit correction) at matched grid resolution (\(N_y = 801\)) is reported
as the extended-domain check of Sec. C.2 (Table C.3); it is consistent with near-saturation of the explicit-history
body-local peak over the tested \(50D\)/\(60D\) windows (\(-0.59\%\) to \(-1.02\%\) between \(50D\) and \(60D\)) whereas
the without-explicit-correction case continues to develop over the extended window (\(+5.12\%\) to \(+10.41\%\), the
later post-overtaking phase). The contrast attenuation across the two domains is quantified in Table C.3.

A harmonized-window coupled space--time refinement series on the wake-interaction reference pair is reported at
\(N_y \in \{801, 961, 1281\}\) (diffusive scaling, \(\Delta t \propto \Delta x^2\)); the \(N_y = 801\) baseline derives
from an earlier code version and is partially bridged to the two current-version refinements. Across \(N_y = 801\),
\(961\), and \(1281\), the recorded signed light-particle contrast on the \(y^* \leq 15.5\) window remains negative
(\(-14.28\%\), \(-15.32\%\), and \(-15.85\%\)), while the successive changes contract from \(1.04\) to \(0.53\)
percentage points; the contrast is therefore stable at the finite-window level under the tested coupled space--time
refinement. Sec. C.2 reports the \(13.8\%\) Majumder-referenced deficit and the \(7.45\%\) \(60D\) endpoint side-by-side
(the per-window denominator conventions are detailed in Sec. 6.2). Further details, including the per-grid values, are
reported in Supplementary Material Sec. S2.4.

\subsubsection{C.6 Marker Spacing and Kernel Support}\label{c.6-marker-spacing-and-kernel-support}

The detailed \(W/\delta\) correlation tabulation supporting Sec.~4.3 (Spearman \(\rho = 0.90\), Kendall \(\tau = 0.80\)
across five datapoints; two-cluster ordering: hat \(W/\delta \in [0.06, 0.10]\) with max slip \(\sim 1.2\)--\(1.6\%\)
\(U_\infty\); P4 \(W/\delta \in [0.19, 0.20]\) with max slip \(\sim 4.7\%\) \(U_\infty\)) is reported in Supplementary
Material Sec.~S2.5. The kernel-support to boundary-layer-thickness scale provides an ordering rationale for the DFC slip
across the tested kernels.

\subsubsection{C.7 Wake-Interaction Particle-Bound Acceleration Peak
Ratios}\label{c.7-wake-interaction-particle-bound-acceleration-peak-ratios}

\noindent \textbf{Table C.4.} Particle-bound acceleration peak ratios in the wake-interaction configuration (cf.~Sec. 5.2.3), on
the reference Velocity-Verlet / explicit-history sampling. Phase boundaries are defined from the inter-particle distance
minimum of each configuration; the per-configuration closest-approach / role-exchange times are \(t^* = 5.512/5.913\)
(Verlet, explicit-history), \(6.010/6.518\) (Verlet, without correction), \(5.560/5.975\) (Euler, 500-step storage
cadence), and \(5.576/5.980\) (Euler, 100-step storage cadence) --- the without-correction arm lags by
\(\approx 0.5\)--\(0.6\) in \(t^*\). The distance minimum is bracketed by the 500-step samples at \(5.22\)--\(5.80\),
and the reported ratios are unchanged for any phase-boundary choice within this bracket. The ratios are formed between
the two density-labeled particles (light over heavy) so that they are unaffected by the upstream/downstream role
exchange; the pre-overtaking maximum is the maximum over the earliest stored records (no step-0 record exists), and the
post-overtaking maxima occur well after the role exchange. Because \(|a^*|\) is normalized by each particle's own
gravitational acceleration scale \(u_g^2/D\), the light/heavy ratios carry a factor-of-two normalization difference
(\(u_{g,\mathrm{heavy}}^2/u_{g,\mathrm{light}}^2 = 2\)) and are not raw acceleration ratios; only the phase-dependent
contrast \(2.54\times\) (\(= 2.81/1.11\)) is normalization-invariant (cf.~Sec. 5.2.3). The Majumder-aligned
incompressible-LBGK / Euler cross-check reproduces the same pattern (\(1.11\times\) / \(2.89\times\), contrast
\(2.60\times\)); across the integrator / formulation / storage-cadence sweep the contrast stays within
\(2.41\)--\(2.60\), a storage-cadence sensitivity band of this auxiliary kinematic diagnostic.

\needspace{4\baselineskip}
\vspace{0.8em}
\sbox0{\scriptsize\setlength{\tabcolsep}{3pt}\renewcommand{\arraystretch}{1.2}%
\begin{tabular}{l l l r}
\toprule
Phase & Comparator & Reference & Ratio \\
\midrule
Pre-overtaking & light / heavy & wake-interaction pair & \(1.11\times\) \\
Post-overtaking & light / heavy & wake-interaction pair & \(2.81\times\) \\
Pre-overtaking → Post-overtaking phase-dependent contrast (light/heavy) & --- & --- & \(2.54\times\) \\
\bottomrule
\end{tabular}}
\ifdim\wd0>\textwidth\noindent\resizebox{\textwidth}{!}{\usebox0}\else\noindent\makebox[\textwidth][c]{\usebox0}\fi
\vspace{0.8em}

\subsection{Appendix D --- Convergence and Wake Verification}\label{appendix-d-convergence-and-wake-verification}

\subsubsection{D.1 Taylor--Green Spatial Convergence}\label{d.1-taylorgreen-spatial-convergence}

The Taylor--Green vortex {[}53{]} provides an analytical solution for spatial convergence-order verification. The
analytical solution is

\[u_x = -u_0 \cos(k x) \sin(k y) e^{-2 \nu k^2 t}, \qquad u_y = +u_0 \sin(k x) \cos(k y) e^{-2 \nu k^2 t}, \tag{D1}\]

\noindent with \(k = \pi / L\), \(u_0 = 0.04\), \(\mathrm{Re} = 10\), and measurement at \(t = L / u_0\) (one convective time).
Grid resolutions \(D \in \{10, 20, 40, 80\}\) cells are tested. Two modes are compared: pure LBM (without IBM) and
IB-LBM with IBM markers whose desired velocity is set to the analytical solution at each time step. The spatial
convergence order is evaluated from the \(L_2\) relative error slope in the \(\log(\Delta x)\)--\(\log(e)\) plane.

\noindent \textbf{Table D.1.} Taylor--Green spatial convergence order (Peskin 4-point delta function).

\needspace{4\baselineskip}
\vspace{0.8em}
\sbox0{\scriptsize\setlength{\tabcolsep}{3pt}\renewcommand{\arraystretch}{1.2}%
\begin{tabular}{r r r r r}
\toprule
\(D\) & Pure LBM \(L_2\) & DF \(L_2\) & MDF \(L_2\) & DFC \(L_2\) \\
\midrule
10 & \(1.634\times 10^{-2}\) & \(4.282\times 10^{-2}\) & \(4.348\times 10^{-2}\) & \(4.345\times 10^{-2}\) \\
20 & \(4.109\times 10^{-3}\) & \(1.069\times 10^{-2}\) & \(1.077\times 10^{-2}\) & \(1.093\times 10^{-2}\) \\
40 & \(9.173\times 10^{-4}\) & \(2.674\times 10^{-3}\) & \(2.696\times 10^{-3}\) & \(2.616\times 10^{-3}\) \\
80 & \(3.188\times 10^{-4}\) & \(6.872\times 10^{-4}\) & \(7.453\times 10^{-4}\) & \(4.795\times 10^{-4}\) \\
Order & 1.920 & 1.988 & 1.960 & 2.157 \\
\bottomrule
\end{tabular}}
\ifdim\wd0>\textwidth\noindent\resizebox{\textwidth}{!}{\usebox0}\else\noindent\makebox[\textwidth][c]{\usebox0}\fi
\vspace{0.8em}

DFC exhibits the highest convergence order (2.157) among the three IB-LBM methods, exceeding the theoretical
second-order rate (Fig. D.1). At the finest resolution (\(D = 80\)), DFC's error (\(4.795 \times 10^{-4}\)) is \(30\%\)
lower than DF and \(36\%\) lower than MDF within the IB-forced family on the Peskin 4-point kernel; the pure-LBM
baseline (without IBM coupling) reaches \(3.188 \times 10^{-4}\), the lowest absolute \(L_2\) within the Peskin 4-point
comparison context. Under the \emph{hat} kernel, hat-DFC at \(D = 80\) attains an absolute \(L_2\) of
\(1.746 \times 10^{-4}\) (Table D.2), lower than the Peskin 4-point pure-LBM baseline of \(3.188 \times 10^{-4}\) on the
same Taylor--Green configuration. This kernel-cross comparison is observed within the present analytical-solution sweep;
the convergence-order ordering (DFC \textgreater{} DF \textgreater{} MDF) is preserved within each kernel. The
super-quadratic DFC rate is driven primarily by the \(D = 40 \to 80\) pair (slope 2.448) and may not persist at finer
resolutions. The pure LBM baseline achieves an order of \(1.920\), confirming that the LBM solver itself attains the
expected second-order accuracy {[}9{]}. All three IB-LBM methods maintain or exceed this convergence-order baseline
under the Peskin 4-point kernel, indicating that the IBM coupling does not degrade the spatial convergence rate; the
absolute \(L_2\) relationship to the pure-LBM reference is kernel-dependent (Peskin 4-point IB-LBM lies above pure LBM,
hat IB-LBM at the finest resolution lies below for DF and DFC).

\begin{figure}
\centering
\includegraphics{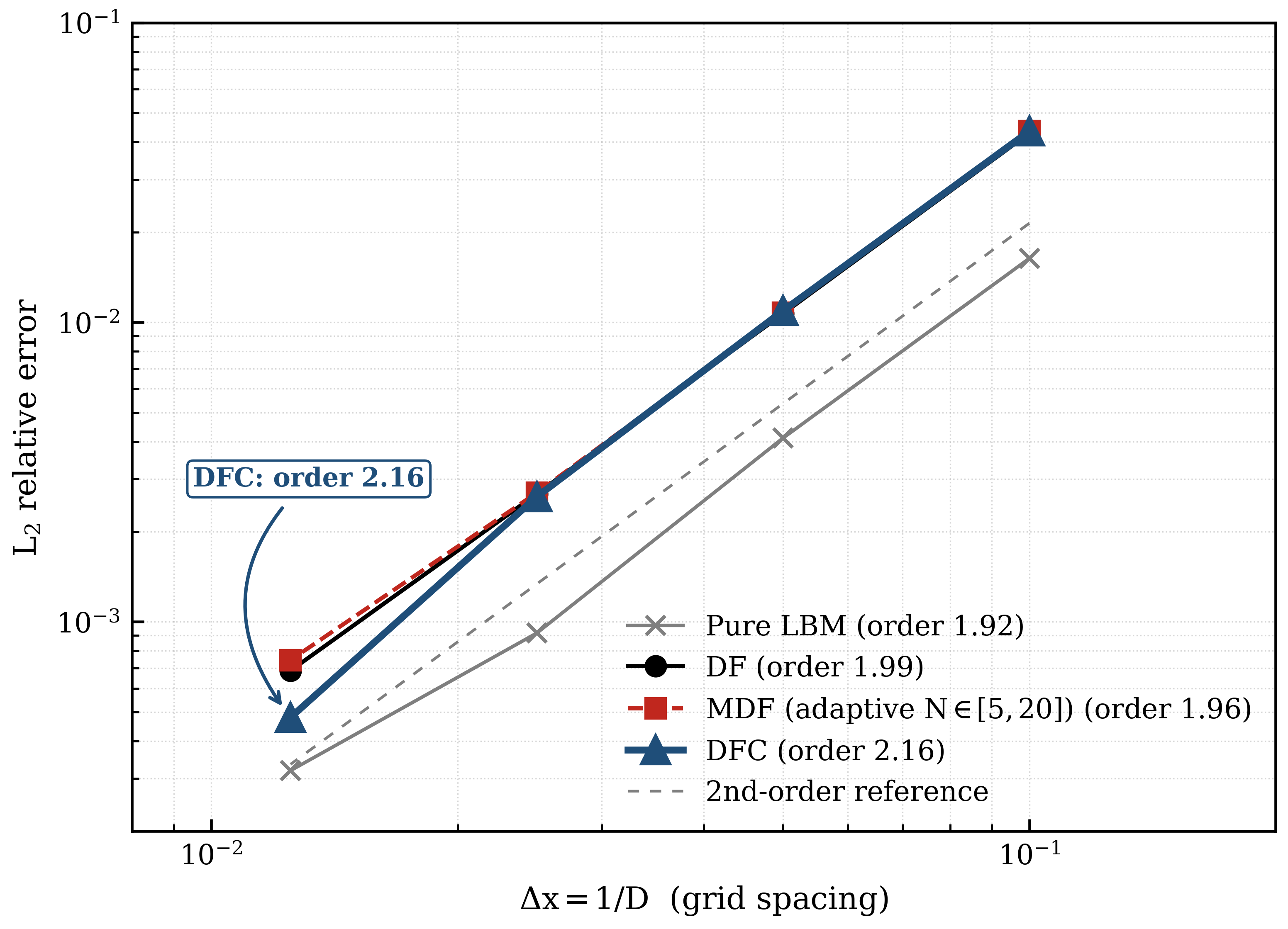}
\caption{\textbf{Fig. D.1.} Taylor--Green spatial convergence (Peskin 4-point delta function). \(L_2\) relative error
against \(\Delta x = L / D\) (with \(L = 1\) throughout this Taylor--Green configuration, so the abscissa label
\(\Delta x = 1/D\) on the figure is equivalent) for \(D \in \{10, 20, 40, 80\}\) on a \(\log\)--\(\log\) plot. Solid
lines show pure LBM, DF, MDF (adaptive \(N \in [5,20]\)), and DFC; the dashed line marks the theoretical second-order
slope. DFC attains slope \(2.16\) and \emph{the lowest \(L_2\) among the three IB-LBM methods} at \(D = 80\)
(\(4.795 \times 10^{-4}\)). The pure-LBM baseline (without IBM coupling) reaches \(3.188 \times 10^{-4}\) at \(D = 80\),
the lowest absolute \(L_2\) within the Peskin 4-point comparison shown here; all three IB-LBM methods (Peskin 4-point)
preserve or exceed its \(1.920\) convergence-order rate.}
\end{figure}

The Taylor--Green convergence analysis displayed here uses the Peskin 4-point delta function. A companion hat-kernel
Taylor--Green sweep on the same four resolutions \(D \in \{10, 20, 40, 80\}\) and three IBM schemes (DF / MDF / DFC) is
reported in Table D.2. The hat-kernel \(L_2\) values are \emph{smaller} in absolute magnitude than the Peskin 4-point
values across all resolutions (the wider Peskin support smears the boundary over more grid points and produces a larger
absolute boundary-coupling error), and the convergence-order ordering is preserved: DFC attains the highest order on
both kernels (Peskin 4-point \(2.157\) vs hat \(2.137\)), DF is intermediate (\(1.988\) vs \(1.876\)), and MDF is the
lowest (\(1.960\) vs \(1.800\)). The hat-kernel order estimates are slightly lower than the Peskin 4-point counterparts,
consistent with the hat kernel's narrower support producing stronger marker-to-marker variability of the spreading sum,
manifesting in the convergence rate as a small order reduction despite the lower absolute error. The hat-kernel results
are consistent with the Sec.~4 mechanism interpretation, which separates the \emph{uniformity} axis (the kernel-reversal
diagnostic) from the \emph{absolute magnitude} axis.

\noindent \textbf{Table D.2.} Taylor--Green spatial convergence with the \emph{hat} delta function (companion to Table D.1 / Fig.
D.1, which reports the Peskin 4-point counterpart). \(L_2\) relative error and convergence-order estimate for the three
IBM schemes (DF / MDF / DFC) at \(D \in \{10, 20, 40, 80\}\), computed on the same TG configuration as Table D.1
(\(k = \pi/L\), \(u_0 = 0.04\), \(\mathrm{Re} = 10\), measurement at \(t = L/u_0\)). \(L_2\) values are computed from
the per-case velocity fields, with the resulting convergence arrays available in the public repository (see Data
Availability Statement).

\needspace{4\baselineskip}
\vspace{0.8em}
\sbox0{\scriptsize\setlength{\tabcolsep}{3pt}\renewcommand{\arraystretch}{1.2}%
\begin{tabular}{c r r r}
\toprule
\(D\) & DF \(L_2\) & MDF \(L_2\) & DFC \(L_2\) \\
\midrule
\(10\) & \(1.385 \times 10^{-2}\) & \(1.592 \times 10^{-2}\) & \(1.466 \times 10^{-2}\) \\
\(20\) & \(3.358 \times 10^{-3}\) & \(3.862 \times 10^{-3}\) & \(3.654 \times 10^{-3}\) \\
\(40\) & \(8.568 \times 10^{-4}\) & \(9.677 \times 10^{-4}\) & \(8.010 \times 10^{-4}\) \\
\(80\) & \(2.860 \times 10^{-4}\) & \(3.947 \times 10^{-4}\) & \(1.746 \times 10^{-4}\) \\
Order & \(1.876\) & \(1.800\) & \(2.137\) \\
\bottomrule
\end{tabular}}
\ifdim\wd0>\textwidth\noindent\resizebox{\textwidth}{!}{\usebox0}\else\noindent\makebox[\textwidth][c]{\usebox0}\fi
\vspace{0.8em}

DFC retains the highest convergence order under the hat kernel (\(2.137\), super-quadratic) and attains the lowest
\(L_2\) among the three IB-LBM methods at \(D = 80\) (\(1.746 \times 10^{-4}\)). The cross-kernel ordering (DFC
\textgreater{} DF \textgreater{} MDF for convergence order; DFC lowest absolute \(L_2\) at \(D = 80\)) is preserved
between Peskin 4-point (Table D.1) and hat (Table D.2), consistent with the Sec.~4 kernel-redistribution mechanism
preserving the convergence-rate ordering even though the absolute boundary-layer error magnitudes differ.

\subsubsection{D.2 Wake Vorticity Comparison}\label{d.2-wake-vorticity-comparison}

A field-level wake-vorticity cross-check (DF / MDF / DFC × Peskin 4-point at \(\mathrm{Re} \in \{100, 200\}\)) is
consistent with the body-local fidelity differences quantified in Sec. 3.1 (Table II) not translating into qualitative
wake-level differences. The full cross-check is reported in Supplementary Material Sec.~S3 Fig. S1. The qualitative wake
structure --- shoulder vortex extent, flow separation, and near-wake symmetry-breaking at \(\mathrm{Re} = 200\) --- is
visually consistent across DF, MDF, and DFC, in agreement with the quantitative \(C_d\) / slip / internal-residual
tabulations in Sec. 3, which indicate that the prescribed-body fidelity differences are local-to-the-boundary phenomena
rather than wake-level discrepancies.

\subsection{Appendix E --- Raw Sedimentation Trajectories and Time
Series}\label{appendix-e-raw-sedimentation-trajectories-and-time-series}

The raw single-particle and two-particle wake-interaction trajectories, velocity histories, and particle-bound
acceleration time series that underlie the Sec.~5 single-particle baseline (Sec.~5.1 Table IV) and the Sec.~5.2
wake-exposed closure sensitivity (Table V internal-mass correction comparison) are summarized in Supplementary Material
Sec.~S4 (subsections Sec.~S4.1--Sec.~S4.4; the phase-resolved paired-difference decomposition is reported in Sec.~S4.5);
the full per-time-series traces (\(y^*(t^*)\), \(v_y^*(t^*)\), wake-interaction phase-plane, and \(|a^*(t)|\)) are
available in the public repository (see Data Availability Statement). The transient-regime summary (single-particle peak
overshoot \(\sim 0.36\%\) at \(\rho_s/\rho_f = 1.1\) and \(\sim 0.58\%\) at \(\rho_s/\rho_f = 1.5\),
\(t^{*}_{99} \sim 10.6\)--\(11.7\)) is tabulated directly in Supplementary Material Sec.~S4.2 Table S10.

\subsection{References}\label{references}

{[}1{]} C. S. Peskin, ``Flow patterns around heart valves: a numerical method,'' \emph{Journal of Computational
Physics}, vol.~10, no. 2, pp.~252--271, 1972.

{[}2{]} R. Mittal and G. Iaccarino, ``Immersed boundary methods,'' \emph{Annual Review of Fluid Mechanics}, vol.~37,
pp.~239--261, 2005.

{[}3{]} C. S. Peskin, ``The immersed boundary method,'' \emph{Acta Numerica}, vol.~11, pp.~479--517, 2002.

{[}4{]} F. Sotiropoulos and X. Yang, ``Immersed boundary methods for simulating fluid--structure interaction,''
\emph{Progress in Aerospace Sciences}, vol.~65, pp.~1--21, 2014.

{[}5{]} W.-X. Huang and F.-B. Tian, ``Recent trends and progress in the immersed boundary method,'' \emph{Proceedings of
the Institution of Mechanical Engineers, Part C: Journal of Mechanical Engineering Science}, vol.~233, no. 23--24,
pp.~7617--7636, 2019.

{[}6{]} R. Verzicco, ``Immersed boundary methods: historical perspective and future outlook,'' \emph{Annual Review of
Fluid Mechanics}, vol.~55, pp.~129--155, 2023.

{[}7{]} X. D. Niu, C. Shu, Y. T. Chew, and Y. Peng, ``A momentum exchange-based immersed boundary-lattice Boltzmann
method for simulating incompressible viscous flows,'' \emph{Physics Letters A}, vol.~354, no. 3, pp.~173--182, 2006.

{[}8{]} C. K. Aidun and J. R. Clausen, ``Lattice-Boltzmann method for complex flows,'' \emph{Annual Review of Fluid
Mechanics}, vol.~42, pp.~439--472, 2010.

{[}9{]} S. Chen and G. D. Doolen, ``Lattice Boltzmann method for fluid flows,'' \emph{Annual Review of Fluid Mechanics},
vol.~30, pp.~329--364, 1998.

{[}10{]} Z.-G. Feng and E. E. Michaelides, ``The immersed boundary-lattice Boltzmann method for solving fluid--particles
interaction problems,'' \emph{Journal of Computational Physics}, vol.~195, no. 2, pp.~602--628, 2004.

{[}11{]} Z.-G. Feng and E. E. Michaelides, ``Proteus: a direct forcing method in the simulations of particulate flows,''
\emph{Journal of Computational Physics}, vol.~202, no. 1, pp.~20--51, 2005.

{[}12{]} M.-C. Lai and C. S. Peskin, ``An immersed boundary method with formal second-order accuracy and reduced
numerical viscosity,'' \emph{Journal of Computational Physics}, vol.~160, no. 2, pp.~705--719, 2000.

{[}13{]} D. Goldstein, R. Handler, and L. Sirovich, ``Modeling a no-slip flow boundary with an external force field,''
\emph{Journal of Computational Physics}, vol.~105, no. 2, pp.~354--366, 1993.

{[}14{]} M. Uhlmann, ``An immersed boundary method with direct forcing for the simulation of particulate flows,''
\emph{Journal of Computational Physics}, vol.~209, no. 2, pp.~448--476, 2005.

{[}15{]} K. Luo, Z. Wang, J. Fan, and K. Cen, ``Full-scale solutions to particle-laden flows: Multidirect forcing and
immersed boundary method,'' \emph{Physical Review E}, vol.~76, no. 6, art. 066709, 2007.

{[}16{]} Z. Wang, J. Fan, and K. Luo, ``Combined multi-direct forcing and immersed boundary method for simulating flows
with moving particles,'' \emph{International Journal of Multiphase Flow}, vol.~34, no. 3, pp.~283--302, 2008.

{[}17{]} W.-P. Breugem, ``A second-order accurate immersed boundary method for fully resolved simulations of
particle-laden flows,'' \emph{Journal of Computational Physics}, vol.~231, no. 13, pp.~4469--4498, 2012.

{[}18{]} S. Tao, Q. He, J. Chen, B. Chen, G. Yang, and Z. Wu, ``A non-iterative immersed boundary-lattice Boltzmann
method with boundary condition enforced for fluid--solid flows,'' \emph{Applied Mathematical Modelling}, vol.~76,
pp.~362--379, 2019.

{[}19{]} D. R. Noble and J. R. Torczynski, ``A lattice-Boltzmann method for partially saturated computational cells,''
\emph{International Journal of Modern Physics C}, vol.~9, no. 8, pp.~1189--1201, 1998.

{[}20{]} S. K. Kang and Y. A. Hassan, ``A comparative study of direct-forcing immersed boundary-lattice Boltzmann
methods for stationary complex boundaries,'' \emph{International Journal for Numerical Methods in Fluids}, vol.~66, no.
9, pp.~1132--1158, 2011.

{[}21{]} S. Tao, Q. He, B. Chen, and F. G. F. Qin, ``Distribution function correction-based immersed boundary lattice
Boltzmann method for thermal particle flows,'' \emph{Computational Particle Mechanics}, vol.~8, pp.~459--469, 2021.

{[}22{]} S. Tao, L. Wang, Q. He, J. Chen, and J. Luo, ``Lattice Boltzmann simulation of complex thermal flows via a
simplified immersed boundary method,'' \emph{Journal of Computational Science}, vol.~65, art. 101878, 2022.

{[}23{]} S. Majumder, A. Ghosh, D. N. Basu, and G. Natarajan, ``Computational assessment of immersed boundary--lattice
Boltzmann method for complex moving boundary problems,'' \emph{Computational Particle Mechanics}, vol.~10, pp.~155--172,
2023.

{[}24{]} L. Wang, Z. Liu, and M. Rajamuni, ``Recent progress of lattice Boltzmann method and its applications in
fluid--structure interaction,'' \emph{Proceedings of the Institution of Mechanical Engineers, Part C: Journal of
Mechanical Engineering Science}, vol.~237, no. 11, pp.~2461--2484, 2023.

{[}25{]} K. Suzuki, E. Falagkaris, T. Krüger, and T. Inamuro, ``Boundary-velocity error and stability of the accelerated
multi-direct-forcing immersed boundary method,'' \emph{arXiv preprint}, arXiv:2507.04986, 2025,
doi:10.48550/arXiv.2507.04986.

{[}26{]} Z. Cheng and A. Wachs, ``An immersed boundary/multi-relaxation time lattice Boltzmann method on adaptive octree
grids for the particle-resolved simulation of particle-laden flows,'' \emph{Journal of Computational Physics}, vol.~471,
art. 111669, 2022.

{[}27{]} M. A. Ferrari, R. de Souza, and A. T. Franco, ``Performance assessment of the immersed boundary method in the
framework of the moment representation lattice Boltzmann method,'' \emph{International Journal for Numerical Methods in
Fluids}, vol.~98, no. 7, pp.~857--870, 2026, doi:10.1002/fld.70071.

{[}28{]} C. Coreixas, G. Wissocq, B. Chopard, and J. Latt, ``Impact of collision models on the physical properties and
the stability of lattice Boltzmann methods,'' \emph{Philosophical Transactions of the Royal Society A}, vol.~378, no.
2175, art. 20190397, 2020.

{[}29{]} I. Cheylan, T. Fringand, J. Jacob, and J. Favier, ``Analysis of the immersed boundary method for turbulent
fluid--structure interaction with lattice Boltzmann method,'' \emph{Journal of Computational Physics}, vol.~492, art.
112418, 2023.

{[}30{]} S. Gsell and J. Favier, ``Direct-forcing immersed-boundary method: A simple correction preventing boundary slip
error,'' \emph{Journal of Computational Physics}, vol.~435, art. 110265, 2021.

{[}31{]} I. Ginzburg, F. Verhaeghe, and D. d'Humières, ``Two-relaxation-time lattice Boltzmann scheme: About
parametrization, velocity, pressure and mixed boundary conditions,'' \emph{Communications in Computational Physics},
vol.~3, no. 2, pp.~427--478, 2008.

{[}32{]} D. d'Humières, I. Ginzburg, M. Krafczyk, P. Lallemand, and L.-S. Luo, ``Multiple-relaxation-time lattice
Boltzmann models in three dimensions,'' \emph{Philosophical Transactions of the Royal Society A}, vol.~360, no. 1792,
pp.~437--451, 2002.

{[}33{]} P. Lallemand and L.-S. Luo, ``Theory of the lattice Boltzmann method: Dispersion, dissipation, isotropy,
Galilean invariance, and stability,'' \emph{Physical Review E}, vol.~61, no. 6, pp.~6546--6562, 2000.

{[}34{]} L.-S. Luo, W. Liao, X. Chen, Y. Peng, and W. Zhang, ``Numerics of the lattice Boltzmann method: Effects of
collision models on the lattice Boltzmann simulations,'' \emph{Physical Review E}, vol.~83, no. 5, art. 056710, 2011.

{[}35{]} G. Wissocq and P. Sagaut, ``Hydrodynamic limits and numerical errors of isothermal lattice Boltzmann schemes,''
\emph{Journal of Computational Physics}, vol.~450, art. 110858, 2022.

{[}36{]} K. Suzuki and T. Inamuro, ``Effect of internal mass in the simulation of a moving body by the immersed boundary
method,'' \emph{Computers \& Fluids}, vol.~49, no. 1, pp.~173--187, 2011.

{[}37{]} Z. Guo, C. Zheng, and B. Shi, ``Discrete lattice effects on the forcing term in the lattice Boltzmann method,''
\emph{Physical Review E}, vol.~65, no. 4, art. 046308, 2002.

{[}38{]} A. De Rosis, R. Huang, and C. Coreixas, ``Universal formulation of central-moments-based lattice Boltzmann
method with external forcing for the simulation of multiphysics phenomena,'' \emph{Physics of Fluids}, vol.~31, no. 11,
art. 117102, 2019, doi:10.1063/1.5124719.

{[}39{]} H. Dütsch, F. Durst, S. Becker, and H. Lienhart, ``Low-Reynolds-number flow around an oscillating circular
cylinder at low Keulegan--Carpenter numbers,'' \emph{Journal of Fluid Mechanics}, vol.~360, pp.~249--271, 1998.

{[}40{]} R. Glowinski, T.-W. Pan, T. I. Hesla, D. D. Joseph, and J. Périaux, ``A fictitious domain approach to the
direct numerical simulation of incompressible viscous flow past moving rigid bodies: application to particulate flow,''
\emph{Journal of Computational Physics}, vol.~169, no. 2, pp.~363--426, 2001.

{[}41{]} Y. Zhang, G. Pan, Y. Zhang, and S. Haeri, ``A relaxed multi-direct-forcing immersed boundary-cascaded lattice
Boltzmann method accelerated on GPU,'' \emph{Computer Physics Communications}, vol.~248, art. 106980, 2020.

{[}42{]} D. J. Tritton, ``Experiments on the flow past a circular cylinder at low Reynolds numbers,'' \emph{Journal of
Fluid Mechanics}, vol.~6, no. 4, pp.~547--567, 1959.

{[}43{]} C. H. K. Williamson, ``Vortex dynamics in the cylinder wake,'' \emph{Annual Review of Fluid Mechanics},
vol.~28, pp.~477--539, 1996.

{[}44{]} S. C. R. Dennis and G.-Z. Chang, ``Numerical solutions for steady flow past a circular cylinder at Reynolds
numbers up to 100,'' \emph{Journal of Fluid Mechanics}, vol.~42, no. 3, pp.~471--489, 1970.

{[}45{]} B. Fornberg, ``A numerical study of steady viscous flow past a circular cylinder,'' \emph{Journal of Fluid
Mechanics}, vol.~98, no. 4, pp.~819--855, 1980.

{[}46{]} J. Park, K. Kwon, and H. Choi, ``Numerical solutions of flow past a circular cylinder at Reynolds numbers up to
160,'' \emph{KSME International Journal}, vol.~12, no. 6, pp.~1200--1205, 1998.

{[}47{]} M. Braza, P. Chassaing, and H. Ha Minh, ``Numerical study and physical analysis of the pressure and velocity
fields in the near wake of a circular cylinder,'' \emph{Journal of Fluid Mechanics}, vol.~165, pp.~79--130, 1986.

{[}48{]} Z.-G. Feng and E. E. Michaelides, ``Robust treatment of no-slip boundary condition and velocity updating for
the lattice-Boltzmann simulation of particulate flows,'' \emph{Computers \& Fluids}, vol.~38, pp.~370--381, 2009.

{[}49{]} M. R. Maxey and J. J. Riley, ``Equation of motion for a small rigid sphere in a nonuniform flow,''
\emph{Physics of Fluids}, vol.~26, no. 4, pp.~883--889, 1983.

{[}50{]} T. R. Auton, J. C. R. Hunt, and M. Prud'homme, ``The force exerted on a body in inviscid unsteady non-uniform
rotational flow,'' \emph{Journal of Fluid Mechanics}, vol.~197, pp.~241--257, 1988.

{[}51{]} J. Wu and C. Shu, ``Implicit velocity correction-based immersed boundary-lattice Boltzmann method and its
applications,'' \emph{Journal of Computational Physics}, vol.~228, no. 6, pp.~1963--1979, 2009.

{[}52{]} Q. Zou and X. He, ``On pressure and velocity boundary conditions for the lattice Boltzmann BGK model,''
\emph{Physics of Fluids}, vol.~9, no. 6, pp.~1591--1598, 1997.

{[}53{]} G. I. Taylor and A. E. Green, ``Mechanism of the production of small eddies from large ones,''
\emph{Proceedings of the Royal Society A}, vol.~158, no. 895, pp.~499--521, 1937.

{[}54{]} F.-B. Tian, H. Luo, L. Zhu, J. C. Liao, and X.-Y. Lu, ``An efficient immersed boundary-lattice Boltzmann method
for the hydrodynamic interaction of elastic filaments,'' \emph{Journal of Computational Physics}, vol.~230, no. 19,
pp.~7266--7283, 2011.

{[}55{]} R. D. Henderson, ``Details of the drag curve near the onset of vortex shedding,'' \emph{Physics of Fluids},
vol.~7, no. 9, pp.~2102--2104, 1995.

{[}56{]} R. Bouard and M. Coutanceau, ``The early stage of development of the wake behind an impulsively started
cylinder for \(40 < \mathrm{Re} < 10^4\),'' \emph{Journal of Fluid Mechanics}, vol.~101, pp.~583--607, 1980.

{[}57{]} P. Koumoutsakos and A. Leonard, ``High-resolution simulations of the flow around an impulsively started
cylinder using vortex methods,'' \emph{Journal of Fluid Mechanics}, vol.~296, pp.~1--38, 1995.

{[}58{]} C. H. K. Williamson, ``Oblique and parallel modes of vortex shedding in the wake of a circular cylinder at low
Reynolds numbers,'' \emph{Journal of Fluid Mechanics}, vol.~206, pp.~579--627, 1989.

{[}59{]} L. Qu, C. Norberg, L. Davidson, S.-H. Peng, and F. Wang, ``Quantitative numerical analysis of flow past a
circular cylinder at Reynolds number between 50 and 200,'' \emph{Journal of Fluids and Structures}, vol.~39,
pp.~347--370, 2013.

{[}60{]} C. Peng and L.-P. Wang, ``Force-amplified, single-sided diffused-interface immersed boundary kernel for correct
local velocity gradient computation and accurate no-slip boundary enforcement,'' \emph{Physical Review E}, vol.~101, no.
5, art. 053305, 2020.

{[}61{]} B. E. Griffith and N. A. Patankar, ``Immersed methods for fluid--structure interaction,'' \emph{Annual Review
of Fluid Mechanics}, vol.~52, pp.~421--448, 2020.

{[}62{]} C. Gruninger and B. E. Griffith, ``Composite B-spline regularized delta functions for the immersed boundary
method: Divergence-free interpolation and gradient-preserving force spreading,'' \emph{Journal of Computational
Physics}, vol.~546, art. 114472, 2026.

{[}63{]} M. Uhlmann and J. Dušek, ``The motion of a single heavy sphere in ambient fluid: A benchmark for
interface-resolved particulate flow simulations with significant relative velocities,'' \emph{International Journal of
Multiphase Flow}, vol.~59, pp.~221--243, 2014.

{[}64{]} Z. Xia, K. W. Connington, S. Rapaka, P. Yue, J. J. Feng, and S. Chen, ``Flow patterns in the sedimentation of
an elliptical particle,'' \emph{Journal of Fluid Mechanics}, vol.~625, pp.~249--272, 2009.

{[}65{]} H. Tabaei Kazerooni, W. Fornari, J. Hussong, and L. Brandt, ``Inertial migration in dilute and semidilute
suspensions of rigid particles in laminar square duct flow,'' \emph{Physical Review Fluids}, vol.~2, no. 8, art. 084301,
2017.

{[}66{]} D. Pu, M. Li, L. Shen, Z. Wang, and Z. Li, ``The effects of channel width on particle sedimentation in fluids
using a coupled lattice Boltzmann-discrete element model,'' \emph{Physics of Fluids}, vol.~35, no. 5, art. 053307, 2023.

{[}67{]} A. Parvan, M. Rahnama, S. Jafari, and E. Jahanshahi Javaran, ``Evaluation of force term in lattice Boltzmann
method with discrete external boundary force for flow over an immersed body,'' \emph{Particulate Science and
Technology}, vol.~38, no. 5, pp.~535--548, 2020, doi:10.1080/02726351.2018.1491486.

{[}68{]} M. Badri Ghomizad, H. Kor, and K. Fukagata, ``A structured adaptive mesh refinement strategy with a sharp
interface direct-forcing immersed boundary method for moving boundary problems,'' \emph{Journal of Fluid Science and
Technology}, vol.~16, no. 2, art. JFST0014, 2021, doi:10.1299/jfst.2021jfst0014.

{[}69{]} S. Ghosh and J. M. Stockie, ``Numerical simulations of particle sedimentation using the immersed boundary
method,'' \emph{Communications in Computational Physics}, vol.~18, no. 2, pp.~380--416, 2015.

{[}70{]} T. Kempe and J. Fröhlich, ``An improved immersed boundary method with direct forcing for the simulation of
particle laden flows,'' \emph{Journal of Computational Physics}, vol.~231, no. 9, pp.~3663--3684, 2012.

{[}71{]} J. Feng, H. H. Hu, and D. D. Joseph, ``Direct simulation of initial value problems for the motion of solid
bodies in a Newtonian fluid Part 1. Sedimentation,'' \emph{Journal of Fluid Mechanics}, vol.~261, pp.~95--134, 1994.

{[}72{]} D. Hui, Z. Xu, W. Wu, G. Zhang, Q. Wu, and M. Liu, ``Drafting, kissing, and tumbling of a pair of particles
settling in non-Newtonian fluids,'' \emph{Physics of Fluids}, vol.~34, no. 2, art. 023301, 2022.

{[}73{]} L. Wang, Z. L. Guo, and J. C. Mi, ``Drafting, kissing and tumbling process of two particles with different
sizes,'' \emph{Computers \& Fluids}, vol.~96, pp.~20--34, 2014.

{[}74{]} A. Eshghinejadfard, A. Abdelsamie, G. Janiga, and D. Thévenin, ``Direct-forcing immersed boundary lattice
Boltzmann simulation of particle/fluid interactions for spherical and non-spherical particles,'' \emph{Particuology},
vol.~25, pp.~93--103, 2016.

{[}75{]} A. J. C. Ladd, ``Numerical simulations of particulate suspensions via a discretized Boltzmann equation. Part 1.
Theoretical foundation,'' \emph{Journal of Fluid Mechanics}, vol.~271, pp.~285--309, 1994.

{[}76{]} C. K. Aidun, Y. Lu, and E.-J. Ding, ``Direct analysis of particulate suspensions with inertia using the
discrete Boltzmann equation,'' \emph{Journal of Fluid Mechanics}, vol.~373, pp.~287--311, 1998.

{[}77{]} E. Xing, G. Liu, Q. Zhang, J. Zhang, and C. Ji, ``Numerical investigation of flow past a cylinder using
cumulant lattice Boltzmann method,'' \emph{Physics of Fluids}, vol.~36, no. 3, art. 035166, 2024.

{[}78{]} Y. Wang, Y. Wu, Y. Zeng, M. Jiang, and Z. Liu, ``An immersed boundary lattice Boltzmann method on
block-structured adaptive grids for the simulation of particle-laden flows on CPUs/GPUs,'' \emph{Computer Physics
Communications}, vol.~314, art. 109674, 2025.

{[}79{]} S. Kemmler, C. Rettinger, U. Rüde, P. Cuéllar, and H. Köstler, ``Efficiency and scalability of fully-resolved
fluid-particle simulations on heterogeneous CPU-GPU architectures,'' \emph{The International Journal of High Performance
Computing Applications}, vol.~39, no. 3, pp.~345--363, 2025, doi:10.1177/10943420241313385.

{[}80{]} J. D. Sterling and S. Chen, ``Stability analysis of lattice Boltzmann methods,'' \emph{Journal of Computational
Physics}, vol.~123, no. 1, pp.~196--206, 1996.

\end{document}